\documentclass[reqno]{amsart}
\usepackage{amssymb}
\usepackage[dvips]{epsfig}
\usepackage{graphicx}
\usepackage{color}
\usepackage{amsmath}
\usepackage{amssymb}
\usepackage{amsfonts}
\usepackage{amsthm}
\usepackage{bbm}
\usepackage{mathrsfs}
\newcommand{\D}{\mathbb D}
\newcommand{\R}{\mathbb R}
\newcommand{\p}{\partial}

\newcommand{\Z}{\mathbb Z}

\newcommand{\C}{\mathbb C}

\newcommand{\ep}{\varepsilon}

\newcommand{\g}{\gamma}
\newcommand{\et}{\eta}

\renewcommand{\lg}{\Lambda}

\newcommand{\N}{\mathbb{N}}
\newcommand{\T}{\mathbb{T}}

\newcommand{\ji}{\langle}
\newcommand{\jd}{\rangle}

\newcommand{\dist}{{\rm dist}}
\newcommand{\diam}{{\rm diam}}
\newcommand{\good}{{\rm good}}

\newtheorem{thm}{Theorem}[section]
\newtheorem{lem}[thm]{Lemma}
\newtheorem{stm}[thm]{Statement}

\theoremstyle{remark}
\newtheorem{rem}{\bf Remark}[section]
\theoremstyle{definition}
\newtheorem{defn}[thm]{Definition}

\numberwithin{equation}{section}

\usepackage[colorlinks=true,pdfstartview=FitV,linkcolor=magenta,citecolor=cyan]{hyperref}
\usepackage{bm}
\begin{document}

	\title[QP Power-law  localization]{Green's function estimates for quasi-periodic operators on $\Z^d$ with power-law long-range hopping}
	
	\author[Shi]{Yunfeng Shi}
	\address[Shi] {School of Mathematics,
		Sichuan University,
		Chengdu 610064,
		China}
	\email{yunfengshi@scu.edu.cn}
	
	\author[Wen]{Li Wen}
	\address[Wen] {School of Mathematics,
		Sichuan University,
		Chengdu 610064,
		China}
	\email{liwen.carol98@gmail.com}
	\date{\today}
	\maketitle
	\begin{abstract}
		We establish quantitative   Green's function  estimates for a class of quasi-periodic (QP)  operators on $\Z^d$ with power-law long-range hopping and analytic cosine type potentials. As applications,  we prove the arithmetic version of  localization, the finite volume version of  $(\frac12-)$-H\"older continuity of the IDS,   and the  absence of eigenvalues (for Aubry dual operators).   
	\end{abstract}
	
	\section{Introduction}
	
	In this paper, we are concerned with  QP   operators 
	\begin{align}\label{model}
		\mathcal{H}(\theta)=\ep \mathcal{W}_\phi+v(\theta+\bm n\cdot\bm \omega)\delta_{\bm n,\bm n'},\ \bm n,\bm n'\in\Z^d,
	\end{align}
	where the off-diagonal  part (i.e., the hopping term) $\mathcal{W}_{\phi}$ is a Toeplitz operator  satisfying 
	\begin{align}\label{wphi}
		(\mathcal{W}_\phi \psi)(\bm n)=\sum_{\bm l\in\Z^d}\phi(\bm n-\bm l)\psi(\bm l),\ \phi(\bm 0)=0,\ |\phi(\bm n)|\le (1+\|\bm n\|)^{-\alpha}
	\end{align}
	with some $\alpha>0$ and $	\|\bm n\|=\sup\limits_{1\le i\le d}|n_i|$.   The potential $v$  is an analytic function defined on 
	\begin{align*}
		\D_R=\{z\in\C/\Z: \ |\Im z|\le R\},\ R>0,
	\end{align*}
	satisfying for some  positive constants $\kappa_1,\kappa_2>0$,  
	\begin{align}\label{vdefn}
		\kappa_1\|z_1-z_2\|_{\T}\|z_1+z_2\|_{\T}\le|v(z_1)-v(z_2)|\le \kappa_2\|z_1-z_2\|_{\T}\|z_1+z_2\|_{\T}\ {\rm for}\ \forall z_1,z_2\in\D_R, 
	\end{align} 
	where
	\begin{align*}
		\|z\|_\T=\sqrt{\|\Re z\|_\T^2+|\Im z|^2},\ \|x\|_{\T}=\inf_{l\in\Z}|l-x|\ {\rm for}\ x\in\R.
	\end{align*}
	We let $\theta\in\T=\R/\Z$, $\bm \omega\in[0,1]^d$, and $
	\bm n\cdot\bm \omega=\sum\limits_{i=1}^{d}n_i\omega_i.$ 
	In the following, we assume that $\bm \omega\in {DC}_{\tau,\g}$ for some $\tau>d$, $\g>0$ with 
	\begin{align}\label{dc}
		{DC}_{\tau,\g}=\left\{\bm\omega\in[0,1]^d:\ \|\bm n\cdot\bm \omega\|_\T\ge\frac{\g}{\|\bm n\|^\tau}\ {\rm for}\ \forall\bm n\in\Z^d\setminus\{\bm 0\}\right\}.
	\end{align}
	The special case of operators with $\mathcal{W}_\phi(m,  n)=\delta_{|m-n|, 1}$ and $v=\cos2\pi\theta$ corresponds to the famous almost Mathieu operator (AMO). 
	
	The present work aims to  establish quantitative Green's function estimates for \eqref{model} via  the multi-scale analysis (MSA) induction. As applications, the arithmetic localization, H\"older continuity of the IDS and absence of eigenvalues (for  Aubry dual operators of \eqref{model}) are proved. %Our proof combines ideas of Bourgain \cite{Bou00}, Cao-Shi-Zhang \cite{CSZ24a} and Berti-Bolle \cite{BB13}.  
	
	Over the past decades, the study of spectral and dynamical properties of discrete QP Schr\"odinger operators has been one of the central themes in mathematical physics. Of particular importance is the phenomenon of Anderson localization (i.e., the pure point spectrum with exponentially decaying eigenfunctions). Determining the nature of the spectrum and the eigenfunctions  information of QP Schr\"odinger operators  can be viewed as a small divisor problem, and substantial progress  has been made following Green’s function estimates method based on  MSA  of Fr\"ohlich-Spencer \cite{FS83}.  %We refer to \cite{CSZ24a, CSZ23, CSZ24b} for more results on the study of Anderson localization for QP Schr\"odinger operators.  
	
	In 2000,  Bourgain \cite{Bou00} first  proved the $(\frac 12-)$-H\"older continuity of the IDS for AMO in the perturbative regime via MSA type Green's function estimates. This result is remarkable   since the important work  of Goldstein-Schlag \cite{GS01} which  is non-perturbative  and  applies to  more general QP potentials does not seem to provide explicit information on the H\"older exponent on the regularity of the IDS at that time. \footnote{Until 2009, Amor \cite{Amo09} obtained the sharp $\frac12$-H\"older continuity  of the IDS for one-dimensional   QP Schr\"odinger
		operators with small analytic potentials  by using the KAM reducibility method of Eliasson \cite{Eli92} (cf. \cite{AJ10} by Avila-Jitomirskaya for a nonperturbative extension).}  However, the main contribution of \cite{Bou00} may be its method: Bourgain established quantitative Green's function estimates via MSA scheme,  and the resonances there  can be completely described by  a pair of zeros of the  Dirichlet determinant via the Weierstrass preparation theorem.  The main idea   originates from \cite{Bou97} in the area  of KAM theory  for Hamiltonian systems,  in which Bourgain revisited the Melnikov's persistency problem  and removed the second Melnikov's conditions  by using  Green's function estimates method  based on the  preparation type theorem.  This  method  (combined  with ideas of Craig-Wayne \cite{CW93}) was  later significantly extended in the breakthrough work  \cite{Bou98} to prove  the existence of  QP  solutions for a class of  2-dimensional nonlinear Schr\"odinger equations (NLS).  Definitely,  the adaptation of this method to  the field of  spectral theory of QP Schr\"odinger operators requires to deal with {\it energy parameters}  that do not appear in the nonlinear KAM setting.  It is significantly difficult to eliminate resonances between  frequencies and energies.   This task  becomes more  challenging  when dealing with  general QP potentials  even in the one-dimensional case (cf. Goldstein-Schlag \cite{GS08}).

	Bourgain's approach in \cite{Bou00} does not restrict  only to the proof of  H\"older continuity of the IDS for AMO. Indeed, in \cite{Bou00}, Bourgain remarked that ``{\it In fact, from this Green's function result, one may also recover the Anderson localization results from \cite{Sin87, FSW90} in the perturbative case. Our method is significantly different in the sense that it relies little on eigenvalues and eigenvalue perturbation theory}.''  However,  it is  nontrivial to prove  Anderson localization via directly  using estimates of \cite{Bou00}:  (1).  The oﬀ-diagonal elements  of Green’s function   in  \cite{Bou00} are only  sub-exponentially  decaying, which is not sufficient  for the proof of Anderson localization; (2). To prove Anderson localization, one has to eliminate the energy   $E\in\R$  in Green’s function estimates  by removing  additional  $\theta$.  In this step, the  {\it symmetry}  property  (which was unknown  in \cite{Bou00}) of  the two zeros (depending on $E$)  of the Dirichlet determinant  becomes essential.  Very recently, in \cite{CSZ24a}, Cao-Shi-Zhang extended fully the method of Bourgain \cite{Bou00} to prove the arithmetic  Anderson localization (i.e., there is certain arithmetic description on the set  of $\theta,\bm \omega$ on which the Anderson localization holds true;  cf. \cite{CSZ24a} for details) for QP Schr\"odinger operators on $\Z^d$ with the cosine potential. Related  results  were  previously obtained by Jitomirskaya-Kachkovskiy \cite{JK16} and Ge-You \cite{GY20} via the reducibility-localization method based on Aubry duality.  In \cite{CSZ24a},  the authors reconstructed   in each MSA induction step enlarged  resonant blocks satisfying   both  the symmetry and translation invariance properties. This suffices for eliminating  energies and double resonances,  then the proof  of  Anderson localization.  However, the arguments of \cite{CSZ24a} rely crucially on the  fast {\it exponential}  decay of  the long-range hopping  and cannot apply to the power-law one  (cf. \cite{Shi22,Liu22} for results on more general sub-exponential long-range hopping QP operators). In addition,  they only considered  the cosine potential in \cite{CSZ24a}  and it is desirable to remove  this restriction. Subsequently,  Cao-Shi-Zhang \cite{CSZ23, CSZ24b} have successfully extended the results of \cite{CSZ24a} to the more general $C^2$-cosine type potentials via  MSA type Green's function estimates based on eigenvalues and eigenvalue perturbation theory,  but required the QP operators to be self-adjoint  (cf.  Liu \cite{Liu22} for recent results on the study of non-self-adjoint QP operators).  The present work extends the method \cite{CSZ24a} to handle  non-self-adjoint QP operators with both power-law long-range hopping and more general analytic cosine type potentials.

	Back to the KAM setting, the  result  of Bourgain \cite{Bou98} was later essentially extended  (cf. chapters 19 and 20 in \cite{Bou05}) to both NLS and nonlinear wave equations  of {\it arbitrary}  space dimensions   by  establishing  large deviation type Green's function estimates, which build on  semi-algebraic geometry arguments and matrix-valued Cartan's estimates originated from the Anderson localization theory of QP Schr\"odinger operators on $\Z^2$ by Bourgain-Goldstein-Schlag \cite{BGS02} (cf. Bourgain \cite{Bou07} and Jitomirskaya-Liu-Shi \cite{JLS20} for the $\Z^d$ case). In order to apply semi-algebraic geometry method,  the nonlinearity of  PDEs in \cite{Bou05} admits some special form of  the polynomial type. Moreover, due to the weak separation  property of  resonant clusters induced by the normal frequencies, the QP solutions obtained in \cite{Bou98,Bou05} are at most Gevrey regular in both time and space variables. Later in \cite{BB13},    Berti-Bolle developed a novel approach again based on the Nash-Moser iteration and MSA type Green's function estimates  to prove the existence of Sobolev type QP solutions for a class of higher dimensional NLS with the finitely differentiable  nonlinearity.  Among others,  the work  \cite{BB13} has  to establish off-diagonal decay estimates on the Green's function  of  linearized operators   with  power-law long-range hopping rather than the  (sub)exponential or finite-range ones. For this purpose, Berti-Bolle proved  a  key coupling lemma to achieve  power-law off-diagonal  decay estimates of Green's functions.  The essential point of Berti-Bolle's proof  lies  in the usage  of {\it tame}  property induced by the power-law weight,  and  this  property is definitely invalid for the  (sub)exponential  weight.  Based on ideas of \cite{BB13},  a series of results  \cite{BB12, BB13, BCP15, BB20} toward the construction of Sobolev QP solutions for higher dimensional Hamiltonian PDEs  have  been  obtained.  As mentioned above, there are important connections between the two areas of Anderson localization and nonlinear KAM theory in the exponential (or analytic)  perturbations  case.   In \cite{BB20}, Berti-Bolle emphasized  that ``{\it The techniques developed in this monograph have deep connections with those used in Anderson-localization theory and we hope that the detailed presentation in this manuscript of all technical aspects of the proofs will allow a deeper interchange between the Anderson-localization and KAM for PDEs scientific communities}.''  It is the other  motivation of the  present work that  studies  spectral problems via the adaptation of  techniques initially developed in the community  of KAM for PDEs and tries to address the above  Berti-Bolle's problem. 
	
	In \cite{Shi21}, Shi first introduced the estimates of \cite{BB13} in the area of random Schr\"odinger operators and proved localization for random operators  with power-law long-range hopping and  the H\"older continuity distribution (maybe singularly continuous). Previously,  in the influential work \cite{AM93}, Aizenman-Molchanov first proved the  power-law localization   via the the famous fractional moment method, which requires the absolute continuity of the  random distribution when applying Simon-Wolff criterion \cite{SW86}.  Recently,  Shi \cite{Shi23} developed a Nash-Moser iteration type diagonalization method to prove localization for power-law long-range QP   operators with monotone potentials in the absence of resonances, which plays an important role in the study of quantum suppression of chaos concerning non-analytic quantum kicked rotor (cf. \cite{SW23}).  The present paper extends the method of \cite{Shi21, CSZ24a} further to investigate  the power-law long-range QP  operators    in the presence of resonances. % going beyond the finite-rank  perturbations  essential of random operators. 

	\subsection{Main results}
	The main result of this paper is a quantitative version of Green's function estimates.

	We first introduce the function class of potentials. For $R>0$, let $\mathscr{V}_R$ denote  the set of analytic functions $v$ on $\D_R=\{z\in\C/\Z:\ |\Im z|\le R\}$ satisfying \eqref{vdefn}. Let $|v|_R=\sup\limits_{z\in\D_R}|v(z)|$ for $v\in\mathscr{V}_R$ and $\mathscr{V}=\bigcup_{R>0}\mathscr{V}_R$.
	\begin{rem}\label{v1}
		\begin{itemize}
			The following functions belong to $\mathscr{V}$ (cf. the Appendix \ref{app} for a detailed proof):
			\item\textit{Example 1}:
			\begin{align*}
				v_1(z)=\cos2\pi z+\lambda_2\cos^2 2\pi z+\cdots+\lambda_n\cos^n 2\pi z,
			\end{align*}
			where $2|\lambda_2|+\cdots+n|\lambda_n|<1$.\\
			\item\textit{Example 2}: 
			\begin{align*}
				v_2(z)=\cos2\pi z+\epsilon f(z),
			\end{align*}
			where $f$ is any  even function defined on $\mathbb D_R$ and $0<|\epsilon|\leq\epsilon_0=\epsilon_0(f)\ll1$.\\
			
			%\item The class $\mathscr V$ is 
		\end{itemize}
	\end{rem}
	
	%\subsubsection{Quantitative Green's function estimates}
	% 	
	
	Given $E\in\C$ and $\lg\subset\Z^d$, the Green's function (if exists) is defined by
	\begin{align*}
		\mathcal{T}_{\lg}^{-1}(E;\theta)=(\mathcal{H}_\lg(\theta)-E)^{-1},\ \mathcal{H}_{\lg}(\theta)=\mathcal{R}_\lg \mathcal{H}(\theta)\mathcal{R}_\lg,
	\end{align*}
	where $\mathcal{H}(\theta)$ is given  by \eqref{model} and $\mathcal R_\Lambda$ denotes the restriction operator.

	Let $\ji\cdot,\cdot\jd$ denote the standard inner on $\ell^2(\Z^d) $ and let  $\mathcal{M}$ be a bounded linear operator on $\ell^2(\Z^d)$.  Write $\mathcal M(\bm m,\bm n)=\ji \delta_{\bm m}, \mathcal M\delta_{\bm n}\jd.$   Define for all $\alpha\ge0$ the Sobolev norm  as 
	\begin{align}\label{norm}
		\|\mathcal{M}\|_\alpha=\sum_{\bm k\in\Z^d}\left(\sup_{\bm l\in\Z^d}|\mathcal{M}(\bm k+\bm l,\bm l)|\right)(1+\|\bm k\|)^\alpha.
	\end{align}
	Typically, we denote by  ${\rm dist}(\cdot, \cdot)$ (resp. ${\rm diam}(\cdot)$) the distance  (resp. the diameter) induced by the norm $\|\cdot\|$ on $\Z^d$.

	Recall that $\bm \omega\in DC_{\tau,\g}$. At the $s$-th iteration step, let $\delta_s^{-1}$ (resp. $N_s$) describe the resonance strength (resp. the size of resonant blocks) defined by
	\begin{align*}
		N_{s+1}=\left[\left(\frac{\g}{\delta_s}\right)^{\frac{1}{30\tau}}\right],\ \frac{\g}{\delta_{s+1}}=\left(\frac{\g}{\delta_s}\right)^{30},\ \delta_0=\ep_0^{\frac{1}{30}},
	\end{align*}
	where $[x]$ denotes the integer part of $x\in\R$. 
	
	Then we have 
	\begin{thm}\label{ge}
		Let $\bm\omega\in DC_{\tau,\g}$. Fix $d<\alpha_0<\tau$ and $\alpha_1>2200\tau$. Let
		\begin{align*}
			v\in\mathscr{V}_R,\ \|\mathcal{W}_{\phi}\|_{\alpha_1+\alpha_0}<+\infty.
		\end{align*}
		Then there is some $\ep_0=\ep_0(\alpha_1,\alpha_0,d,\tau,\g,v,R,\phi)>0$ so that for $0<|\ep|\leq \ep_0$ and $E\in v(\D_{R/2})$, there exists a sequence 
		\begin{align*}
			\{\theta_s=\theta_s(E)\}_{s=0}^{s'}\subset\C\ (s'\in\N\cup\{+\infty\})
		\end{align*}
		with the following properties: Fix any $\theta\in \T$, if a subset $\lg\subset\Z^d$ is $s$-$\good$ (cf.  $(\bm e)_s$ of Statement \ref{state} for the definition of $s$-$\good$ set, and Section \ref{qgfe} for the definitions of $\{\theta_s\}_{s=0}^{s'}$, sets $P_s$, $Q_s$, $\tilde{\Omega}_{\bm k}^s$ and $\zeta_s>0$), then
		\begin{align*}
			\|\mathcal{T}_\lg^{-1}(E; \theta)\|_0&<\delta_{s}^{-\frac{2}{15}}\times\sup\limits_{\{\bm k\in P_s:\ \tilde{\Omega}_{\bm k}^s\subset\lg\}}(\|\theta+\bm k\cdot\bm \omega-\theta_s\|_{\T}^{-1}\cdot\|\theta+\bm k\cdot\bm\omega+\theta_s\|_{\T}^{-1}),
		\end{align*}
		and for $\alpha\in(0,\alpha_1],$
		\begin{align}\label{pwdec}
			\|\mathcal{T}_{\lg}^{-1}(E; \theta)\|_{\alpha}&<\zeta_s^{\alpha}\delta_{s}^{-\frac{14}{3}}.
		\end{align}
		In particular, for any finite set $\lg\subset\Z^d$, there exists some $\tilde{\lg}$ satisfying
		\begin{align}\label{tl}
			\lg\subset\tilde{\lg}\subset\{\bm k\in\Z^d:\ \dist(\bm k,\lg)\le 50N_s^{5}\}
		\end{align}
		so that if 
		\begin{align*}
			\min_{\bm k\in\tilde{\lg}^*}\min_{\sigma=\pm 1}(\|\theta+\bm k\cdot\bm\omega+\sigma\theta_s\|_{\T})>\delta_s,
		\end{align*}
		then
		\begin{align*}
			\|\mathcal{T}_{\tilde{\lg}}^{-1}(E; \theta)\|_0&<\delta_{s}^{-\frac{32}{15}},
		\end{align*}
		and for $\alpha\in(0,\alpha_1], $
		\begin{align*}
			\|\mathcal{T}_{\tilde{\lg}}^{-1}(E; \theta)\|_{\alpha}&<\zeta_s^{\alpha}\delta_{s}^{-\frac{14}{3}},
		\end{align*}
		where
		\begin{align*}
			\tilde{\lg}^*=\left\{\bm k\in\frac{1}{2}\Z^d:\ \dist(\bm k,\tilde{\lg})\le\frac{1}{2}\right\}.
		\end{align*}
	\end{thm}
	\begin{rem}
		\ \\
		\begin{itemize}
			\item  In \eqref{pwdec}, we have 
			$${\rm diam}(\Omega_{\bm k}^s)=\zeta_s\ll  ({\rm diam}(\tilde\Omega_{\bm k}^s))^{2/3}\leq ({\rm diam}(\Lambda))^{2/3},$$
			which leads to power-law off-diagonal decay  of $\mathcal{T}_{\lg}^{-1}(E; \theta).$
			\item The lower bound $\alpha_1>2200\tau$ may  be improved. 
			\item Let us refer to Section \ref{qgfe} for a complete description of  Green's function estimates.
		\end{itemize}
	\end{rem}

	\subsubsection{Arithmetic  localization}
	
	In this part, we  state our arithmetic  type localization results. We assume $\mathcal H(\theta)$ is self-adjoint for $\theta\in\T$. 
	
	We first introduce the power-law (spectral) localization result.
	\begin{thm}\label{apl}
		Let $d<\tau_1<\tau$ and  define  
		\begin{align*}
			\Theta_{\tau_1}=\{\theta\in\T: \ \|2\theta+\bm n\cdot\bm\omega\|_{\T}\le\frac{1}{\|\bm n\|^{\tau_1}}\ {\rm holds\  for\  finitely\  many\ } \bm n\in\Z^d\}.
		\end{align*}
		Under the assumptions  of Theorem \ref{ge},   %there exists some $$\ep_0=\ep_0(\alpha_1,\alpha_0,d,\tau,\g,v,R,\phi,\gamma_1)>0$$ such that,  for all $0<|\ep|<\ep_0$ and 
		$\mathcal{H}(\theta)$ has pure point spectrum with power-law decay eigenfunctions (i.e.,  power-law localization) for  $\theta\in\T\setminus \Theta_{\tau_1}$. 
	\end{thm}
	\begin{rem}
		The investigations of the localization for QP Schr\"odinger operators have attracted  great attention in  both one dimension (cf. e.g.,  \cite{FSW90, Jit94, Eli97, Jit99, BG00,  BJ02, Kle05, AYZ17,  JL18, JSY19,  HS22, Liu23, JL24, Han24, GJ24, FV25})  and higher dimensions (cf. e.g., \cite{CD93, Din97, BGS02, Bou07, JLS20, GY20, Shi22, Liu22,  GYZ23, CSZ23, CSZ24a, CSZ24b}).  %We also mention the recent work  \cite{CG22}  that proved  the  localization for QP operators with power-law long-range hopping and the cosine potential via reducibility-localization type method. 
	\end{rem}
	
	Our result on dynamical localization is 
	\begin{thm}\label{adl}
		Let $d<\tau_1<\tau$  and define for $A\in (0,1),$
		\begin{align}\label{TA}
			\Theta_{\tau_1,A}^*=\left\{\theta\in\T:\ \|2\theta+\bm n\cdot\bm\omega\|_{\T}>\frac{A}{\|\bm n\|^{\tau_1}}\  {\rm for} \ \forall\bm n\in\Z^d\setminus\{\bm 0\}\right\}.
		\end{align}
		%$\mathcal{H}(\theta)$ be given by \eqref{model} and self-adjoint. 
		Under the assumptions  of Theorem \ref{ge}, % there exists some $\ep_0=\ep_0(\alpha_1,\alpha_0,d,\tau,\g,v,R,\phi)>0$ such that for all $0<|\ep|<\ep_0$, the following statement holds true. Denote for $A>0$,Then for any $A>0$, $\theta\in\Theta_{\tau_1,A}^*$ and 
		we have for 
		$p\in(0,\frac{\alpha_1}{60}-2d)$ and $\theta\in \Theta_{\tau_1,A}^*$, %$H(\theta)$ satisfies the polynomially dynamical localization, i.e., there exists some $C(\alpha_1,q,d)>0$ such that for any $\bm x\in\Z^d$,
		\begin{align*}
			&\sup_{t\in\R}\sum_{\bm n\in\Z^d}(1+\|\bm n\|)^p|\ji e^{\sqrt{-1}t\mathcal{H}(\theta)}\delta_{\bm 0},\delta_{\bm n}\jd|\\
			\le& C \max(A^{-\frac{29(p+2d)}{\tau}},\ep_0^{-\frac{29(p+2d)}{\tau}}),
		\end{align*}
		where $C=C(\alpha_1, p, d)>0.$
	\end{thm}
	\begin{rem}
		For more results on the study of dynamical localization for QP operators, we refer  to \cite{CSZ23, CSZ24b} and references therein. 
	\end{rem}
	\subsubsection{$\left(\frac{1}{2}-\right)$-H\"older continuity of the IDS}
	We now  consider  the finite volume version of H\"older continuity of the IDS for $\mathcal{H}(\theta)$. We also assume $\mathcal H(\theta)$ is self-adjoint for $\theta\in\T$. 
	
	For a finite set $\lg$, denote by $\#\lg$ its cardinality. Let
	\begin{align*}
		\mathscr{N}_\lg (E;\theta)=\frac{1}{\#\lg}\#\{\lambda\in\sigma(\mathcal H_\lg(\theta)): \ \lambda\le E\}
	\end{align*} 
	and define the IDS as 
	\begin{align}\label{ids}
		\mathscr{N}(E)=\lim\limits_{N\rightarrow\infty}\mathscr{N}_{\lg_N}(E;\theta), 
	\end{align}
	where $\lg_N=\{\bm k\in\Z^d: \ \|\bm k\|\le N\}$ for $N>0$. It is well known that the limit in \eqref{ids} exists and is independent of $\theta$ for a.e. $\theta$.
	
	We have 
	\begin{thm}\label{tids}
		Let $\mu\in\left[\frac{3400\tau}{2\alpha_1+3097\tau},\frac{1}{2}\right)$. Under the assumptions  of Theorem \ref{ge},      we have for  $0<\et<\et_0(\alpha_1,\alpha_0,d,\tau,\g,v,R,\phi,\mu)$ and  sufficiently large $N$ (depending on $\et$),
		\begin{align*}
			\sup_{\theta\in\T,\ E\in\R}(\mathscr{N}_{\lg_N}(E+\et;\theta)-\mathscr{N}_{\lg_N}(E-\et;\theta))\le\et^{\frac{1}{2}-\mu}.
		\end{align*} 
		In particular, the IDS is $(\frac12-\mu)$-H\"older continuous. 
	\end{thm}
	\begin{rem}
		We refer to \cite{Liu22, CSZ23, CSZ24a, CSZ24b} and references therein for more results on the study of regularity of the IDS. 
	\end{rem}
	\subsubsection{Absence of eigenvalues}
	Let
	\begin{align*}
		h:\ \R^d/\Z^d=\T^d\rightarrow\R
	\end{align*}
	and let  $\hat{h}=\{\hat{h}(\bm n)\}_{\bm n\in\Z^d}$ satisfy 
	\begin{align*}
		\hat{h}(\bm n)=\int_{\T^d} h(\bm \theta)e^{-2\pi\sqrt{-1}\bm n\cdot\bm \theta}d\bm \theta.
	\end{align*}
	
	%We first study the multi-dimensional operator with polynomial long-range hopping and an analytic potential.
	%Assume that $u$ is finite-smooth, i.e., $u(\theta)\in C^r(\T^d,\R)$ satisfies 
	%\begin{align}\label{hu}
	%$\|\hat{u}\|_{\alpha_1+\alpha_0}<+\infty.
	%\end{align}
	
	We study the Aubry  dual operators of \eqref{model}, which read as  
	\begin{align}\label{md2}
		\tilde{\mathcal{H}}(\bm x)=\mathcal{W}_{\hat{v}}+\ep u(\bm x+l\bm \omega)\delta_{l, l'},\ \bm x\in\T^d,\  l, l'\in\Z,
	\end{align}
	where $u(\bm x)=\sum\limits_{\bm n\in\Z^d}\phi(\bm n)e^{2\pi\sqrt{-1}\bm n\cdot\bm x}.$
	
	We have 
	\begin{thm}\label{aps}
		Let $\tilde{\mathcal{H}}(\bm x)$ be defined by \eqref{md2}. Under the assumptions of Theorem \ref{ge}, we have  that $\tilde{\mathcal{H}}(\bm x)$ has no eigenvalues  for all $\bm x\in\T^d$.
	\end{thm}
	\begin{rem}
		The operator $\tilde{\mathcal{H}}(\bm x)$ can be  non-self-adjoint  with  the  exponential long-range hopping.  The potential $u$ in \eqref{md2} is finitely differentiable. This theorem extends the result of \cite{Shi22} from Gevrey regular potentials to the finitely smooth  ones  permitting a  fixed Diophantine frequency. 
	\end{rem}
	\subsection{Ideas of the proof and new ingredients}
	Our proof of Theorem \ref{ge} is based on a MSA type induction and combines ideas from \cite{Bou00, CSZ24a, Shi21, BB13}.  Once Theorem \ref{ge} was  established, the proofs of arithmetic localization, H\"older continuity of the IDS and absence of eigenvalues just follow  from  standard arguments.

	Fix $
	\theta\in\D_{R/2},\ E\in v(\D_{R/2}). 
	$ 
	%for some  $\theta_0\in\D_{R/2}$. Consider
	%f\begin{align*} 
		%f	\mathcal{T}(E;\theta)=\mathcal{H}(\theta)-E=\mathcal{D}+\ep \mathcal{W}_\phi,
		%f\end{align*}
	%fwhere
	%f\begin{align*}
		%f	\mathcal{D}=\mathcal{D}_{\bm n}\delta_{\bm n,\bm n'},\ \mathcal{D}_{\bm n}=v(\theta+\bm n\cdot\bm\omega)-E.
		%f\end{align*}
	%fFor simplicity, we may omit the dependence of $\mathcal{T}(E;\theta)$ on $E,\theta$ and that of $\mathcal{W}_\phi$ on $\phi$, respectively.
	We use a MSA  induction scheme to establish  quantitative estimates on Green's functions. Of particular importance is the analysis of resonances, which will be described by zeros of certain functions appearing as perturbations of some quadratic polynomials. Roughly speaking, at the $s$-th iteration step, the set $Q_s\subset\frac{1}{2}\Z^d$ of singular sites will be  described by a pair of symmetric zeros of certain functions, i.e.,
	\begin{align*}
		Q_s=\{\bm k\in P_s:\ \min_{\sigma=\pm1}\|\theta+\bm k\cdot\bm\omega+\sigma\theta_s(E)\|_{\T}<\delta_s\}.
	\end{align*}
	While Green's functions restricted on $Q_s$ can not be generally well controlled, the algebraic structure of $Q_s$ combined with the Diophantine condition of $\bm\omega$ can  lead to fine separation property of singular sites. As a result, one can cover $Q_s$ with a new generation of resonant blocks ${\Omega}_{\bm k}^{s+1}\ (\bm k\in P_{s+1})$ and enlarged resonant blocks $\tilde{\Omega}_{\bm k}^{s+1}\ (\bm k\in P_{s+1})$. It turns out that one can control $\|\mathcal{T}_{\tilde{\Omega}_{\bm k}^{s+1}}^{-1}(E; \theta)\|_0$ via zeros $\pm\theta_{s+1}(E)$ of some new functions which are also perturbations of quadratic polynomials in the sense that
	\begin{align*}
		\|\mathcal{T}_{\tilde{\Omega}_{\bm k}^{s+1}}^{-1}(E; \theta)\|_0<\delta_{s+1}^{-\frac{1}{15}}\|\theta+\bm k\cdot\bm\omega-\theta_{s+1}(E)\|_{\T}^{-1}\cdot\|\theta+\bm k\cdot\bm\omega+\theta_{s+1}(E)\|_{\T}^{-1}.
	\end{align*} 
	The key point is that some $\mathcal{T}_{\tilde{\Omega}_{\bm k}^{s+1}}^{-1}(E; \theta)$ may become    controllable at the $(s+1)$-th step while $\tilde{\Omega}_{\bm k}^{s+1}$ intersects $Q_s$. Moreover, the completely uncontrollable singular sites form the $(s+1)$-th singular ones, i.e.,
	\begin{align*}
		Q_{s+1}=\{\bm k\in P_{s+1}:\ \min_{\sigma=\pm1}\|\theta+\bm k\cdot\bm\omega+\sigma\theta_{s+1}(E)\|_{\T}<\delta_{s+1}\}.
	\end{align*}

	%To obtain quantitative Green's functions, we first construct in each induction step the resonant blocks (i.e., $\Omega_{\bm k}^s$) and enlarged resonant blocks (i.e., $\tilde\Omega_{\bm k}^s$). Due to the Morse type assumption on the potential, we can control Green's functions on enlarged resonant blocks via  a pair of symmetrical roots of the Dirichlet determinant obtained via Cramer's rule.  In \cite{Bou00}, Bourgain used the Weierstrass preparation theorem to represent the roots of the  Dirichlet determinant  as that of some polynomials of degree at most two, while in \cite{CSZ24a}, the roots were obtained via the Rouch\'e's theorem. The present work would like to use  the method of \cite{CSZ24a}.  Definitely, once the roots of the Dirichlet determinant  were found, the resonant and enlarged resonant blocks can be constructed again via the  idea of \cite{CSZ24a}: One can ensure the symmetry and translation invariance  properties of the enlarged resonant blocks, which  depends crucially on  the separation  property (which is described  arithmetically via the two roots)  of resonant points,  and  we can identify the locations of resonances. 
	
	The reduction of finding zeros (i.e., $\pm\theta_s(E)$) of the Dirichlet determinant  to that of perturbations of certain polynomials of degree at most 2 relies on Green's function estimates on enlarged resonant blocks of previous induction steps and the Schur complement argument. In the case of Laplacian or  (sub)exponential long-range hopping, this reduction can be completed via just  iterating the resolvent identities, which turns out  to be quite standard. However, if the hopping  has a slower power-law decay, this argument of iterating  resolvent identities cannot work. To overcome this difficulty, we use ideas from \cite{BB13,Shi21}, in which Green's function estimates can be established  via perturbing  the left inverses. This {\it constructive}  method  can avoid  multiple steps of iterations.  Once the zeros were determined via this novel approach  that {\it directly guesses   out} the estimates  of  Green's functions on enlarged  resonant blocks via Green's function estimates on smaller induction scales enlarged resonant blocks,  it remains to handle estimates of Green's functions on more general {\it good}   subsets that contain no {\it bad} enlarged resonant blocks of certain induction scales. Again in this procedure, we do not iterate the resolvent identities, but only focus on the left inverses.

	While our method here is motivated by \cite{BB13, Shi21}, there are some major differences: (1). Since we are dealing with spectral problems, the resonances between energies and frequencies are inevitable, and we have to establish estimates valid for all Diophantine frequencies, all phases  $\theta\in\T$ and all $E\in v(\mathbb D_{R/2})$. In \cite{BB13}, the separation  property of resonant blocks which is essential for the off-diagonal decay estimates   was obtained by removing additional tangential  frequencies along the Diophantine directions;  (2). The key coupling lemma (cf. Proposition 4.1, page 249 of \cite{BB13}) cannot be used in the present:  Since we are in  the case of both QP potentials and non-self-adjoint operators, there is no prior control on the growth of the operator norm (by using eigenvalues variations argument)  of Green's functions on general blocks.  To overcome this difficulty, we  first establish Green's function estimates on {\it good} enlarged resonant blocks (namely, on which the operator norm of the Green's function can be controlled via roots of the Dirichlet determinant) via  taking account of  estimates obtained in  {\it all} previous induction scales. Then we cover general {\it good} sets with {\it good} enlarged resonant blocks and  there is no need to deal with {\it bad} resonant blocks in this case; (3). Note that the definition of Sobolev norm \eqref{norm}  in the present is different from that of Berti-Bolle \cite{BB13}, in which they  used the $\ell^2$-type norm (cf. Definition 3.2, page 244  of  \cite{BB13}).  We do think it is more suitable to use the  $\ell^1$-type norm of \eqref{norm} in the study of spectral problems, while the $\ell^2$-type one may be better in the study of nonlinear PDEs.  Apart from this consideration, we also give self-contained   proofs  of  all needed estimates concerning our definition.   
	
	We believe the method developed in the present paper may have potential applications in  proving localization type results for QP   Schr\"odinger operators with more general potentials beyond the  analytic cosine type ones.

	\subsection{Structure of the paper}
	The paper is organized as follows.  We introduce some important properties  concerning Sobolev type norms in \S \ref{Sob}.  In \S \ref{qgfe}, we establish the quantitative Green's function estimates via the MSA induction.  In \S \ref{Loc}--\S \ref{Abs},  we will apply quantitative Green's function estimates to prove power-law localization, dynamical localization, the H\"older continuity of the IDS and absence of eigenvalues, respectively. Some useful estimates are given in the appendix.

	\section{The notation}
	\begin{itemize}
		\item Given $A,B\ge0$, we write $A\lesssim B$ (resp. $A\gtrsim B$) if there is some $C=C(\alpha_1,\alpha_0,d,\tau,\g,v,R,\phi)>0$ depending on $\alpha_1,\alpha_0,d,\tau,\g,v,R,\phi$ so that $A\le CB$ (resp. $A\ge CB$). We also define
		\begin{align*}
			A\sim B\Leftrightarrow\frac{1}{C}<\frac{A}{B}<C. 
		\end{align*}
		
		\item The determinant of a matrix $M$ is denoted by $\det M$.
		
		\item If $a\in\R$, let $\|a\|_{\T}=\dist(a,\Z)=\inf\limits_{l\in\Z}|l-a|$. For $z=a+\sqrt{-1}b\in\C$ with $a,b\in\R$, define $\|z\|_{\T}=\sqrt{\|a\|_{\T}^2+|b|^2}$.
		
		\item For $\bm n\in\R^d$, let 
		\begin{align*}
			\|\bm n\|=\sup_{1\le i\le d}|n_i|.
		\end{align*}
		Denoted by $\dist(\cdot,\cdot)$ the distance induced by $\|\cdot\|$ on $\R^d$, and define
		\begin{align*}
			\diam(\lg)=\sup_{\bm k,\bm k'\in\lg}\|\bm k-\bm k'\|.
		\end{align*}
		Given $\bm n\in\Z^d$, $\lg'\subset\frac{1}{2}\Z^d$ and $L>0$, define
		\begin{align*}
			\lg_L(\bm n)=\{\bm k\in\Z^d:\ \|\bm k-\bm n\|\le L\}
		\end{align*}
		and
		\begin{align*}
			\lg_L(\lg')=\{\bm k\in\Z^d:\ \dist(\bm k,\lg')\le L\}.
		\end{align*}
		In particular, write $\lg_L=\lg_L(\bm 0)$.
		
		\item $\{\delta_{\bm x}\}_{\bm x\in\Z^d}$ is the standard basis of $\ell^2(\Z^d)$.
		
		\item $\mathcal{I}$ typically denotes the identity operator.
		
		\item $\mathcal{R}_\lg$ is the restriction operator  with  $\lg\subset\Z^d$.
		
		\item Let $\mathcal{T}:\ \ell^2(\Z^d)\to\ell^2(\Z^d)$ be a linear operator. Denote by $\ji\cdot,\cdot\jd$ the standard inner product on $\ell^2(\Z^d)$. Set $\mathcal{T}(\bm x,\bm y)=\ji \delta_{\bm x}, \mathcal{T}\delta_{\bm y}\jd$. The spectrum of operator $\mathcal{T}$ is denoted by $\sigma(\mathcal{T})$. Finally, we define $\mathcal{T}^{\{\bm k\}}=\mathcal{R}_{\{\bm k\}}\mathcal{T}$, $\mathcal{T}_\lg=\mathcal{R}_\lg \mathcal{T}\mathcal{R}_\lg$ and $\mathcal{T}_{\lg}^*$ the adjugate operator of $\mathcal{T}_{\lg}$, where $\{\bm k\},\lg\subset\Z^d.$ 
		
		\item For $\lg_1,\lg_2\subset\Z^d$, denote by $\mathbf{M}_{\lg_2}^{\lg_1}$ the set of all operators $\mathcal{R}_{\lg_1}\mathcal{M}\mathcal{R}_{\lg_2}$ with $\mathcal{M}:\ \ell^2(\Z^d)\to\ell^2(\Z^d)$ being   linear operators.
		
		\item Define for $\alpha\ge0$ the Sobolev norm of $\psi=\{\psi(\bm k)\}\in\C^{\Z^d}$ as
		\begin{align*}
			\|\psi\|_\alpha=\sum_{\bm k\in\Z^d}|\psi(\bm k)|(1+\|\bm k\|)^\alpha.
		\end{align*}

	\end{itemize}
	
	\section{Preliminaries}\label{Sob}
	In this section, we will introduce some important facts on Sobolev norms  \eqref{norm}.  %Note that the definition in the present is slightly different from that of Berti-Bolle \cite{BB13}, in which they  used the $\ell^2$-type norm (cf. Definition 3.2, page 244  of  \cite{BB13}).  We do think it is more suitable to use a $\ell^1$-type norm of \eqref{norm} in the study of spectral problems, while the $\ell^2$-type one may be better in the study of nonlinear PDEs.  Apart from this consideration, we also give self-contained (and short)   proofs  of  all needed estimates concerning our definition. 

	\subsection{Tame property}
	The norm defined by \eqref{norm} has the following important tame property.
	\begin{lem}[Tame property]\label{tp}
		For any $n\ge1$ and $\alpha\ge0$,  we have 
		\begin{align}\label{tame}
			\|\prod_{i=1}^n \mathcal{M}_i\|_\alpha\le K(n,\alpha)\sum_{i=1}^n\left(\prod_{j\ne i}\|\mathcal{M}_j\|_0\right)\|\mathcal{M}_i\|_\alpha, 
		\end{align}
		where
		\begin{align}\label{kns}
			K(n,\alpha)=n^{\max(0,\alpha-1)}\ge1.
		\end{align}
	\end{lem}
	\begin{proof}
		We refer to the Appendix \ref{app} for a detailed proof.
	\end{proof}
	\subsection{Smoothing property}
	The smoothing property plays an essential role in our estimates. In the present,  we have
	\begin{lem}[Smoothing property]\label{spl}
		For $\alpha\ge \alpha'\ge0$, $N\ge0$, we have 
		\begin{align}\label{smo1}
			\mathcal{M}(\bm k,\bm k')=0\ \text{\rm for}\ \|\bm k-\bm k'\|\le N\ \Rightarrow\ \|\mathcal{M}\|_{\alpha'}\le (1+N)^{-(\alpha-\alpha')}\|\mathcal{M}\|_{\alpha},
		\end{align}
		and
		\begin{align}\label{smo2}
			\mathcal{M}(\bm k,\bm k')=0\ \text{\rm for}\ \|\bm k-\bm k'\|\ge N\ \Rightarrow\ \|\mathcal{M}\|_{\alpha}\le (1+N)^{\alpha-\alpha'}\|\mathcal{M}\|_{\alpha'}.
		\end{align}
	\end{lem}
	\begin{proof}
		By the definition \eqref{norm}, if $\mathcal{M}(\bm k,\bm k')=0\ \text{for}\ \|\bm k-\bm k'\|\le N$,  we have
		\begin{align*}
			\|\mathcal{M}\|_{\alpha'}&=\sum_{\|\bm k\|> N}\left(\sup_{\bm l\in\Z^d}|\mathcal{M}(\bm k+\bm l,\bm l)|\right)(1+\|\bm k\|)^{\alpha'}\\
			&\le(1+N)^{-(\alpha-\alpha')}\sum_{\|\bm k\|> N}\left(\sup_{\bm l\in\Z^d}|\mathcal{M}(\bm k+\bm l,\bm l)|\right)(1+\|\bm k\|)^{\alpha}\\
			&=(1+N)^{-(\alpha-\alpha')}\|\mathcal{M}\|_{\alpha}.
		\end{align*}
		Similarly, we can obtain
		\begin{align*}
			\|\mathcal{M}\|_{\alpha}\le (1+N)^{\alpha-\alpha'}\|\mathcal{M}\|_{\alpha'} 
		\end{align*}
		if $\mathcal{M}(\bm k,\bm k')=0\ \text{for}\ \|\bm k-\bm k'\|\ge N$.
	\end{proof}
	We also need the rows estimate  %In the next lemma we bound the $\alpha$-norm of a operator in terms of the $(\alpha+\alpha_0)$-norm of its diagonal lines.
	\begin{lem}[Rows estimate]
		Let $\mathcal{M}\in\mathbf{M}_C^B$. Then for all $\alpha_0>d$, we have 
		\begin{align}\label{re}
			\|\mathcal{M}\|_\alpha\le B_1(\alpha_0)\max\limits_{\bm k\in B}\|\mathcal{M}^{\{\bm k\}}\|_{\alpha+\alpha_0},
		\end{align}
		where $B_1(\alpha_0)=\sum\limits_{\bm k\in\Z^d}(1+\|\bm k\|)^{-\alpha_0}$.
	\end{lem}
	\begin{proof}
		For $\bm k\in B,\bm i\in\Z^d$ and $\alpha_0>d$,  we get 
		\begin{align*}
			|\mathcal{M}(\bm k,\bm i)|\le\frac{\|\mathcal{M}^{\{\bm k\}}\|_{\alpha+\alpha_0}}{(1+\|\bm k-\bm i\|)^{\alpha+\alpha_0}}\le\frac{\max\limits_{\bm k\in B}\|\mathcal{M}^{\{\bm k\}}\|_{\alpha+\alpha_0}}{(1+\|\bm k-\bm i\|)^{\alpha+\alpha_0}}.
		\end{align*}
		As a consequence,
		\begin{align*}
			\|\mathcal{M}\|_\alpha&=\sum_{\bm l\in\Z^d}\left(\sup_{\bm k\in B}|\mathcal{M}(\bm k,\bm k-\bm l)|\right)(1+\|\bm l\|)^\alpha\\
			&\le\max\limits_{\bm k\in B}\|\mathcal{M}^{\{\bm k\}}\|_{\alpha+\alpha_0}\left(\sum_{\bm l\in\Z^d}(1+\|\bm l\|)^{-\alpha_0}\right)\\
			&=B_1(\alpha_0)\max\limits_{\bm k\in B}\|\mathcal{M}^{\{\bm k\}}\|_{\alpha+\alpha_0}.
		\end{align*}
	\end{proof}
	\subsection{Perturbation argument}
	It will be convenient to use the notion of left invertible operators.
	\begin{defn}[Left inverse]
		An  operator $\mathcal{M}\in\mathbf{M}_B^C$ is left invertible if there is some $\mathcal{N}\in\mathbf{M}_C^B$ such that $\mathcal{N}\mathcal{M}=\mathcal{I}_B$. Then $\mathcal{N}$ is called a \textit{left inverse} of $\mathcal{M}$.
	\end{defn}
	We shall often use the following perturbation lemma concerning left invertible operators.
	\begin{lem}[Perturbation argument]\label{pa} If $\mathcal{M}\in\mathbf{M}_B^C$ has a left inverse $\mathcal{N}\in\mathbf{M}_C^B$, then for all $\mathcal{P}\in\mathbf{M}_B^C$ with $\|\mathcal{N}\|_0\|\mathcal{P}\|_0\le\frac{1}{2}$, the operator $\mathcal{M}+\mathcal{P}$ has a left inverse $\mathcal{N}_{\mathcal{P}}$  satisfying 
		\begin{align}
			\label{pa0}\|\mathcal{N}_{\mathcal{P}}\|_0&\le2\|\mathcal{N}\|_0,\\
			\label{paa}\|\mathcal{N}_{\mathcal{P}}\|_\alpha&\le B_2(\alpha)\left(\|\mathcal{N}\|_\alpha+\|\mathcal{N}\|_0^2\|\mathcal{P}\|_\alpha\right),
		\end{align}
		where $B_2(\alpha)=K(2,\alpha)\left(3+\sum_{i=1}^{\infty}\frac{K(2i,\alpha)}{2^{i-1}}\right)$ and $K(n,\alpha)$ is defined by  \eqref{kns}.
	\end{lem}
	\begin{proof}
		Since $\|\mathcal{N}\mathcal{P}\|_0\le\|\mathcal{N}\|_0\|\mathcal{P}\|_0\le\frac{1}{2}$ by  \eqref{tame}, we have via  using the Neumann series argument 
		\begin{align*}
			(\mathcal{I}_B+\mathcal{N}\mathcal{P})^{-1}=\mathcal{I}_B+\sum_{i=1}^{\infty}(-\mathcal{N}\mathcal{P})^{i}, 
		\end{align*}
		and again by \eqref{tame}, 
		\begin{align*}
			\|(\mathcal{I}_B+\mathcal{N}\mathcal{P})^{-1}\|_0&\le 1+\sum_{i=1}^{\infty}(\|\mathcal{N}\mathcal{P}\|_0)^{i}\le2,\\
			\|(\mathcal{I}_B+\mathcal{N}\mathcal{P})^{-1}\|_\alpha&\le 1+\sum_{i=1}^{\infty}K(2i,\alpha)\left(\|\mathcal{N}\|_0^i\|\mathcal{P}\|_0^{i-1}\|\mathcal{P}\|_\alpha+\|\mathcal{P}\|_0^i\|\mathcal{N}\|_0^{i-1}\|\mathcal{N}\|_\alpha\right)\\
			&\le1+\sum_{i=1}^{\infty}\frac{K(2i,\alpha)}{2^{i-1}}\left(\|\mathcal{N}\|_0\|\mathcal{P}\|_\alpha+\|\mathcal{N}\|_\alpha\|\mathcal{P}\|_0\right).
		\end{align*}

		Set $\mathcal{N}_{\mathcal{P}}=(\mathcal{I}_B+\mathcal{N}\mathcal{P})^{-1}\mathcal{N}$. We  get
		\begin{align*}
			\mathcal{N}_{\mathcal{P}}(\mathcal{M}+\mathcal{P})&=(\mathcal{I}_B+\mathcal{N}\mathcal{P})^{-1}(\mathcal{N}\mathcal{M}+\mathcal{N}\mathcal{P})\\
			&=(\mathcal{I}_B+\mathcal{N}\mathcal{P})^{-1}(\mathcal{I}_B+\mathcal{N}\mathcal{P})=\mathcal{I}_B,
		\end{align*}
		which implies $\mathcal{N}_{\mathcal{P}}$ is a left inverse of $\mathcal{M}+\mathcal{P}$. According to \eqref{tame}, we obtain
		\begin{align*}
			\|\mathcal{N}_{\mathcal{P}}\|_0&\le\|(\mathcal{I}_B+\mathcal{N}\mathcal{P})^{-1}\|_0\|\mathcal{N}\|_0\le 2\|\mathcal{N}\|_0,\\
			\|\mathcal{N}_{\mathcal{P}}\|_\alpha&\le K(2,\alpha)\left(\|(\mathcal{I}_B+\mathcal{N}\mathcal{P})^{-1}\|_0\|\mathcal{
				N}\|_\alpha+\|(\mathcal{I}_B+\mathcal{N}\mathcal{P})^{-1}\|_\alpha\|\mathcal{N}\|_0\right)\\
			&\le K(2,\alpha)\left(\left(2+\sum_{i=1}^{\infty}\frac{K(2i,\alpha)}{2^{i}}\right)\|N\|_\alpha+\|\mathcal{N}\|_0+\sum_{i=1}^{\infty}\frac{K(2i,\alpha)}{2^{i-1}}\|\mathcal{N}\|_0^2\|P\|_\alpha\right)\\
			&\le K(2,\alpha)\left(3+\sum_{i=1}^{\infty}\frac{K(2i,\alpha)}{2^{i-1}}\right)\left(\|\mathcal{N}\|_\alpha+\|\mathcal{N}\|_0^2\|\mathcal{P}\|_\alpha\right).
		\end{align*}
		
		This finishes the proof. 
	\end{proof}
	
	\subsection{Hadamard  type estimate}
	We  need to estimate $0$-norm of the  inverse of some operator $\mathcal{S}_{\lg}$. By the Cramer's rule,   $\mathcal{S}_{\lg}^{-1}=(\det \mathcal{S}_{\lg})^{-1}S_{\lg}^*$, where $\mathcal{S}_{\lg}^*(\bm i,\bm j)$ is a determinant for $\bm i,\bm j\in\lg$. Therefore,  we can apply Hadamard's inequality to estimate $0$-norm of $\mathcal{S}_{\lg}^*$ and thus  that of $\mathcal{S}_{\lg}^{-1}$.
	\begin{lem}[Hadamard's estimate]\label{chi} 
		Let $\mathcal{S}:\ \ell^2(\Z^d)\to\ell^2(\Z^d)$ be a linear operator and let $\lg$ be a finite subset of $\Z^d$. Then for any $\bm i,\bm j\in\lg$, we have 
		\begin{align*}
			|\ji \delta_{\bm i}, \mathcal{S}_{\lg}^* \delta_{\bm j}\jd|\le \|\mathcal{S}_{\lg}\|_0^{\#\lg-1}.
		\end{align*}
		Moreover,  
		\begin{align*}
			\|\mathcal{S}_{\lg}^*\|_0\le(\#\lg)^2\|\mathcal{S}_{\lg}\|_0^{\#\lg-1}. 
		\end{align*}
	\end{lem}
	\begin{proof}
		We refer to the Appendix \ref{app} for a detailed proof.
	\end{proof}
	
	\subsection{Off-diagonal decay of Green's functions}
	%To illustrate localization results and absence of eigenvalues, we need the following lemma:
	\begin{lem}
	Assume $\lg'\subset\lg\subset\Z^d$ and $\mathcal{F}:\ \ell^2(\Z^d)\rightarrow\ell^2(\Z^d)$ is a linear operator. If $\bm n\in \lg$ and $\dist(\bm n,\lg')\ge \frac{1}{100}\diam(\lg)$, then we have 
	\begin{align}\label{oddgf}
		\sum_{\bm n'\in\lg'}|\mathcal{F}_{\lg}(\bm n,\bm n')|\lesssim(\diam(\lg))^{-\alpha_1}\|\mathcal{F}_{\lg}\|_{\alpha_1}.
	\end{align} 
	\end{lem}
\begin{proof}
	Since \eqref{norm} and $\dist(\bm n,\lg')\ge \frac{1}{100}\diam(\lg)$, we have
	\begin{align*}
		\sum_{\bm n'\in\lg'}|\mathcal{F}_{\lg}(\bm n,\bm n')|&\lesssim(\diam(\lg))^{-\alpha_1}\sum_{\bm n'\in\lg'}|\mathcal{F}_{\lg}(\bm n,\bm n')|(1+\|\bm n-\bm n'\|)^{\alpha_1}\\
		&\le(\diam(\lg))^{-\alpha_1}\sum_{\bm n'\in\lg}|\mathcal{F}_{\lg}(\bm n,\bm n')|(1+\|\bm n-\bm n'\|)^{\alpha_1}\\
		&\le(\diam(\lg))^{-\alpha_1}\|\mathcal{F}_{\lg}\|_{\alpha_1}.
	\end{align*}
\end{proof}

	\section{Quantitative Green's function estimates}\label{qgfe}
	In this section, we fix
	\begin{align*}
		\theta\in\D_{R/2},\ E\in v(\D_{R/2}).
	\end{align*}
	Write
	\begin{align}\label{theta0}
		E=v(\theta_0)
	\end{align}
	for some  $\theta_0\in\D_{R/2}$. Consider
	\begin{align}\label{T}
		\mathcal{T}(E;\theta)=\mathcal{H}(\theta)-E=\mathcal{D}+\ep \mathcal{W}_\phi,
	\end{align}
	where
	\begin{align*}
		\mathcal{D}=\mathcal{D}_{\bm n}\delta_{\bm n,\bm n'},\ \mathcal{D}_{\bm n}=v(\theta+\bm n\cdot\bm\omega)-E.
	\end{align*}
	
	For simplicity, we may omit the dependence of $\mathcal{T}(E;\theta)$ on $E,\theta$ and that of $\mathcal{W}_\phi$ on $\phi$, respectively.

	Now we introduce  the statement of our main result on the  MSA type Green's function estimates. Define  the induction parameters 
	\begin{align}\label{indpa}
		N_{s+1}=\left[\left(\frac{\g}{\delta_s}\right)^{\frac{1}{30\tau}}\right],\ \frac{\g}{\delta_{s+1}}=\left(\frac{\g}{\delta_s}\right)^{30},\ \delta_0=\ep_0^{\frac{1}{30}}.
	\end{align} 
	
	We first introduce the following induction  statement.
	\begin{stm}[called $\mathscr{P}_s\ (s\ge1)$]\label{state}
	\end{stm}
	Let
	\begin{align*}
		Q_{s-1}^{\pm}&=\{\bm k\in P_{s-1}:\ \|\theta+\bm k\cdot\bm\omega\pm\theta_{s-1}\|_{\T}<\delta_{s-1}\},\ Q_{s-1}=Q_{s-1}^+\cup Q_{s-1}^-,\\
		\tilde{Q}_{s-1}^{\pm}&=\{\bm k\in P_{s-1}:\ \|\theta+\bm k\cdot\bm\omega\pm\theta_{s-1}\|_{\T}<\delta_{s-1}^{\frac{2}{3}}\},\ \tilde{Q}_{s-1}=\tilde{Q}_{s-1}^+\cup \tilde{Q}_{s-1}^-.
	\end{align*}
	We distinguish two cases:
	\begin{align}\label{c1}
		(\bm C1)_{s-1}:\ \dist(\tilde{Q}_{s-1}^-,Q_{s-1}^+)>100N_s^3
	\end{align}
	and
	\begin{align}\label{c2}
		(\bm C2)_{s-1}:\ \dist(\tilde{Q}_{s-1}^-,Q_{s-1}^{+})\le100N_s^3.
	\end{align}
	
	Let
	\begin{align*}
		\Z^d\ni \bm l_{s-1}=\left\{\begin{array}{lc}
			\bm 0,& \text{if \eqref{c1} holds true},\\
			\bm i_{s-1}-\bm j_{s-1}, & \text{if \eqref{c2} holds true},
		\end{array}\right.
	\end{align*}
	where $\bm i_{s-1}\in Q_{s-1}^+$ and $\bm j_{s-1}\in \tilde{Q}_{s-1}^-$ such that $\|\bm i_{s-1}-\bm j_{s-1}\|\le 100N_s^3$ in $(\bm C2)_{s-1}$. Set $\Omega_{\bm k}^0=\{\bm k\}\ (\bm k\in\Z^d)$. Let $\lg\subset\Z^d$ be a finite set. We say $\lg$ is $(s-1)$-$\good$ iff
	\begin{align*}
		\left\{\begin{array}{l}
			\bm k'\in Q_{s'},\ \tilde{\Omega}_{\bm k'}^{s'}\subset\lg,\ \tilde{\Omega}_{\bm k'}^{s'}\subset\Omega_{\bm k}^{s'+1}\Rightarrow\tilde{\Omega}_{\bm k}^{s'+1}\subset\lg,\ \text{for}\ s'<s-1,\\
			\{\bm k\in P_{s-1}:\ \tilde{\Omega}_{\bm k}^{s-1}\subset\lg\}\cap Q_{s-1}=\emptyset.
		\end{array}\right.
	\end{align*}
	
	Then we have 
	
	$(\bm a)_s:$ There is $P_s\subset\frac{1}{2}\Z^d$ so that the following holds true. In the case of $(\bm C1)_{s-1}$, we have
	\begin{align}\label{as1}
		P_s=Q_{s-1}\subset\left\{\bm k\in\Z^d+\frac{1}{2}\sum_{i=0}^{s-1}\bm l_i:\ \min_{\sigma=\pm1}\|\theta+\bm k\cdot\bm\omega+\sigma\theta_{s-1}\|_{\T}<\delta_{s-1}\right\}.
	\end{align}
	For the case of $(\bm C2)_{s-1}$, we have
	\begin{align}\label{as2}
		\begin{array}{l}
			P_s\subset\left\{\bm k\in\Z^d+\frac{1}{2}\sum\limits_{i=0}^{s-1}\bm l_i:\ \|\theta+\bm k\cdot\bm\omega\|_{\T}<3\delta_{s-1}^{\frac{2}{3}}\right\},\\
			\text{or}\ P_s\subset\left\{\bm k\in\Z^d+\frac{1}{2}\sum\limits_{i=0}^{s-1}\bm l_i:\ \|\theta+\bm k\cdot\bm\omega+\frac{1}{2}\|_{\T}<3\delta_{s-1}^{\frac{2}{3}}\right\}.
		\end{array}
	\end{align}
	For every $\bm k\in P_s$, we can find a resonant block $\Omega_{\bm k}^s\subset\Z^d$ and the  enlarged resonant  block $\tilde{\Omega}_{\bm k}^s\subset\Z^d$ with the following properties. If \eqref{c1} holds true, then
	\begin{align*}
		&\lg_{N_s}(\bm k)\subset\Omega_{\bm k}^s\subset\lg_{N_s+50N_{s-1}^{5}}(\bm k),\\
		&\lg_{N_s^3}(\bm k)\subset\tilde{\Omega}_{\bm k}^s\subset\lg_{N_s^3+50N_{s-1}^{5}}(\bm k),
	\end{align*}
	and if \eqref{c2} holds true, then
	\begin{align*}
		&\lg_{N_s^3}(\bm k)\subset\Omega_{\bm k}^s\subset\lg_{N_s^3+50N_{s-1}^5}(\bm k),\\
		&\lg_{N_s^5}(\bm k)\subset\tilde{\Omega}_{\bm k}^s\subset\lg_{N_s^5+50N_{s-1}^5}(\bm k).
	\end{align*}
	These resonant blocks are constructed to satisfy the following two properties:\\ 
	$(\bm a1)_s$: 
	\begin{align}\label{a1s}
		\left\{\begin{array}{l}
			\Omega_{\bm k}^s\cap\tilde{\Omega}_{\bm k'}^{s'}\ne\emptyset\ (s'<s)\Rightarrow\tilde{\Omega}_{\bm k'}^{s'}\subset\Omega_{\bm k}^s,\\
			\tilde{\Omega}_{\bm k}^s\cap\tilde{\Omega}_{\bm k'}^{s'}\ne\emptyset\ (s'<s)\Rightarrow\tilde{\Omega}_{\bm k'}^{s'}\subset\tilde{\Omega}_{\bm k}^s,\\
			\dist(\tilde{\Omega}_{\bm k}^s,\tilde{\Omega}_{\bm k'}^s)>10\diam(\tilde{\Omega}_{\bm k}^s)\ \text{for}\ \bm k\ne\bm k'\in P_s.
		\end{array}\right.
	\end{align}\\
	$(\bm a2)_s$:  The translation of $\tilde{\Omega}_{\bm k}^s$
	\begin{align*}
		\tilde{\Omega}_{\bm k}^s-\bm k\subset\Z^d+\frac{1}{2}\sum_{i=0}^{s-1}\bm l_i,
	\end{align*}
	is independent of $\bm k\in P_s$ and symmetrical about the origin.  
	
	We denote 
	\begin{align}\label{zetas}
		\zeta_s=\diam(\Omega_{\bm k}^s), \ \tilde{\zeta}_s=\diam(\tilde{\Omega}_{\bm k}^s).
	\end{align}
	%We sometimes simplify $$ and $\diam(\tilde{\Omega}_{\bm k}^s)$ by $\zeta_s$ and $\tilde{\zeta}_s$ respectively .
	
	$(\bm b)_s$:  $Q_{s-1}$ is covered by $\Omega_{\bm k}^s\ (\bm k\in P_s)$ in the sense that for every $\bm k'\in Q_{s-1}$, there exists a $\bm k\in P_s$ such that
	\begin{align}\label{311}
		\tilde{\Omega}_{\bm k'}^{s-1}\subset \Omega_{\bm k}^s.
	\end{align}
	
	$(\bm c)_s$:  For each $\bm k\in P_s$, $\tilde{\Omega}_{\bm k}^s$ contains a subset $A_{\bm k}^s\subset\Omega_{\bm k}^s$ with $\# A_{\bm k}^s\le 2^s$ such that $\tilde{\Omega}_{\bm k}^s\setminus A_{\bm k}^s$ is $(s-1)$-$\good$. Moreover, $A_{\bm k}^s-\bm k$ is independent of $\bm k$ and is symmetrical about the origin.
	
	$(\bm d)_s$:  There is a $\theta_s=\theta_s(E)\in\C$ with the following properties. Replacing $\theta+\bm n\cdot\bm\omega$ by $z+(\bm n-\bm k)\cdot\bm\omega$ and restricting $z$ in
	\begin{align}\label{zs}
		\left\{z\in\C:\ \min_{\sigma=\pm1}\|z+\sigma\theta_s\|_{\T}<\delta_s^{\frac{1}{2}}\right\},
	\end{align}
	we write %see that $\mathcal{T}_{\tilde{\Omega}_{\bm k}^s}$ becomes
	\begin{align*}
		\mathcal{M}_s(z)=\mathcal{T}_{\tilde{\Omega}_{\bm k}^s-\bm k}(z)=((v(z+\bm n\cdot\bm\omega)-E)\delta_{\bm n,\bm n'}+\ep \mathcal{W})_{\bm n\in\tilde{\Omega}_{\bm k}^s-\bm k}.
	\end{align*}
	Then $(\mathcal{M}_s(z))_{(\tilde{\Omega}_{\bm k}^s\setminus A_{\bm k}^s)-\bm k}$ is invertible and we can define the Schur complement
	\begin{align*}
		\mathcal{S}_s(z)&=(\mathcal{M}_s(z))_{A_{\bm k}^s-\bm k}-\mathcal{R}_{A_{\bm k}^s-\bm k}\mathcal{M}_s(z)\mathcal{R}_{(\tilde{\Omega}_{\bm k}^s\setminus A_{\bm k}^s)-\bm k}((\mathcal{M}_s(z))_{(\tilde{\Omega}_{\bm k}^s\setminus A_{\bm k}^s)-\bm k})^{-1}\\
		&\ \ \times \mathcal{R}_{(\tilde{\Omega}_{\bm k}^s\setminus A_{\bm k}^s)-\bm k}\mathcal{M}_s(z)\mathcal{R}_{A_{\bm k}^s-\bm k}.
	\end{align*}
	Moreover, if $z$ belongs to the set in  \eqref{zs}, then we have
	\begin{align}\label{ss}
		\|\mathcal{S}_s(z)\|_0<2|v|_R+\sum_{l=0}^{s-1}\delta_l<4|v|_R
	\end{align}
	and
	\begin{align}\label{detss}
		\left|\det \mathcal{S}_s(z)\right|\gtrsim\delta_{s}^{\frac{2}{75}}\|z-\theta_s\|_{\T}\cdot\|z+\theta_s\|_{\T}. 
	\end{align}
	Combining the Schur complement lemma (cf. Lemma \ref{scl}), we get
	\begin{align}
		\label{tb0}\|\mathcal{T}_{\tilde{\Omega}_{\bm k}^s}^{-1}\|_0&<\delta_s^{-\frac{1}{15}}\|\theta+\bm k\cdot\bm\omega-\theta_s\|_{\T}^{-1}\cdot\|\theta+\bm k\cdot\bm\omega+\theta_s\|_{\T}^{-1}. 
	\end{align}
	
	$(\bm e)_s$:  Let
	\begin{align*}
		Q_{s}^{\pm}&=\{\bm k\in P_{s}:\ \|\theta+\bm k\cdot\bm\omega\pm\theta_{s}\|_{\T}<\delta_{s}\},\ Q_{s}=Q_{s}^+\cup Q_{s}^-,\\
		\tilde{Q}_{s}^{\pm}&=\{\bm k\in P_{s}:\ \|\theta+\bm k\cdot\bm\omega\pm\theta_{s}\|_{\T}<\delta_{s}^{\frac{2}{3}}\},\ \tilde{Q}_{s}=\tilde{Q}_{s}^+\cup \tilde{Q}_{s}^-.
	\end{align*}
	For $\bm k\in P_s\setminus Q_s$ and $\alpha\in(0,\alpha_1]$, we have
	\begin{align}
		\label{tba}\|\mathcal{T}_{\tilde{\Omega}_{\bm k}^s}^{-1}\|_\alpha&<  \zeta_s^{\alpha}\delta_s^{-\frac{7}{3}}. 
	\end{align}
	We say a finite set $\lg\subset\Z^d$ is $s$-$\good$ iff 
	\begin{align*}
		\left\{\begin{array}{l}
			\bm k'\in Q_{s'},\ \tilde{\Omega}_{\bm k'}^{s'}\subset\lg,\ \tilde{\Omega}_{\bm k'}^{s'}\subset\Omega_{\bm k}^{s'+1}\Rightarrow\tilde{\Omega}_{\bm k}^{s'+1}\subset\lg\ \text{for}\ s'<s,\\
			\{\bm k\in P_s:\ \tilde{\Omega}_{\bm k}^s\subset\lg\}\cap Q_s=\emptyset.
		\end{array}\right.
	\end{align*}
	Assume that $\lg$ is $s$-$\good$. Then 
	\begin{align}
		\nonumber\|\mathcal{T}_\lg^{-1}\|_0&<\delta_{s}^{-\frac{2}{15}}\times\sup\limits_{\{\bm k\in P_s:\ \tilde{\Omega}_{\bm k}^s\subset\lg\}}(\|\theta+\bm k\cdot\bm\omega-\theta_s\|_{\T}^{-1}\cdot\|\theta+\bm k\cdot\bm\omega+\theta_s\|_{\T}^{-1})\\
		\label{tsg01}&<\delta_s^{-\frac{32}{15}},
	\end{align}
	and for $\alpha\in(0,\alpha_1],$
	\begin{align}
		\label{tsg1}\|\mathcal{T}_{\lg}^{-1}\|_{\alpha}<\zeta_s^{\alpha}\delta_s^{-\frac{14}{3}}.
	\end{align}
	
	$(\bm f)_s$:  We have
	\begin{align}\label{fs}
		\left\{\bm k\in\Z^d+\frac{1}{2}\sum_{i=0}^{s-1}\bm l_i:\ \min_{\sigma=\pm1}\|\theta+\bm k\cdot\bm\omega+\sigma\theta_s\|_{\T}<10\delta_s^{\frac{2}{3}}\right\}\subset P_s.
	\end{align}
	
	The main theorem of this section is  
	\begin{thm}\label{ind}
		Let $\bm\omega\in DC_{\tau,\g}$. Then there is some $\ep_0=\ep_0(\alpha_1,\alpha_0,d,\tau,\g,v,R,\phi)>0$ so that for $0<|\ep|<\ep_0$, the statement $\mathscr{P}_s$ holds true  for all $s\ge1$.
	\end{thm}
	
	The following three  subsections are devoted to proving  Theorem \ref{ind}.
	
	\subsection{The initial step}
	Recalling $v(\theta_0)=E$, $\mathcal{D}_{\bm n}=v(\theta+\bm n\cdot\bm\omega)-E$ and \eqref{vdefn}, we have
	\begin{align*}
		|\mathcal{D}_{\bm n}|=|v(\theta+\bm n\cdot\bm\omega)-v(\theta_0)|\ge \kappa_1
		\|\theta+\bm n\cdot\bm\omega+\theta_0\|_{\T}\cdot\|\theta+\bm n\cdot\bm\omega-\theta_0\|_{\T}.
	\end{align*}
	
	Denote %$\delta_0=\ep_0^{\frac{1}{30}}$ and 
	\begin{align*}
		P_0=\Z^d,\ Q_0=\{\bm k\in P_0:\ \min(\|\theta+\bm k\cdot\bm\omega+\theta_0\|_{\T},\|\theta+\bm k\cdot\bm\omega-\theta_0\|_{\T})<\delta_0\}.
	\end{align*}
	
	We say a finite set $\lg\subset \Z^d$ is $0$-$\good$ iff $\lg\cap Q_0=\emptyset$.
	
	\begin{lem}\label{dt}
		If the set $\lg\subset\Z^d$ is $0$-$\good$, we have
		\begin{align}\label{0g}
			\|\mathcal{T}_\lg^{-1}\|_{\alpha_1+\alpha_0}\lesssim\delta_0^{-2}.
		\end{align}
	\end{lem}
	\begin{proof}
		Assuming that  $\lg$ is $0$-good, for $\bm n\in\lg$,  we have
		\begin{align*}
			|\mathcal{D}_{\bm n}|\ge\kappa_1\|\theta+\bm n\cdot\bm\omega+\theta_0\|_{\T}\cdot\|\theta+\bm n\cdot\bm\omega-\theta_0\|_{\T}\ge\kappa_1\delta_0^{2}.
		\end{align*}
		Then
		\begin{align*}
			\|\mathcal{D}_{\lg}^{-1}\|_0=\|\mathcal{D}_{\lg}^{-1}\|_{\alpha_1+\alpha_0}\le\kappa_1^{-1}\delta_0^{-2}.
		\end{align*}
		Applying \eqref{tame} implies 
		\begin{align*}
			\|\ep \mathcal{D}_{\lg}^{-1}\mathcal{W}_{\lg}\|_{\alpha_1+\alpha_0}\le 2^{\alpha_1+\alpha_0}\ep\|\mathcal{D}_{\lg}^{-1}\|_{\alpha_1+\alpha_0}\|\mathcal{W}\|_{\alpha_1+\alpha_0}<\frac{1}{2}.
		\end{align*}
		Thus
		\begin{align*}
			(\mathcal{I}_{\lg}+\ep \mathcal{D}_{\lg}^{-1}\mathcal{W}_{\lg})^{-1}=\mathcal{I}_{\lg}+\sum_{i=1}^{\infty}(-\ep \mathcal D_{\lg}^{-1}\mathcal W_{\lg})^i
		\end{align*}
		and 
		\begin{align*}
			\mathcal{T}_{\lg}^{-1}=(\mathcal{I}_\lg+\ep \mathcal{D}_{\lg}^{-1}\mathcal{W}_{\lg})^{-1}\mathcal{D}_\lg^{-1}.
		\end{align*}
		Using \eqref{tame} again yields 
		\begin{align*}
			\|(\mathcal{I}_{\lg}+\ep \mathcal{D}_{\lg}^{-1}\mathcal{W}_{\lg})^{-1}\|_{\alpha_1+\alpha_0}&\le 1+\sum_{i=1}^\infty i^{\alpha_1+\alpha_0}\|\ep \mathcal{D}_{\lg}^{-1}\mathcal{W}_{\lg}\|_{\alpha_1+\alpha_0}^i\\
			&\le1+\sum_{i=1}^\infty\frac{i^{\alpha_1+\alpha_0}}{2^i}\lesssim1,
		\end{align*}
		and thus 
		\begin{align*}
			\|\mathcal{T}_\lg^{-1}\|_{\alpha_1+\alpha_0}\le2^{\alpha_1+\alpha_0}\|(\mathcal{I}_{\lg}+\ep \mathcal{D}_{\lg}^{-1}\mathcal{W}_{\lg})^{-1}\|_{\alpha_1+\alpha_0}\|D_{\lg}^{-1}\|_{\alpha_1+\alpha_0}\lesssim\delta_0^{-2}.
		\end{align*}
	\end{proof}
	
	\subsection{Verification of $\mathscr{P}_1$}\label{vo1}
	If $\lg\cap Q_0\ne\emptyset$, then the Neumann series argument of the previous subsection does not work. Thus we use the resolvent identity argument to estimate $\mathcal{T}_\lg^{-1}$ whenever  $\lg$ is $1$-$\good$.
	
	We outline the main steps of the proof. First, we construct resonant blocks $\Omega_{\bm k}^1\ (\bm k\in P_1)$ to cover $Q_0$. Second, we use the Schur complement lemma (cf. Lemma \ref{scl} in the Appendix \ref{app}), Cramer's rule and Hadamard's inequality  to get 
	\begin{align*}
		\|\mathcal{T}_{\tilde{\Omega}_{\bm k}^1}^{-1}\|_0&<\delta_0^{-2}\|\theta+\bm k\cdot\bm\omega-\theta_1\|_{\T}^{-1}\cdot\|\theta+\bm k\cdot\bm\omega+\theta_1\|_{\T}^{-1}, 
	\end{align*}
	where $\tilde{\Omega}_{\bm k}^1$ is the enlarged resonant block and $\theta_1$ is obtained via Rouch\'e's theorem as one  root of the Dirichlet determinant equation $\det \mathcal{T}_{\tilde{\Omega}_{\bm k}^1}(z-\bm k\cdot\bm\omega)=0$. Next, we use smoothing property, rows estimate and perturbation argument to get for $\alpha\in(0,\alpha_1],$
	\begin{align*}
		\|\mathcal{T}_{\tilde{\Omega}_{\bm k}^1}^{-1}\|_\alpha&<\zeta_1^{\alpha}\delta_1^{-\frac{1}{3}}\|\theta+\bm k\cdot\bm\omega-\theta_1\|_{\T}^{-1}\cdot\|\theta+\bm k\cdot\bm\omega+\theta_1\|_{\T}^{-1}
	\end{align*}
	via  a {\it constructive}  procedure. Finally, we combine the estimates of $\mathcal{T}_{\tilde{\Omega}_{\bm k}^1}^{-1}$ to obtain  that of $\mathcal{T}_\lg^{-1}$ in a similar way for more general $1$-$\good$ $\lg$.
	
	%Recall that
	%\begin{align*}
	%	1<b_1<b_2=\frac{3}{2}<b_3=\frac{8}{5}<b_4<c=3,\ f=\frac{5}{3},\ g=30.
	%\end{align*}
	Recall that 
	\begin{align*}
		N_1=\left[\left(\frac{\g}{\delta_0}\right)^{\frac{1}{30\tau}}\right]
	\end{align*}
	and define (cf. \eqref{theta0})
	\begin{align*}
		Q_0^{\pm}&=\{\bm k\in\Z^d:\ \|\theta+\bm k\cdot\bm\omega\pm\theta_0\|_\T<\delta_0\},\ Q_0=Q_0^+\cup Q_0^-,\\
		\tilde{Q}_0^{\pm}&=\{\bm k\in\Z^d:\ \|\theta+\bm k\cdot\bm\omega\pm\theta_0\|_\T<\delta_0^{\frac{2}{3}}\},\ \tilde{Q}_0=\tilde{Q}_0^+\cup \tilde{Q}_0^-.
	\end{align*}
	
	We distinguish the verification into three steps.
	
	\begin{itemize}
		\item[\textbf{Step 1}]: \textbf{Estimates of $\|\mathcal{T}_{\tilde{\Omega}_{\bm k}^1}^{-1}\|_0$}.
	\end{itemize}
	In this step, we will find $\theta_1=\theta_1(E)$ so that 
	\begin{align*}
		\|\mathcal{T}_{\tilde{\Omega}_{\bm k}^1}^{-1}\|_0&<\delta_0^{-2}\|\theta+\bm k\cdot\bm\omega-\theta_1\|_{\T}^{-1}\cdot\|\theta+\bm k\cdot\bm\omega+\theta_1\|_{\T}^{-1}.
	\end{align*}
	We  again divide the discussions   into two cases.
	\begin{itemize}
		\item[\textbf{Case 1}]:  {The case $(\bm C1)_0$ occurs}, i.e., 
	\end{itemize}
	\begin{align}\label{-+1}
		\text{dist}\left(\tilde{Q}_0^-,Q_0^+\right)>100N_1^3.
	\end{align}
	\begin{rem}\label{dsq1}
		We have in fact
		\begin{align*}
			\text{dist}\left(\tilde{Q}_0^-,Q_0^+\right)=\text{dist}\left(\tilde{Q}_0^+,Q_0^-\right).
		\end{align*}
		Thus \eqref{-+1} also implies
		\begin{align*}
			\text{dist}\left(\tilde{Q}_0^+,Q_0^-\right)>100N_1^3.
		\end{align*}
		We refer to the Appendix \ref{app} for a detailed proof.
	\end{rem}
	Assuming \eqref{-+1} holds true, we define
	\begin{align}\label{P11}
		P_1=Q_0=\{\bm k\in\Z^d:\ \min(\|\theta+\bm k\cdot\bm\omega+\theta_0\|_{\T},\|\theta+\bm k\cdot\bm\omega-\theta_0\|_{\T})<\delta_0\}.
	\end{align}
	Associate each  $\bm k\in P_1$ with  $\Omega_{\bm k}^1:=\lg_{N_1}(\bm k)$ and  $\tilde{\Omega}_{\bm k}^1:=\lg_{N_1^3}(\bm k)$. Then $\tilde{\Omega}_{\bm k}^1-\bm k\subset\Z^d$ is independent of $\bm k\in P_1$ and symmetrical about the origin. If $\bm k\ne\bm k'\in P_1$,  then 
	\begin{align*}
		\|\bm k-\bm k'\|\ge\min\left(100N_1^3,\left(\frac{\g}{2\delta_0}\right)^{\frac{1}{\tau}}\right)\ge100N_1^3.
	\end{align*}
	Thus
	\begin{align*}
		\text{dist}\left(\tilde{\Omega}_{\bm k}^1,\tilde{\Omega}_{\bm k'}^1\right)>10\tilde{\zeta}_1\ \text{for}\ \bm k\ne \bm k'\in P_1.
	\end{align*}
	For $\bm k\in Q_0^-$, we consider
	\begin{align*}
		\mathcal{M}_1(z):=\mathcal{T}_{\tilde{\Omega}_{\bm k}^1-\bm k}(z)=\left((v(z+\bm n\cdot\bm\omega)-E)\delta_{\bm n,\bm n'}+\ep \mathcal{W}\right)_{\bm n\in\tilde{\Omega}_{\bm k}^1-\bm k}
	\end{align*}
	defined in
	\begin{align}\label{z0-}
		\left\{z\in\C:\ |z-\theta_0|<\delta_0^{\frac{18}{19}}\right\}.
	\end{align}
	For $\bm n\in\left(\tilde{\Omega}_{\bm k}^1-\bm k\right)\setminus\{\bm 0\}$, we have for $0<\delta_0\ll 1$, 
	\begin{align*}
		\|z+\bm n\cdot\bm\omega-\theta_0\|_{\T}&\ge\|\bm n\cdot\bm\omega\|_{\T}-|z-\theta_0|\\
		&\ge\frac{\g}{(N_1^3)^{\tau}}-\delta_0^{\frac{18}{19}}\\
		&\gtrsim\delta_0^{\frac{1}{10}}.
	\end{align*}
	For $\bm n\in\tilde{\Omega}_{\bm k}^1-\bm k$, we have
	\begin{align*}
		\|z+\bm n\cdot\bm\omega+\theta_0\|_{\T}&\ge\|\theta+(\bm n+\bm k)\cdot\bm\omega+\theta_0\|_{\T}-|z-\theta_0|-\|\theta+\bm k\cdot\bm\omega-\theta_0\|_{\T}\\
		&\ge\delta_0^{\frac{2}{3}}-\delta_0^{\frac{18}{19}}-\delta_0>\frac{1}{2}\delta_0^{\frac{2}{3}}.
	\end{align*}
	Hence for $\bm n\in\left(\tilde{\Omega}_{\bm k}^1-\bm k\right)\setminus\{\bm 0\}$,
	\begin{align*}
		|v(z+\bm n\cdot\bm\omega)-E|\gtrsim \delta_0^{\frac{23}{30}}\gg\ep.
	\end{align*}
	By Neumann series argument,  we have
	\begin{align}\label{m1z01}
		\left\|\left((\mathcal{M}_1(z))_{(\tilde{\Omega}_{\bm k}^1-\bm k)\setminus\{\bm 0\}}\right)^{-1}\right\|_0\lesssim\delta_0^{-\frac{23}{30}}.
	\end{align}
	Now,  we can apply the Schur complement lemma to establish desired estimates. By Lemma \ref{scl}, $(\mathcal{M}_1(z))^{-1}$ is controlled by the Schur complement (of $(\tilde{\Omega}_{\bm k}^1-\bm k)\setminus\{\bm 0\}$)
	\begin{align*}
		\mathcal{S}_1(z)&=(\mathcal{M}_1(z))_{\{\bm 0\}}-\mathcal{R}_{\{\bm 0\}}\mathcal{M}_1(z)\mathcal{R}_{(\tilde{\Omega}_{\bm k}^1-\bm k)\setminus\{\bm 0\}}\left((\mathcal{M}_1(z))_{(\tilde{\Omega}_{\bm k}^1-\bm k)\setminus\{\bm 0\}}\right)^{-1}\\
		&\ \ \times\mathcal{R}_{(\tilde{\Omega}_{\bm k}^1-\bm k)\setminus\{\bm 0\}}\mathcal{M}_1(z)\mathcal{R}_{\{\bm 0\}}\\
		&=v(z)-E+r(z)=g(z)((z-\theta_0)+r_1(z)),
	\end{align*}
	where $g(z)$ and $r_1(z)$ are analytic functions in the set of \eqref{z0-} satisfying $|g(z)|>\kappa_1\|z+\theta_0\|_{\T}>\frac{\kappa_1}{2}\delta_0^{\frac{2}{3}}$ and $|r_1(z)|\lesssim \ep^2\delta_0^{-\frac{43}{30}}\ll\delta_0^{\frac{18}{19}}$. Since
	\begin{align*}
		|r_1(z)|<|z-\theta_0|\ \text{for}\ |z-\theta_0|=\delta_0^{\frac{18}{19}},
	\end{align*}
	using the Rouch\'e's  theorem and maximum modulus principle implies  
	\begin{align*}
		(z-\theta_0)+r_1(z)=0
	\end{align*}
	has a unique root $\theta_1$ in  the set of \eqref{z0-} satisfying
	\begin{align}\label{t0-t1}
		|\theta_0-\theta_1|=|r_1(\theta_1)|<\ep,\ |(z-\theta_0)+r_1(z)|\sim|z-\theta_1|.
	\end{align}
	Moreover, $\theta_1$ is the unique root of $\det \mathcal{M}_1(z)=0$ in the set of  \eqref{z0-}. Since $\|z+\theta_0\|_{\T}>\frac{1}{2}\delta_0^{\frac{2}{3}}$ and $|\theta_0-\theta_1|<\ep$, we get
	\begin{align*}
		\|z+\theta_1\|_{\T}\sim\|z+\theta_0\|_{\T}.
	\end{align*}
	Then by \eqref{m1z01} and \eqref{sc}, we have  for $z$ being in the set of \eqref{z0-},
	\begin{align}
		|\mathcal{S}_1(z)|&\gtrsim\|z+\theta_1\|_{\T}\cdot\|z-\theta_1\|_{\T}, \label{S1}\\
		\nonumber\|(\mathcal{M}_1(z))^{-1}\|_0&<4\left(1+\left\|\left(\mathcal{M}_1(z)_{(\tilde{\Omega}_{\bm k}^1-\bm k)\setminus\{\bm 0\}}\right)^{-1}\right\|_0\right)^2(1+|\mathcal{S}_1(z)|^{-1})\\
		&< \delta_0^{-2}\|z+\theta_1\|_{\T}^{-1}\cdot\|z-\theta_1\|_{\T}^{-1}.\label{M-1}
	\end{align}
	Now, for $\bm k\in Q_0^+$, we consider $\mathcal{M}_1(z)$ in the set 
	\begin{align}\label{z0+}
		\left\{z\in\C:\ |z+\theta_0|<\delta_0^{\frac{18}{19}}\right\}.
	\end{align}
	The similar argument shows that $\det \mathcal{M}_1(z)=0$ has a unique root $\theta_1'$ in the set of \eqref{z0+}. We will show $\theta_1+\theta_1'=0$. In fact, by Lemma \ref{ef}, $\det \mathcal{M}_1(z)$ is an even function of $z$. Then the uniqueness of the root implies $\theta_1'=-\theta_1$. Thus for $z$ being in the set of \eqref{z0+}, both \eqref{S1} and \eqref{M-1} hold true as well. Finally, \eqref{S1} and \eqref{M-1} remain valid for
	\begin{align}\label{z0-+}
		\{z\in\C:\ \min\limits_{\sigma=\pm1}|z+\sigma\theta_0|<\delta_0^{\frac{18}{19}}\}.
	\end{align}
	From \eqref{P11}, we have $\theta+\bm k\cdot\bm\omega$ belongs to the set of \eqref{z0-+}. Thus for $\bm k\in P_1$, we get
	\begin{align}
		\nonumber\|\mathcal{T}_{\tilde{\Omega}_{\bm k}^1}^{-1}\|_0&=\|(\mathcal{M}_1(\theta+\bm k\cdot\bm\omega))^{-1}\|_0\\
		&<\delta_0^{-2}\|\theta+\bm k\cdot\bm\omega+\theta_1\|_{\T}^{-1}\cdot\|\theta+\bm k\cdot\bm\omega-\theta_1\|_{\T}^{-1}\label{T-11}.
	\end{align}
	
	\begin{itemize}
		\item[\textbf{Case 2}]:  {The case $(\bm C2)_0$ occurs},  i.e., 
	\end{itemize}
	\begin{align}\label{-+2}
		\text{dist}\left(\tilde{Q}_0^-,Q_0^+\right)\le100N_1^3.
	\end{align}
	Then there exist $\bm i_0\in Q_0^+$ and $\bm j_0\in\tilde{Q}_0^-$ with $\|\bm i_0-\bm j_0\|\le100N_1^3$ such that
	\begin{align*}
		\|\theta+\bm i_0\cdot\bm \omega+\theta_0\|_{\T}<\delta_0,\ \|\theta+\bm j_0\cdot\bm \omega-\theta_0\|_{\T}<\delta_0^{\frac{2}{3}}.
	\end{align*}
	Set $\bm l_0=\bm i_0-\bm j_0$. Then 
	\begin{align*}
		\|\bm l_0\|=\text{dist}\left(\tilde{Q}_0^-,Q_0^+\right)=\text{dist}\left(\tilde{Q}_0^+,Q_0^-\right).
	\end{align*}
	Define
	\begin{align*}
		O_1=Q_0^-\cup(Q_0^+-\bm l_0).
	\end{align*}
	For $\bm k\in Q_0^+$, we have
	\begin{align*}
		\|\theta+(\bm k-\bm l_0)\cdot\bm \omega-\theta_0\|_{\T}&<\|\theta+\bm k\cdot\bm\omega+\theta_0\|_{\T}+\|\bm l_0\cdot\bm\omega+2\theta_0\|_{\T}\\
		&<\delta_0+\delta_0+\delta_0^{\frac{2}{3}}<2\delta_0^{\frac{2}{3}}.
	\end{align*}
	Thus
	\begin{align*}
		O_1\subset\left\{\bm o\in\Z^d:\ \|\theta+\bm o\cdot\bm \omega-\theta_0\|_{\T}<2\delta_0^{\frac{2}{3}}\right\}.
	\end{align*}
	For every $\bm o\in O_1$, define its mirror point
	\begin{align*}
		\bm o^*=\bm o+\bm l_0.
	\end{align*}
	Next, define
	\begin{align}\label{P12}
		P_1=\left\{\frac{1}{2}(\bm o+\bm o^*):\ \bm o\in O_1\right\}=\left\{\bm o+\frac{\bm l_0}{2}:\ \bm o\in O_1\right\}.
	\end{align}
	Associate each  $\bm k\in P_1$ with  $\Omega_{\bm k}^1:=\lg_{100N_1^3}(\bm k)$ and   $\tilde{\Omega}_{\bm k}^1:=\lg_{N_1^{5}}(\bm k)$. Thus
	\begin{align*}
		Q_0\subset\bigcup_{\bm k\in P_1}\Omega_{\bm k}^1
	\end{align*}
	and $\tilde{\Omega}_{\bm k}^1-\bm k\subset\Z^d+\frac{\bm l_0}{2}$ is independent of $\bm k\in P_1$ and symmetrical about origin. Notice that
	\begin{align*}
		\min&\left(\left\|\frac{\bm l_0}{2}\cdot\bm\omega+\theta_0\right\|_{\T},\left\|\frac{\bm l_0}{2}\cdot\bm \omega+\theta_0-\frac{1}{2}\right\|_{\T}\right)\\
		&=\frac{1}{2}\|\bm l_0\cdot\bm\omega+2\theta_0\|_{\T}\\
		&\le\frac{1}{2}(\|\theta+\bm i_0\cdot\bm\omega+\theta_0\|_{\T}+\|\theta+\bm j_0\cdot\bm \omega-\theta_0\|_{\T})<\delta_0^{\frac{2}{3}}.
	\end{align*}
	Since $\delta_0\ll1$, only one of
	\begin{align*}
		\left\|\frac{\bm l_0}{2}\cdot\bm\omega+\theta_0\right\|_{\T}<\delta_0^{\frac{2}{3}}\ \text{and}\ \left\|\frac{\bm l_0}{2}\cdot\bm \omega+\theta_0-\frac{1}{2}\right\|_{\T}<\delta_0^{\frac{2}{3}}
	\end{align*}
	holds true. First, we consider the case of 
	\begin{align}\label{l01}
		\left\|\frac{\bm l_0}{2}\cdot\bm\omega+\theta_0\right\|_{\T}<\delta_0^{\frac{2}{3}}.
	\end{align}
	Let $\bm k\in P_1$. Since $\bm k=\frac{1}{2}(\bm o+\bm o^*)=\bm o+\frac{\bm l_0}{2}$ (for some $\bm o\in O_1$), we have
	\begin{align}\label{t+ko1}
		\|\theta+\bm k\cdot\bm\omega\|_{\T}\le\|\theta+\bm o\cdot\bm\omega-\theta_0\|_{\T}+\left\|\frac{\bm l_0}{2}\cdot\bm\omega+\theta_0\right\|_{\T}<3\delta_0^{\frac{2}{3}}.
	\end{align}
	Thus if $\bm k\ne \bm k'\in P_1$, we obtain
	\begin{align*}
		\|\bm k-\bm k'\|\ge\left(\frac{\g}{6\delta_0^{\frac{2}{3}}}\right)^{\frac{1}{\tau}}\sim N_1^{20}\gg 100N_1^{5},
	\end{align*}
	which implies
	\begin{align*}
		\text{dist}\left(\tilde{\Omega}_{\bm k}^1,\tilde{\Omega}_{\bm k'}^1\right)>10\zeta_1\ \text{for}\ \bm k\ne\bm k'\in P_1.
	\end{align*}
	Consider
	\begin{align*}
		\mathcal{M}_1(z):=\mathcal{T}_{\tilde{\Omega}_{\bm k}^1-\bm k}(z)=\left((v(z+\bm n\cdot\bm\omega)-E)\delta_{\bm n,\bm n'}+\ep \mathcal{W}\right)_{\bm n\in\tilde{\Omega}_{\bm k}^1-\bm k}
	\end{align*}
	in the set of 
	\begin{align}\label{z02}
		\left\{z\in\C:\ |z|<\delta_0^{\frac{5}{8}}\right\}.
	\end{align}
	For $\bm n\ne\pm\frac{\bm l_0}{2}$ and $\bm n\in\tilde{\Omega}_{\bm k}^1-\bm k$, we have
	\begin{align*}
		\|\bm n\cdot\bm\omega\pm\theta_0\|_{\T}&\ge\left\|\left(\bm n\mp\frac{\bm l_0}{2}\right)\cdot\bm \omega\right\|_{\T}-\left\|\frac{\bm l_0}{2}\cdot\bm\omega+\theta_0\right\|_{\T}\\
		&>\frac{\g}{(2N_1^{5})^\tau}-\delta_0^{\frac{2}{3}}\gtrsim\delta_0^{\frac{1}{6}}.
	\end{align*}
	Thus for $z$ being in the set of \eqref{z02} and $\bm n\ne\pm\frac{\bm l_0}{2}$, we have
	\begin{align*}
		\|z+\bm n\cdot\bm\omega\pm\theta_0\|_{\T}\ge\|\bm n\cdot\bm\omega\pm\theta_0\|_{\T}-|z|\gtrsim\delta_0^{\frac{1}{6}}.
	\end{align*}
	Hence for $\bm n\in(\tilde{\Omega}_{\bm k}^1-\bm k)\setminus\left\{\pm\frac{\bm l_0}{2}\right\},$ we have 
	\begin{align*}
		|v(z+\bm n\cdot\bm\omega)-E|\gtrsim \delta_0^{\frac{1}{3}}\gg\ep.
	\end{align*}
	Using Neumann series argument concludes
	\begin{align}\label{M-12}
		\left\|\left((\mathcal{M}_1(z))_{(\tilde{\Omega}_{\bm k}^1-\bm k)\setminus\left\{\pm\frac{\bm l_0}{2}\right\}}\right)^{-1}\right\|_0\lesssim\delta_0^{-\frac{1}{3}}.
	\end{align}
	Thus by Lemma \ref{scl}, $(\mathcal{M}_1(z))^{-1}$ is controlled by   the Schur complement of $(\tilde{\Omega}_{\bm k}^1-\bm k)\setminus\left\{\pm\frac{\bm l_0}{2}\right\}$, i.e.,
	\begin{align*}
		\mathcal{S}_1(z)&=(\mathcal{M}_1(z))_{\left\{\pm\frac{\bm l_0}{2}\right\}}-\mathcal{R}_{\left\{\pm\frac{\bm l_0}{2}\right\}}\mathcal{M}_1(z)\mathcal{R}_{(\tilde{\Omega}_{\bm k}^1-\bm k)\setminus\left\{\pm\frac{\bm l_0}{2}\right\}}\\
		&\ \ \times\left((\mathcal{M}_1(z))_{(\tilde{\Omega}_{\bm k}^1-\bm k)\setminus\left\{\pm\frac{\bm l_0}{2}\right\}}\right)^{-1}\mathcal{R}_{(\tilde{\Omega}_{\bm k}^1-\bm k)\setminus\left\{\pm\frac{\bm l_0}{2}\right\}}\mathcal{M}_1(z)\mathcal{R}_{\left\{\pm\frac{\bm l_0}{2}\right\}},
	\end{align*}
	where 
	\begin{align*}
		\left\|\mathcal{R}_{\left\{\pm\frac{\bm l_0}{2}\right\}}\mathcal{M}_1(z)\mathcal{R}_{(\tilde{\Omega}_{\bm k}^1-\bm k)\setminus\left\{\pm\frac{\bm l_0}{2}\right\}}\right\|_0&\lesssim\ep,\\
		\left\|\mathcal{R}_{(\tilde{\Omega}_{\bm k}^1-\bm k)\setminus\left\{\pm\frac{\bm l_0}{2}\right\}}\mathcal{M}_1(z)\mathcal{R}_{\left\{\pm\frac{\bm l_0}{2}\right\}}\right\|_0&\lesssim\ep.
	\end{align*}
	Then by Lemma \ref{det1} and \eqref{M-12},  we  get
	\begin{align}
		\nonumber\|\mathcal{S}_1(z)\|_0&\le\left\|(\mathcal{M}_1(z))_{\left\{\pm\frac{\bm l_0}{2}\right\}}\right\|_0+O(\ep^2\delta_0^{-\frac{1}{3}})\\
		\label{s10}&\le 2|v|_R+\delta_0<4|v|_R,
	\end{align}
	and
	\begin{align*}
		\det \mathcal{S}_1(z)&=\det\left((\mathcal{M}_1(z))_{\left\{\pm\frac{\bm l_0}{2}\right\}}\right)+O(\ep^2\delta_0^{-\frac{1}{3}})\\
		&=\left(v\left(z+\frac{\bm l_0}{2}\cdot\bm\omega\right)-E\right)\left(v\left(z-\frac{\bm l_0}{2}\cdot\bm\omega\right)-E\right)+O(\ep^2\delta_0^{-\frac{1}{3}}).
	\end{align*}
	In the case of $\bm l_0=\bm0$, the argument is easier and we omit the discussion. In the following, we deal with $\bm l_0\ne\bm 0$. By \eqref{l01} and \eqref{z02}, we have
	\begin{align*}
		\left\|z+\frac{\bm l_0}{2}\cdot\bm\omega-\theta_0\right\|_{\T}&\ge\|\bm l_0\cdot\bm\omega\|_{\T}-\left\|\frac{\bm l_0}{2}\cdot\bm\omega+\theta_0\right\|_{\T}-|z|\\
		&>\frac{\g}{(100N_1^3)^\tau}-\delta_0^{\frac{2}{3}}-\delta_0^{\frac{5}{8}}\\
		&\gtrsim\delta_0^{\frac{1}{10}},
	\end{align*}
	and
	\begin{align*}
		\left\|z-\frac{\bm l_0}{2}\cdot\bm\omega+\theta_0\right\|_{\T}&\ge\|\bm l_0\cdot\bm\omega\|_{\T}-\left\|\frac{\bm l_0}{2}\cdot\bm\omega+\theta_0\right\|_{\T}-|z|\\
		&>\frac{\g}{(100N_1^3)^\tau}-\delta_0^{\frac{2}{3}}-\delta_0^{\frac{5}{8}}\\
		&\gtrsim\delta_0^{\frac{1}{10}}.
	\end{align*}
	Let $z_1$ satisfy
	\begin{align}\label{z1}
		z_1\equiv\frac{\bm l_0}{2}\cdot\bm\omega+\theta_0\ (\text{mod}\ \Z),\ |z_1|=\left\|\frac{\bm l_0}{2}\cdot\bm\omega+\theta_0\right\|_{\T}<\delta_0^{\frac{2}{3}}.
	\end{align}
	Then
	\begin{align*}
		|\det \mathcal{S}_1(z)|&\gtrsim \left\|z+\frac{\bm l_0}{2}\cdot\bm\omega-\theta_0\right\|_{\T}\cdot\left\|z-\frac{\bm l_0}{2}\cdot\bm\omega+\theta_0\right\|_{\T}\cdot|(z-z_1)(z+z_1)+r_1(z)|\\
		&\gtrsim\delta_0^{\frac{1}{5}}|(z-z_1)(z+z_1)+r_1(z)|,
	\end{align*}
	where $r_1(z)$ is an analytic function in the set of  \eqref{z02} with 
	\begin{align}\label{r1}
		|r_1(z)|\lesssim\ep^2\delta_0^{-\frac{8}{15}}\ll\ep\ll\delta_0^{\frac{5}{8}}.
	\end{align}
	Applying Rouch\'e's  theorem shows that 
	\begin{align*}
		(z-z_1)(z+z_1)+r_1(z)=0
	\end{align*}
	has exact two roots $\theta_1$ and $\theta_1'$ in the set of \eqref{z02}, which are perturbations of $\pm z_1$. If
	\begin{align*}
		|z_1-\theta_1|>|r_1(\theta_1)|^{\frac{1}{2}}\ \text{and}\ |z_1+\theta_1|>|r_1(\theta_1)|^{\frac{1}{2}},
	\end{align*}
	then
	\begin{align*}
		|r_1(\theta_1)|=|z_1-\theta_1|\cdot|z_1+\theta_1|>|r_1(\theta_1)|,
	\end{align*}
	which is a contradiction. Without loss of generality, we assume that
	\begin{align*}
		|z_1-\theta_1|\le|r_1(\theta_1)|^{\frac{1}{2}}\le\ep^{\frac{1}{2}}.
	\end{align*}
	Notice that
	\begin{align*}
		\left\{|z|<\delta_0^{\frac{5}{8}}:\ \det \mathcal{M}_1(z)=0\right\}=\left\{|z|<\delta_0^{\frac{5}{8}}:\ \det \mathcal{S}_1(z)=0\right\}
	\end{align*}
	and $\det \mathcal{M}_1(z)$ is an even function (cf. Lemma \ref{ef}) of $z$. Thus
	\begin{align*}
		\theta_1'=-\theta_1.
	\end{align*}
	Moreover, since \eqref{z1} and \eqref{r1}, we get for $|z|=\delta_0^{\frac{5}{8}},$
	\begin{align*}
		\frac{|r_1(z)-r_1(\theta_1)|}{|(z-z_1)(z+z_1)+r_1(\theta_1)|}\le2\ep^2\delta_0^{-\frac{107}{60}},
	\end{align*}
	which combined with $\theta_1^2-z_1^2+r_1(\theta_1)=0$ shows 
	\begin{align*}
		\frac{|(z-z_1)(z+z_1)+r_1(z)|}{|(z-\theta_1)(z+\theta_1)|}&=\frac{|(z-z_1)(z+z_1)+r_1(z)|}{|(z-z_1)(z+z_1)+r_1(\theta_1)|}\\
		&\in\left[1-2\ep^2\delta_0^{-\frac{107}{60}},1+2\ep^2\delta_0^{-\frac{107}{60}}\right]. 
	\end{align*}
	By the maximum modulus principle,  we have
	\begin{align*}
		|(z-z_1)(z+z_1)+r_1(z)|\sim|(z-\theta_1)(z+\theta_1)|.
	\end{align*}
	Thus for $z$ being in the set of  \eqref{z02}, we have
	\begin{align}\label{detS11}
		|\det \mathcal{S}_1(z)|\gtrsim\delta_0^{\frac{1}{5}}\|z-\theta_1\|_{\T}\cdot\|z+\theta_1\|_{\T}. 
	\end{align}
	By  the Cramer's rule, Lemma \ref{chi}, \eqref{s10} and \eqref{detS11}, we  obtain
	\begin{align}\label{S1-11}
		\|(\mathcal{S}_1(z))^{-1}\|_0=\frac{\|(\mathcal{S}_1(z))^*\|_0}{|\det \mathcal{S}_1(z)|}\lesssim \delta_0^{-\frac{1}{5}}\|z-\theta_1\|_{\T}^{-1}\cdot\|z+\theta_1\|_{\T}^{-1}.
	\end{align}
	Recalling \eqref{M-12} and \eqref{S1-11}, we get since Lemma \ref{scl}
	\begin{align}\label{M1z-11}
		\nonumber\|(\mathcal{M}_1(z))^{-1}\|_0&<4\left(1+\left\|\left((\mathcal{M}_1(z))_{(\tilde{\Omega}_{\bm k}^1-\bm k)\setminus\left\{\pm\frac{\bm l_0}{2}\right\}}\right)^{-1}\right\|_0\right)^2(1+\|(\mathcal{S}_1(z))^{-1}\|_0)\\
		&<\delta_0^{-2}\|z-\theta_1\|_{\T}^{-1}\cdot\|z+\theta_1\|_{\T}^{-1}.
	\end{align}
	Thus for \eqref{l01}, both \eqref{S1-11} and \eqref{M1z-11} are established for $z$ belonging to
	\begin{align*}
		\{z\in\C:\ \|z\|_{\T}<\delta_0^{\frac{5}{8}}\}
	\end{align*}
	since $\mathcal{M}_1(z)$ is $1$-periodic (in $z$). By \eqref{t+ko1}, for $\bm k\in P_1$, we also have
	\begin{align*}
		\|\mathcal{T}_{\tilde{\Omega}_{\bm k}^1}^{-1}\|_0&=\|(\mathcal{M}_1(\theta+\bm k\cdot\bm\omega))^{-1}\|_0\\
		&<\delta_0^{-2}\|\theta+\bm k\cdot\bm\omega-\theta_1\|_{\T}^{-1}\cdot\|\theta+\bm k\cdot\bm\omega+\theta_1\|_{\T}^{-1}.
	\end{align*}
	For the case of 
	\begin{align}\label{l02}
		\left\|\frac{\bm l_0}{2}\cdot\bm\omega+\theta_0-\frac{1}{2}\right\|_{\T}<\delta_0^{\frac{2}{3}},
	\end{align}
	we have for $\bm k\in P_1$,
	\begin{align}\label{t+ko2}
		\left\|\theta+\bm k\cdot\bm\omega-\frac{1}{2}\right\|_{\T}<3\delta_0^{\frac{2}{3}}.
	\end{align}
	Consider
	\begin{align*}
		\mathcal{M}_1(z):=\mathcal{T}_{\tilde{\Omega}_{\bm k}^1-\bm k}(z)=\left((v(z+\bm n\cdot\bm\omega)-E)\delta_{\bm n,\bm n'}+\ep \mathcal{W}\right)_{\bm n\in\tilde{\Omega}_{\bm k}^1-\bm k}
	\end{align*}
	in
	\begin{align}\label{z03}
		\left\{z\in\C:\ \left|z-\frac{1}{2}\right|<\delta_0^{\frac{5}{8}}\right\}.
	\end{align}
	The similar argument shows that $\det \mathcal{M}_1(z)=0$ has two roots $\theta_1$ and $\theta'_1$  in the set of  \eqref{z03} such that \eqref{s10}--\eqref{M1z-11} hold true  for $z$ being in the set of \eqref{z03}.
	Hence if  \eqref{l02}, then \eqref{s10}--\eqref{M1z-11} hold for $z$ being in
	\begin{align*}
		\{z\in\C:\ \left\|z-\frac{1}{2}\right\|_{\T}<\delta_0^{\frac{5}{8}}\}.
	\end{align*}
	By \eqref{t+ko2}, for $\bm k\in P_1$, we also have
	\begin{align}
		\nonumber\|\mathcal{T}_{\tilde{\Omega}_{\bm k}^1}^{-1}\|_0&=\|(\mathcal{M}_1(\theta+\bm k\cdot\bm\omega))^{-1}\|_0\\
		\label{to1-1}&<\delta_0^{-2}\|\theta+\bm k\cdot\bm\omega-\theta_1\|_{\T}^{-1}\cdot\|\theta+\bm k\cdot\bm\omega+\theta_1\|_{\T}^{-1}.
	\end{align}
	
	Therefore,  we have established desired estimates on $\|\mathcal{T}_{\tilde{\Omega}_{\bm k}^1}^{-1}\|_0$ for both cases of $(\bm C1)_0$ and $(\bm C2)_0$.\\
	
	\  \\
	
	\begin{itemize}
		\item[\textbf{Step 2}]: \textbf{Estimates of $\|\mathcal{T}_{\tilde{\Omega}_{\bm k}^1}^{-1}\|_{\alpha}$ for $\alpha\in(0, \alpha_1].$} 
	\end{itemize}
	The main result of this step is Theorem \ref{psq1} below,  which says that even though there is some singular sites of $0$-scale  in the block $\tilde{\Omega}_{\bm k}^1$, $\mathcal{T}_{\tilde{\Omega}_{\bm k}^1}^{-1}$ can become controllable if there is no $1$-scale singular sites. Recalling
	\begin{align*}
		\left(\frac{\g}{\delta_1}\right)=\left(\frac{\g}{\delta_0}\right)^{30}
	\end{align*}
	and
	\begin{align*}
		Q_1^{\pm}=\{\bm k\in P_1:\ \|\theta+\bm k\cdot\bm\omega\pm\theta_1\|_{\T}<\delta_1\},\ Q_1=Q_1^+\cup Q_1^-,
	\end{align*}
	we have  
	\begin{thm}\label{psq1}
		For $\bm k\in P_1\setminus Q_1$ and $\alpha\in(0,\alpha_1]$,  we have
		\begin{align}\label{tg1a}
			\|\mathcal{T}_{\tilde{\Omega}_{\bm k}^1}^{-1}\|_\alpha<\zeta_1^{\alpha}\delta_1^{-\frac{7}{3}}.
		\end{align}
	\end{thm}
	The proof of theorem \ref{psq1} builds on  several lemmas. The next Lemmas \ref{mn} and \ref{tz} say that we can construct $\mathcal{T}_{\tilde{\Omega}_{\bm k}^1}^{-1}$ by some operators with good controls  of $\alpha$-norm directly. In each of them, we use the smoothing property (cf. Lemma \ref{spl}) to control norms and use the perturbation argument (cf. Lemma \ref{pa}) to construct inverse operators.
	
	For $\bm k\in P_1$, we define $A_{\bm k}^1\subset \Omega_{\bm k}^1$ to be 
	\begin{align*}
		A_{\bm k}^1:=\left\{\begin{array}{lc}
			\{\bm k\} & \text{case}\ (\bm C1)_0\\
			\{\bm o\}\cup\{\bm o^*\} & \text{case}\ (\bm C2)_0
		\end{array}\right.,
	\end{align*}
	where $\bm k=\frac{1}{2}(\bm o+\bm o^*)$ for some $\bm o\in O_1$ in the case of  $(\bm C2)_0$. 
	
	Let $G=\tilde{\Omega}_{\bm k}^1\setminus\Omega_{\bm k}^1$ and $B=\Omega_{\bm k}^1$. Thus $X:=\tilde{\Omega}_{\bm k}^1=B\cup G$, $G\cap A_{\bm k}^1=\emptyset$ and $A_{\bm k}^1\subset B$. 
	We have 
	\begin{lem}\label{mn}
		Let $0<\ep_0=\ep_0(\alpha_1,\alpha_0,d,\tau,\g,v,R,\phi)\ll1$. Then there exist $\mathcal{M}\in\mathbf{M}_X^G$ and $\mathcal{N}\in\mathbf{M}_B^G$ satisfying
		\begin{align}\label{eq2}
			\mathcal{M}\mathcal{T}_X=\mathcal{R}_G+\mathcal{N} 
		\end{align}
		with the following estimates: 
		\begin{align*}
			\|\mathcal{N}\|_0\lesssim N_1^{-\alpha_1+\alpha_0},\ \|\mathcal{M}\|_0\lesssim \delta_0^{-2},
		\end{align*}
		and for $\alpha\in(0,\alpha_1]$,
		\begin{align}\label{emn1}
			\|\mathcal{N}\|_\alpha\lesssim\delta_0^{28}<1,\ \|\mathcal{M}\|_\alpha\lesssim\delta_0^{-2}. 
		\end{align}

	\end{lem}

	\begin{proof}
		From our construction, we have
		\begin{align*}
			Q_0\subset \bigcup_{\bm k\in P_1}A_{\bm k}^1\subset\bigcup_{\bm k\in P_1}\Omega_{\bm k}^1.
		\end{align*}
		Thus
		\begin{align*}
			(\tilde{\Omega}_{\bm k}^1\setminus A_{\bm k}^1)\cap Q_0=\emptyset,
		\end{align*}
		which shows that $\tilde{\Omega}_{\bm k}^1\setminus A_{\bm k}^1$ is $0$-good. Since \eqref{to1-1} and $\bm k\notin Q_1$, we have
		\begin{align}\label{tk0}
			\|\mathcal{T}_{\tilde{\Omega}_{\bm k}^1}^{-1}\|_0<\delta_0^{-2}\|\theta+\bm k\cdot\bm\omega-\theta_1\|_{\T}^{-1}\cdot\|\theta+\bm k\cdot\bm\omega+\theta_1\|_{\T}^{-1}\lesssim\delta_1^{-\frac{31}{15}}.
		\end{align}
		
		Fix $\bm l\in G$. Then there exists a $0$-$\good$ set $U_{\bm l}=\lg_{\frac{N_1}{2}}(\bm l)\cap X$ such that $\bm l\in U_{\bm l}$, $\dist(\bm l,X\setminus U_{\bm l})\ge\frac{N_1}{2}.$  Define $\mathcal{Q}_{\bm l}=\ep \mathcal{T}_{U_{\bm l}}^{-1}\mathcal{R}_{U_{\bm l}}\mathcal{W}\mathcal{R}_{X\setminus U_{\bm l}}\in\mathbf{M}_{X\setminus U_{\bm l}}^{U_{\bm l}}$. Using \eqref{tame} leads to 
		\begin{align}\label{1qa1}
			\|\mathcal{Q}_{\bm l}\|_{\alpha_1}\lesssim \ep \|\mathcal T_{U_{\bm l}}^{-1}\|_{\alpha_1}\|\mathcal W_X\|_{\alpha_1}\lesssim \delta_0^{28}<1.
		\end{align}
		By \eqref{tame} and \eqref{0g},  we have for $\alpha\in(0,\alpha_1],$
		\begin{align}
			\nonumber\|\mathcal{Q}_{\bm l}\|_{\alpha+\alpha_0}&\lesssim\ep \left(\|\mathcal T_{U_{\bm l}}^{-1}\|_{\alpha+\alpha_0}\|\mathcal  W_X\|_{0}+\|\mathcal  T_{U_{\bm l}}^{-1}\|_{0}\|\mathcal  W_X\|_{\alpha+\alpha_0}\right)\\
			\label{1qa+}&\lesssim\ep \|\mathcal  W_X\|_{\alpha_1+\alpha_0}\|\mathcal  T_{U_{\bm l}}^{-1}\|_{\alpha_1+\alpha_0}\lesssim \delta_0^{28}<1.
		\end{align}
		
		We now vary $\bm l\in G$. Define $\mathcal{K},\mathcal{L}\in\mathbf{M}_{X}^{G}$ as 
		\begin{align*}
			\mathcal{K}(\bm l,\bm l')=\left\{\begin{array}{ll}
				0, & \text{for}\ \bm l'\in U_{\bm l}\\
				\mathcal{Q}_{\bm l}(\bm l,\bm l'), &\text{for}\ \bm  l'\in X\setminus U_{\bm l},
			\end{array}\right.
		\end{align*}
		and
		\begin{align*}
			\mathcal{L}(\bm l,\bm l')=\left\{\begin{array}{ll}
				\mathcal{T}_{U_{\bm l}}^{-1}(\bm l,\bm l'), & \text{for}\ \bm l'\in U_{\bm l}\\
				0, &\text{for}\ \bm l'\in X\setminus U_{\bm l}.
			\end{array}\right.
		\end{align*}
		Direct computation yields 
		\begin{align}\label{eq1}
			\mathcal{L}\mathcal{T}_X=\mathcal{R}_G+\mathcal{K}.
		\end{align}
		We estimate $\mathcal{K}\in\mathbf{M}_{X}^{G}$. Fix $\bm l\in G$. Note that if $\bm l'\in X\setminus U_{\bm l}$, then $\|\bm l-\bm l'\|\ge\frac{N_1}{2}$. This implies $\mathcal{K}^{\{\bm l\}}(\bm l,\bm l')=0$ for $\|\bm l-\bm l'\|<\frac{N_1}{2}$. By \eqref{smo1}, \eqref{re} and \eqref{1qa1}, we obtain
		\begin{align}
			\nonumber\|\mathcal{K}\|_0&\lesssim \sup_{\bm l\in G}\|\mathcal{K}^{\{\bm l\}}\|_{\alpha_0}\lesssim N_1^{-\alpha_1+\alpha_0}\sup_{\bm l\in G}\|\mathcal{K}^{\{\bm l\}}\|_{\alpha_1}\\
			\label{k0}&\le N_1^{-\alpha_1+\alpha_0}\sup_{\bm l\in G}\|\mathcal{Q}_{\bm l}\|_{\alpha_1}\le N_1^{-\alpha_1+\alpha_0}.
		\end{align}
		Similarly, by recalling \eqref{re} and \eqref{1qa+}, we obtain for $\alpha\in(0,\alpha_1], $ 
		\begin{align}\label{ka}
			\|\mathcal{K}\|_{\alpha}&\lesssim \sup_{\bm l\in G}\|\mathcal{K}^{\{\bm l\}}\|_{\alpha+\alpha_0}\le \sup_{\bm l\in G}\|\mathcal{Q}_{\bm l}\|_{\alpha+\alpha_0}\lesssim \delta_0^{28}<1.
		\end{align}
		We then estimate $\mathcal{L}\in\mathbf{M}_X^G$.  By \eqref{re} and \eqref{0g}, we have for $\alpha\in[0,\alpha_1]$,
		\begin{align}
			\label{la}\|\mathcal{L}\|_{\alpha}&\lesssim\sup_{\bm l\in G}\|\mathcal{L}^{\{\bm l\}}\|_{\alpha+\alpha_0}\le\sup_{\bm l\in G}\|T_{U_{\bm l}}^{-1}\|_{\alpha+\alpha_0}\lesssim \delta_0^{-2}.
		\end{align}
		Since $N_1\gg1$ and $-\alpha_1+\alpha_{0}<0$, we have 
		\begin{align*}
			\|\mathcal{K}\|_0\le\frac{1}{2}.
		\end{align*}
		Recalling Lemma \ref{pa}, we have that $\mathcal{I}_G+\mathcal{K}_G$ is invertible and
		\begin{align}\label{I+G}
			\|(\mathcal{I}_G+\mathcal{K}_G)^{-1}\|_{\alpha}&\lesssim \|\mathcal{K}\|_{\alpha}<1\ \text{for $\alpha\in[0,\alpha_1]$}.
		\end{align}
		From \eqref{eq1}, we have
		\begin{align*}
			(\mathcal{I}_G+\mathcal{K}_G)^{-1}\mathcal{L}\mathcal{T}_X=\mathcal{R}_G+(\mathcal{I}_G+\mathcal{K}_G)^{-1}\mathcal{K} \mathcal{R}_B.
		\end{align*}

		Let
		\begin{align*}
			\mathcal{M}=(\mathcal{I}_G+\mathcal{K}_G)^{-1}\mathcal{L}\in\mathbf{M}_X^G,\ \mathcal{N}=(\mathcal{I}_G+\mathcal{K}_G)^{-1}\mathcal{K} \mathcal{R}_B\in\mathbf{M}_B^G.
		\end{align*}
		Then we have
		\begin{align*}
			\mathcal{M}\mathcal{T}_X=\mathcal{R}_G+\mathcal{N}.
		\end{align*}
		Combining \eqref{tame} and \eqref{k0}--\eqref{I+G}  implies 
		\begin{align*}
			\|\mathcal{N}\|_0&\le\|(\mathcal{I}_G+\mathcal{K}_G)^{-1}\|_0\|\mathcal{K}\|_0\lesssim N_1^{-\alpha_1+\alpha_0},\\
			\|\mathcal{M}\|_0&\le\|(\mathcal{I}_G+\mathcal{K}_G)^{-1}\|_0\|\mathcal{L}\|_0\lesssim \delta_0^{-2}, 
		\end{align*}
		and for $\alpha\in(0,\alpha_1]$,
		\begin{align*}
			\|\mathcal{N}\|_\alpha&\lesssim \left(\|(\mathcal{I}_G+\mathcal{K}_G)^{-1}\|_\alpha\|\mathcal{K}\|_0+\|(\mathcal{I}_G+\mathcal{K}_G)^{-1}\|_0\|\mathcal{K}\|_\alpha\right)\lesssim\delta_0^{28}<1,\\
			\|\mathcal{M}\|_\alpha&\lesssim\left(\|(\mathcal{I}_G+\mathcal{K}_G)^{-1}\|_\alpha\|\mathcal{L}\|_0+\|(\mathcal{I}_G+\mathcal{K}_G)^{-1}\|_0\|\mathcal{L}\|_\alpha\right)\lesssim\delta_0^{-2}.
		\end{align*}
		
		We finish the proof. 
	\end{proof}
	
	We have  further 
	\begin{lem}\label{tz}
		Let \begin{align}\label{Z}
			\mathcal{T}'=\mathcal{T}_X\mathcal{R}_B-\mathcal{T}_X\mathcal{R}_G \mathcal{N}\in\mathbf{M}_B^X,\ \mathcal{Z}=\mathcal{I}_X-\mathcal{T}_X\mathcal{R}_G \mathcal{M}\in\mathbf{M}_X^X. 
		\end{align}
		Then 
		\begin{align}\label{ztt}
			\mathcal{Z}\mathcal{T}_X=\mathcal{T}' 
		\end{align}
		and for $\alpha\in[0, \alpha_1],$
		\begin{align}\label{ez}
			\|\mathcal{T}'\|_\alpha\lesssim1,\ \|\mathcal{Z}\|_\alpha\lesssim\delta_0^{-2}.
		\end{align}
		Moreover, $\mathcal{R}_B\mathcal{T}_X^{-1}$ is a left inverse of $\mathcal{T}'$.
	\end{lem}
	\begin{proof}
		Recalling \eqref{eq2} and \eqref{Z}, we get
		\begin{align*}
			\mathcal{Z}\mathcal{T}_X&=\mathcal{T}_X-\mathcal{T}_X \mathcal{R}_G\mathcal{M}\mathcal{T}_X=\mathcal{T}_X(\mathcal{I}_X-\mathcal{R}_G-\mathcal{R}_G\mathcal{N})\\
			&=\mathcal{T}_X \mathcal{R}_B-\mathcal{T}_X \mathcal{R}_G\mathcal{N}=\mathcal{T}'.
		\end{align*}
		
		We will prove  \eqref{ez}. % for $\mathcal{T}'$ and $\mathcal{Z}$.
		For $\alpha\in[0,\alpha_1]$, we obtain
		\begin{align*}
			\|\mathcal{T}_X\|_\alpha\le\|\ep \mathcal{W}_X\|_\alpha+\|\mathcal{D}_X\|\lesssim1.		
		\end{align*}
		From \eqref{tame}, we have for $\alpha\in[0,\alpha_1], $
		\begin{align*}
			\|\mathcal{T}'\|_\alpha&\lesssim \|\mathcal{T}_X\|_\alpha+\|\mathcal{T}_X\|_\alpha\|\mathcal{N}\|_0+\|\mathcal{T}_X\|_0\|\mathcal{N}\|_\alpha\lesssim1,\\
			\|\mathcal{Z}\|_\alpha&\lesssim \|\mathcal{T}_X\|_\alpha\|\mathcal{M}\|_0+\|\mathcal{T}_X\|_0\|\mathcal{M}\|_\alpha\lesssim\delta_0^{-2}.
		\end{align*}

		Finally, direct computation shows
		\begin{align*}
			\mathcal{R}_B\mathcal{T}_X^{-1}R_X\mathcal{T}=\mathcal{R}_B\mathcal{T}_X^{-1}\mathcal{T}_X(\mathcal{R}_B-\mathcal{R}_G\mathcal{N})=\mathcal{I}_B-\mathcal{R}_B\mathcal{R}_G\mathcal{N}=\mathcal{I}_B,
		\end{align*}
		which implies $\mathcal{R}_B \mathcal{T}_X^{-1} R_X$ is a left inverse of $\mathcal{T}'$.
	\end{proof}
	
	Now $\mathcal{T}'\in\mathbf{M}_B^X$ has a left inverse with $0$-norm $O(\delta_1^{-\frac{31}{15}})$. We will utilize Lemma \ref{pa} to obtain  another left inverse $\mathcal{V}$ of $\mathcal{T}'$  so that   for $\alpha\in[0,\alpha_1], $
	\begin{align*}
		\|\mathcal{V}\|_\alpha\lesssim \zeta_1^{\alpha}\delta_1^{-\frac{31}{15}}.
	\end{align*}
	The details are given by 
	\begin{lem}\label{t'v1}
		The operator $\mathcal{T}'$ defined in \eqref{Z} has a left inverse $\mathcal{V}$ satisfying  for $\alpha\in[0,\alpha_1], $
		\begin{align}\label{ev1}
			\|\mathcal{V}\|_\alpha\lesssim\zeta_1^{\alpha}\delta_1^{-\frac{31}{15}}.
		\end{align}
	\end{lem}		
	\begin{proof}		
		We introduce $Y=\lg_{\zeta_1}(\bm k)$  and let  $\mathcal{E}=\mathcal{R}_{Y}\mathcal{T}'\in\mathbf{M}_B^Y$.
		
		We claim that $\mathcal{E}$ has a left inverse $\mathcal{V}$ satisfying 
		\begin{align*}
			\|\mathcal{V}\|_0<2\delta_1^{-\frac{31}{15}}.
		\end{align*}
		In fact, let $\|\bm l-\bm l'\|\le \frac{\diam(Y)}{2}$ and $\mathcal{P}=\mathcal{T}'-\mathcal{E}$, which implies \begin{align*}
			\mathcal{P}(\bm l,\bm l')=0.
		\end{align*}
		Then recalling \eqref{smo1} and \eqref{ez}, we obtain for $\alpha\in[0,\alpha_1],$
		\begin{align*}
			\|\mathcal{P}\|_\alpha&\lesssim (\diam (Y))^{-\alpha_1+\alpha}\|\mathcal{P}\|_{\alpha_1}\le (\diam (Y))^{-\alpha_1+\alpha}\|\mathcal{T}'\|_{\alpha_1}\lesssim \zeta_1^{-\alpha_1+\alpha}.
		\end{align*}
		Thus by \eqref{indpa}, \eqref{tk0} and $\alpha_1>2200\tau$, we have
		\begin{align*}
			\|\mathcal{P}\|_0\|\mathcal{R}_B\mathcal{T}_X^{-1}\|_0&\le\|\mathcal{P}\|_0\|\mathcal{T}_X^{-1}\|_0\lesssim \zeta_1^{-\alpha_1}\delta_1^{-\frac{31}{15}}\\
			&\lesssim N_1^{-\alpha_1+1860\tau}<\frac{1}{2}.
		\end{align*}
		Hence it follows from the Lemma \ref{pa} that $\mathcal{E}$ has a left inverse $\mathcal{V}\in\mathbf{M}_Y^B$ satisfying 
		\begin{align}\label{v01} 
			\|\mathcal{V}\|_0\le 2\|\mathcal{R}_B\mathcal{T}_X^{-1}\|_0<2\delta_1^{-\frac{31}{15}}.
		\end{align}
		
		We now establish $\alpha$-norm estimates. If $\|\bm l-\bm l'\|\ge 2\diam(Y)$, we have
		\begin{align*}
			\mathcal{V}(\bm l,\bm l')=0.
		\end{align*}
		Using \eqref{smo2} and \eqref{v01} yields for $\alpha\in[0,\alpha_1]$,
		\begin{align*}
			\|\mathcal{V}\|_\alpha\lesssim(\diam(Y))^{\alpha}\|\mathcal{V}\|_0\lesssim\zeta_1^{\alpha}\delta_1^{-\frac{31}{15}}.
		\end{align*}

		Finally, we can obtain 
		\begin{align*}
			\mathcal{V}\mathcal{P}=\mathcal{R}_B\mathcal{V}\mathcal{R}_{Y}(\mathcal{T}'-\mathcal{R}_Y \mathcal{T}')=0.
		\end{align*}
		Since  $\mathcal{T}'=\mathcal{E}+\mathcal{P}$ and $\mathcal{V}$ is a left inverse of $\mathcal{E}$, $\mathcal{V}$ is a left inverse of $\mathcal{T}'$. 
	\end{proof}
	
	We are ready to prove Theorem \ref{psq1}: 
	\begin{proof}[Proof of Theorem \ref{psq1}]
		Since \eqref{eq2}, \eqref{ztt} and Lemma \ref{t'v1}, we obtain
		\begin{align*}
			\left(\mathcal{M}-\mathcal{N}\mathcal{V}\mathcal{Z}+\mathcal{V}\mathcal{Z}\right)\mathcal{T}_X&=\mathcal{R}_G+\mathcal{N}+(\mathcal{R}_B-\mathcal{N})\mathcal{V}\mathcal{T}'\\
			&=\mathcal{R}_G\oplus\mathcal{R}_B=\mathcal{I}_X,
		\end{align*}
		which implies
		\begin{align*}
			\mathcal{T}_X^{-1}=\mathcal{M}-\mathcal{N}\mathcal{V}\mathcal{Z}+\mathcal{V}\mathcal{Z}.
		\end{align*}
		Hence 
		\begin{align*}
			\mathcal{R}_B\mathcal{T}_X^{-1}=\mathcal{V}\mathcal{Z},\ \mathcal{R}_G \mathcal{T}_X^{-1}=\mathcal{M}-\mathcal{N}\mathcal{R}_B\mathcal{T}_X^{-1}.
		\end{align*}
		Then for $\alpha\in(0,\alpha_1]$, we get  by using \eqref{tame}, \eqref{emn1}, \eqref{ez} and \eqref{ev1}  that 
		\begin{align*}
			\|\mathcal{R}_B\mathcal{T}_X^{-1}\|_\alpha&\lesssim\left(\|\mathcal{V}\|_\alpha\|\mathcal{Z}\|_0+\|\mathcal{V}\|_0\|\mathcal{Z}\|_\alpha\right)\\
			&\lesssim\zeta_1^{\alpha}\delta_0^{-2}\delta_1^{-\frac{31}{15}},\\
			\|\mathcal{R}_G \mathcal{T}_X^{-1}\|_\alpha&\lesssim \|\mathcal{M}\|_\alpha+\|\mathcal{N}\|_\alpha\|\mathcal{R}_B \mathcal{T}_X^{-1}\|_0+\|\mathcal{N}\|_0\|\mathcal{R}_B \mathcal{T}_X^{-1}\|_\alpha\\
			&\lesssim\zeta_1^{\alpha}\delta_0^{-2}\delta_1^{-\frac{31}{15}}.
		\end{align*}
		Thus for any $\alpha\in(0,\alpha_1]$, we have
		\begin{align*}
			\|\mathcal{T}_X^{-1}\|_\alpha&\le\|\mathcal{R}_B \mathcal{T}_X^{-1}\|_\alpha+\|\mathcal{R}_G \mathcal{T}_X^{-1}\|_\alpha\lesssim \zeta_1^{\alpha}\delta_0^{-2}\delta_1^{-\frac{31}{15}}<\zeta_1^{\alpha}\delta_1^{-\frac{7}{3}},
		\end{align*} 
		which concludes the proof of Theorem \ref{psq1}. 
	\end{proof}
	
	\ \\

	\begin{itemize}
		\item[\textbf{Step 3}]: \textbf{Estimates of $\|\mathcal{T}_{\lg}^{-1}\|_{\alpha}$ for general $1$-good $\lg$}.
	\end{itemize}
	In this step, we will complete the verification of $\mathscr{P}_1$.
	We recall that a finite set $\lg\subset\Z^d$ is 1-$\good$ iff 
	\begin{align*}
		\left\{\begin{array}{l}
			\lg\cap Q_0\cap\Omega_{\bm k}^1\ne\emptyset\Rightarrow\tilde{\Omega}_{\bm k}^1\subset\lg,\\
			\{\bm k\in P_1:\ \tilde{\Omega}_{\bm k}^1\subset\lg\}\cap Q_1=\emptyset.
		\end{array}\right.
	\end{align*}
	
	We will combine the estimates of $\mathcal{T}_{\tilde{\Omega}_{\bm k}^1}^{-1}$, smoothing property and rows estimate  to finish this verification.
	\begin{thm}\label{1g}
		If $\lg$ is $1$-$\good$, then
		\begin{align*}
			\|\mathcal T_\lg^{-1}\|_0&<\delta_1^{-\frac{2}{15}}\times\sup\limits_{\{\bm k\in P_1:\ \tilde{\Omega}_{\bm k}^1\subset\lg\}}\left(\|\theta+\bm k\cdot\bm\omega-\theta_1\|_{\T}^{-1}\cdot\|\theta+\bm k\cdot\bm\omega+\theta_1\|_{\T}^{-1}\right),
		\end{align*}
		and for $\alpha\in(0,\alpha_1],$
		\begin{align*}
			\|\mathcal T_\lg^{-1}\|_\alpha&<\zeta_1^{\alpha}\delta_1^{-\frac{14}{3}}.
		\end{align*}
	\end{thm}
	\begin{proof}
		%We can prove Theorem \ref{1g} now. 
		Define
		\begin{align*}
			\tilde{P}_1=\{\bm k\in P_1:\ \lg\cap\Omega_{\bm k}^1\cap Q_0\ne\emptyset\}.
		\end{align*} 
		For $\bm l\in\lg$,  let 
		\begin{align*}
			U_{\bm l}=\left\{\begin{array}{ll}
				\lg_{\frac{1}{2}N_1}(\bm l)\cap\lg, &\text{if}\ \bm l\notin\bigcup_{\bm k\in \tilde{P}_1}\Omega_{\bm k}^1,\\
				\tilde{\Omega}_{\bm k}^1, &\text{if}\ \bm l\in\bigcup_{\bm k\in \tilde{P}_1}\Omega_{\bm k}^1.
			\end{array}\right.
		\end{align*}
		Denote  $\mathcal{Q}_{\bm l}=\ep \mathcal{T}_{U_{\bm l}}^{-1}\mathcal{R}_{U_{\bm l}}\mathcal{W}\mathcal{R}_{\lg\setminus U_{\bm l}}\in\mathbf{M}_{\lg\setminus U_{\bm l}}^{U_{\bm l}}$. By varying  $\bm l\in\lg$, we are led to considering   
		\begin{align*}
			\mathcal{K}(\bm l,\bm l')=\left\{\begin{array}{ll}
				0, & \text{for}\ \bm l'\in U_{\bm l}\\
				\mathcal{Q}_{\bm l}(\bm l,\bm l'), &\text{for}\ \bm l'\in \lg\setminus U_{\bm l},
			\end{array}\right.
		\end{align*}
		and
		\begin{align*}
			\mathcal{L}(\bm l,\bm l')=\left\{\begin{array}{ll}
				\mathcal{T}_{U_{\bm l}}^{-1}(\bm l,\bm l'), & \text{for}\ \bm l'\in U_{\bm l}, \\
				0, &\text{for}\ \bm l'\in \lg\setminus U_{\bm l}.
			\end{array}\right.
		\end{align*}
		Direct computation shows 
		\begin{align}\label{eq3}
			\mathcal{L}\mathcal{T}_\lg=\mathcal{I}_\lg+\mathcal{K}.
		\end{align}
		
		We will estimate $\mathcal{K}\in\mathbf{M}_{\lg}^{\lg}$ and $\mathcal{L}\in\mathbf{M}_\lg^\lg$ in the  following cases.\\
		(1) Assume  $\bm l\notin\bigcup_{\bm k\in \tilde{P}_1}\Omega_{\bm k}^1$.  Then $U_{\bm l}$ is $0$-$\good$. Similar to the proof of Lemma \ref{mn},  we obtain  
		\begin{align*}
			&\|\mathcal{K}^{\{\bm l\}}\|_{\alpha_0}\lesssim N_1^{-\alpha_1+\alpha_0},\\
		\end{align*}
		and for $\alpha\in(0,\alpha_1],$
		\begin{align*}
			&\|\mathcal{K}^{\{\bm l\}}\|_{\alpha+\alpha_0}\lesssim \delta_0^{28}<1,\\
			&\|\mathcal{L}^{\{\bm l\}}\|_{\alpha+\alpha_0}\lesssim\delta_0^{-2}.
		\end{align*}
		\ \\ 
		(2) Assume  $\bm l\in\bigcup_{\bm k\in \tilde{P}_1}\Omega_{\bm k}^1$.  Then there exists  some $\bm k\in\tilde{P}_1$ such that $\bm l\in \Omega_{\bm k}^1$.  By \eqref{tame} and \eqref{tg1a}, we get 
		\begin{align}\label{qla1}
			\|\mathcal{Q}_{\bm l}\|_{\alpha_1}\lesssim \ep\|\mathcal{T}_{U_{\bm l}}^{-1}\|_{\alpha_1}\|\mathcal{W}_\lg\|_{\alpha_1}\lesssim\zeta_1^{\alpha_1}\delta_1^{-\frac{7}{3}}.
		\end{align}
		Combining \eqref{tame}, \eqref{smo2} and \eqref{tg1a},  we have  for $\alpha\in(0,\alpha_1],$ 
		\begin{align*}
			\|\mathcal{Q}_{\bm l}\|_{\alpha+\alpha_0}&\lesssim \left(\|\mathcal{T}_{U_{\bm l}}^{-1}\|_{\alpha+\alpha_0}\|\mathcal{W}_\lg\|_{0}+\|\mathcal{T}_{U_{\bm l}}^{-1}\|_{0}\|\mathcal{W}_\lg\|_{\alpha+\alpha_0}\right)\\
			&\lesssim \tilde{\zeta}_1^{\alpha_0}\|\mathcal{T}_{U_{\bm l}}^{-1}\|_{\alpha}\|\mathcal{W}_\lg\|_{\alpha_1+\alpha_0}\\
			&\lesssim\zeta_1^{\alpha} \tilde{\zeta}_1^{\alpha_0}\delta_1^{-\frac{7}{3}}.
		\end{align*}
		Note that if $\bm l'\in \lg\setminus U_{\bm l}$, then $\|\bm l-\bm l'\|\ge\frac{\zeta_1}{2}$. This implies $\mathcal{K}^{\{\bm l\}}(\bm l,\bm l')=0$ for $\|\bm l-\bm l'\|<\frac{\zeta_1}{2}$. By \eqref{smo1} and \eqref{qla1},  we obtain
		\begin{align*}
			\|\mathcal{K}^{\{\bm l\}}\|_{\alpha_0}&\lesssim \tilde{\zeta}_1^{-\alpha_1+\alpha_0}\|\mathcal{K}^{\{\bm l\}}\|_{\alpha_1}\\
			&\lesssim \tilde{\zeta}_1^{-\alpha_1+\alpha_0}\|\mathcal{Q}_{\bm l}\|_{\alpha_1}\\
			&\lesssim \zeta_1^{\alpha_1} \tilde{\zeta}_1^{-\alpha_1+\alpha_0}\delta_1^{-\frac{7}{3}}.
		\end{align*}
		Similarly, for $\alpha\in(0,\alpha_1]$, we obtain 
		\begin{align*}
			\|\mathcal{K}^{\{\bm l\}}\|_{\alpha+\alpha_0}\le\|\mathcal{Q}_{\bm l}\|_{\alpha+\alpha_0}\lesssim\zeta_1^{\alpha} \tilde{\zeta}_1^{\alpha_0}\delta_1^{-\frac{7}{3}}.
		\end{align*}
		By the definition of $U_{\bm l}$, if $\|\bm l-\bm l'\|>2\tilde{\zeta}_1$, then $\bm l'\notin U_{\bm l}$. This implies $\mathcal{L}^{\{\bm l\}}(\bm l,\bm l')=0$ for $\|\bm l-\bm l'\|>2\tilde{\zeta}_1$. By \eqref{smo2}, \eqref{to1-1} and \eqref{tg1a},  we have 
		\begin{align*}
			\|\mathcal{L}^{\{\bm l\}}\|_{\alpha_0}&\lesssim \tilde{\zeta}_1^{\alpha_0}\|\mathcal{L}^{\{\bm l\}}\|_0\le \tilde{\zeta}_1^{\alpha_0}\|\mathcal{T}_{U_{\bm l}}^{-1}\|_0\\
			&\le\tilde{\zeta}_1^{\alpha_0}\delta_0^{-2}\|\theta+\bm k\cdot\bm\omega-\theta_1\|_{\T}^{-1}\cdot\|\theta+\bm k\cdot\bm\omega+\theta_1\|_{\T}^{-1}, 
		\end{align*}
		and for $\alpha\in(0,\alpha_1],$
		\begin{align*}
			\|\mathcal{L}^{\{\bm l\}}\|_{\alpha+\alpha_0}&\lesssim \tilde{\zeta}_1^{\alpha_0}\|\mathcal{L}^{\{\bm l\}}\|_{\alpha}\le \tilde{\zeta}_1^{\alpha_0}\|\mathcal{T}_{U_{\bm l}}^{-1}\|_{\alpha}\\
			&\lesssim\zeta_1^{\alpha} \tilde{\zeta}_1^{\alpha_0}\delta_1^{-\frac{7}{3}}.
		\end{align*}
		Next, using rows estimate (cf.  \eqref{re}) yields  for $\alpha\in(0,\alpha_1], $
		\begin{align*}
			\|\mathcal{K}\|_0\lesssim &\sup_{\bm l\in \lg}\|\mathcal{K}^{\{\bm l\}}\|_{\alpha_0}\lesssim\max\left(N_1^{-\alpha_1+\alpha_0}, \zeta_1^{\alpha_1} \tilde{\zeta}_1^{-\alpha_1+\alpha_0}\delta_1^{-\frac{7}{3}}\right),\\
			\|\mathcal{K}\|_{\alpha}\lesssim& \sup_{\bm l\in \lg}\|\mathcal{K}^{\{\bm l\}}\|_{\alpha+\alpha_0}\lesssim\zeta_1^{\alpha} \tilde{\zeta}_1^{\alpha_0}\delta_1^{-\frac{7}{3}},
		\end{align*}
		and
		\begin{align}
			\nonumber\|\mathcal{L}\|_{0}&\lesssim \sup_{\bm l\in \lg}\|\mathcal{L}^{\{\bm l\}}\|_{\alpha_0}\\
			\label{l0l}&\lesssim\tilde{\zeta}_1^{\alpha_0}\delta_0^{-2}
			\times\sup\limits_{\{\bm k\in P_1:\ \tilde{\Omega}_{\bm k}^1\subset\lg\}}\left(\|\theta+\bm k\cdot\bm\omega-\theta_1\|_{\T}^{-1}\cdot\|\theta+\bm k\cdot\bm\omega+\theta_1\|_{\T}^{-1}\right),\\
			\label{lal}\|\mathcal{L}\|_{\alpha}&\lesssim\sup_{\bm l\in \lg}\|\mathcal{L}^{\{\bm l\}}\|_{\alpha+\alpha_0}\lesssim\zeta_1^{\alpha} \tilde{\zeta}_1^{\alpha_0}\delta_1^{-\frac{7}{3}}.
		\end{align}

		Finally, we use  tame estimate and Lemma \ref{pa} to finish the proof. Since $N_1\gg1$ and $\alpha_1>2200\tau$, we have
		\begin{align*}
			\|\mathcal{K}\|_0\le\frac{1}{2}.
		\end{align*}
		Recalling Lemma \ref{pa}, we have that $\mathcal{I}_\lg+\mathcal{K}$ is invertible and  
		\begin{align}\label{il+k}
			\|(\mathcal{I}_\lg+\mathcal{K})^{-1}\|_{\alpha}&\lesssim\min\{1,\|\mathcal{K}\|_\alpha\}\ \text{for}\ \alpha\in[0,\alpha_1].
		\end{align}
		From \eqref{eq3}, we have
		\begin{align*}
			\mathcal{T}_\lg^{-1}=(\mathcal{I}_\lg+\mathcal{K})^{-1}\mathcal{L}.
		\end{align*}
		Recalling \eqref{tame} and \eqref{l0l}--\eqref{il+k},  we have
		\begin{align*}
			\|\mathcal{T}_\lg^{-1}\|_0&\lesssim\|(\mathcal{I}_\lg+\mathcal{K})^{-1}\|_0\|\mathcal{L}\|_0\\
			&<\delta_1^{-\frac{2}{15}}\times\sup\limits_{\{\bm k\in P_1:\ \tilde{\Omega}_{\bm k}^1\subset\lg\}}\left(\|\theta+\bm k\cdot\bm\omega-\theta_1\|_{\T}^{-1}\cdot\|\theta+\bm k\cdot\bm\omega+\theta_1\|_{\T}^{-1}\right)
		\end{align*}
		and for $\alpha\in(0,\alpha_1]$,
		\begin{align*}
			\|\mathcal T_\lg^{-1}\|_\alpha&\lesssim\left(\|(\mathcal{I}_\lg+\mathcal{K})^{-1}\|_\alpha\|\mathcal{L}\|_0+\|(\mathcal{I}_\lg+\mathcal{K})^{-1}\|_0\|\mathcal{L}\|_\alpha\right)\\
			&\lesssim \zeta_1^{\alpha}\tilde{\zeta}_1^{2\alpha_0}\delta_0^{-2}\delta_1^{-\frac{7}{3}}\times\sup\limits_{\{\bm k\in P_1:\ \tilde{\Omega}_{\bm k}^1\subset\lg\}}\left(\|\theta+\bm k\cdot\bm\omega-\theta_1\|_{\T}^{-1}\cdot\|\theta+\bm k\cdot\bm\omega+\theta_1\|_{\T}^{-1}\right)\\
			&<\zeta_1^{\alpha}\delta_1^{-\frac{14}{3}}.
		\end{align*}
	\end{proof}
	
	\subsection{The proof of Theorem \ref{ind}: (from $\mathscr{P}_s$ to $\mathscr{P}_{s+1}$)}
	We have finished the proof of $\mathscr{P}_1$ in the Subsection \ref{vo1}. Assume that $\mathscr{P}_s$ holds true. In order to complete the proof of Theorem \ref{ind}, it suffices to establish $\mathscr{P}_{s+1}$.
	
	We outline the main steps of the proof. First, we construct resonant blocks $\Omega_{\bm k}^{s+1}\ (\bm k\in P_{s+1})$ to cover $Q_{s}$. Second ,we use the Schur complement lemma (cf. Lemma \ref{scl} in the Appendix \ref{app}), resolvent identity, Cramer's rule and Hadamard's inequality to get
	\begin{align*}
		\|\mathcal{T}_{\tilde{\Omega}_{\bm k}^{s+1}}^{-1}\|_0<\delta_{s+1}^{-\frac{1}{15}}\|\theta+\bm k\cdot\bm\omega-\theta_{s+1}\|_{\T}^{-1}\cdot\|\theta+\bm k\cdot\bm\omega+\theta_{s+1}\|_{\T}^{-1},
	\end{align*}
	where $\tilde{\Omega}_{\bm k}^{s+1}$ is $(s+1)$-th enlarged resonant block and $\theta_{s+1}$ is obtained via Rouch\'e's theorem as one root of the Dirichlet determinant equation $\det\mathcal{T}_{\tilde{\Omega}_{\bm k}^{s+1}}(z-\bm k\cdot\bm\omega)=0$. Next, we use smoothing property, rows estimate and perturbation argument to get for $\alpha\in(0,\alpha_1]$,
	\begin{align*}
		\|\mathcal{T}_{\tilde{\Omega}_{\bm k}^{s+1}}^{-1}\|_{\alpha}<\zeta_{s+1}^{\alpha}\delta_{s+1}^{-\frac{1}{3}}\|\theta+\bm k\cdot\bm\omega-\theta_{s+1}\|_{\T}^{-1}\cdot\|\theta+\bm k\cdot\bm\omega+\theta_{s+1}\|_{\T}^{-1}
	\end{align*}
	via  a {\it constructive}  procedure. Finally, we combine the estimates of $\mathcal{T}_{\tilde{\Omega}_{\bm k}^{s+1}}^{-1}$ to obtain  that of $\mathcal{T}_\lg^{-1}$ in a similar way for more general $(s+1)$-$\good$ $\lg$.
	
	Recall that
	\begin{align*}
		Q_{s}^{\pm}&=\{\bm k\in P_{s}:\ \|\theta+\bm k\cdot\bm\omega\pm\theta_{s}\|_{\T}<\delta_{s}\},\ Q_{s}=Q_{s}^+\cup Q_{s}^-,\\
		\tilde{Q}_{s}^{\pm}&=\{\bm k\in P_{s}:\ \|\theta+\bm k\cdot\bm\omega\pm\theta_{s}\|_{\T}<\delta_{s}^{\frac{2}{3}}\},\ \tilde{Q}_{s}=\tilde{Q}_{s}^+\cup \tilde{Q}_{s}^-.
	\end{align*}
	
	We distinguish the verification into three steps.
	
	\begin{itemize}
		\item[\textbf{Step 1}]: \textbf{Estimates of $\|\mathcal{T}_{\tilde{\Omega}_{\bm k}^{s+1}}^{-1}\|_0$}.
	\end{itemize}
	In this step, we will use the resolvent indentity and Rouch\'e's theorem to find $\theta_{s+1}=\theta_{s+1}(E)$ so that
	\begin{align}
		\label{s+1T-1}\|\mathcal{T}_{\tilde{\Omega}_{\bm k}^{s+1}}^{-1}\|_0&<\delta_{s+1}^{-\frac{1}{15}}\|\theta+\bm k\cdot\bm\omega-\theta_{s+1}\|_{\T}^{-1}\cdot\|\theta+\bm k\cdot\bm\omega+\theta_{s+1}\|_{\T}^{-1}.
	\end{align}
	We again divide the discussion into two cases.
	
	\begin{itemize}
		\item[\textbf{Case 1}]: The case $(\bm C1)_s$ occurs, i.e.,
	\end{itemize}
	\begin{align}\label{dqs1-}
		\dist(\tilde{Q}_s^-,Q_s^+)>100N_{s+1}^3.
	\end{align}
	\begin{rem}
		We can prove similar to Remark \ref{dsq1} that
		\begin{align*}
			\dist(\tilde{Q}_s^-,Q_s^+)=\dist(\tilde{Q}_s^+,Q_s^-).
		\end{align*}
		Thus \eqref{dqs1-} also implies that
		\begin{align}\label{dqs1+}
			\dist(\tilde{Q}_s^+,Q_s^-)>100N_{s+1}^3.
		\end{align}
		%The proof is similar to that of Remark \ref{dsq1} and we omit the details.
	\end{rem}
	
	By \eqref{fs} and the definitions  of $Q_s^{\pm}$ and $\tilde{Q}_s^{\pm}$, we obtain
	\begin{align}
		\label{qspm}Q_s^{\pm}&=\left\{\bm k\in\Z^d+\frac{1}{2}\sum_{i=0}^{s-1}\bm l_i:\ \|\theta+\bm k\cdot\bm\omega\pm\theta_{s}\|_{\T}<\delta_{s}\right\},\\
		\nonumber\tilde{Q}_{s}^{\pm}&=\left\{\bm k\in\Z^d+\frac{1}{2}\sum_{i=0}^{s-1}\bm l_i:\ \|\theta+\bm k\cdot\bm\omega\pm\theta_{s}\|_{\T}<\delta_{s}^{\frac{2}{3}}\right\}.
	\end{align}
	
	Assuming \eqref{dqs1-} holds true, we define
	\begin{align}\label{ps+1}
		P_{s+1}=Q_s,\ \bm l_s=\bm 0.
	\end{align}
	By\eqref{qspm}, we have
	\begin{align}\label{ps+11}
		P_{s+1}\subset\left\{\bm k\in\Z^d+\frac{1}{2}\sum_{i=0}^{s}\bm l_i:\ \min_{\sigma=\pm1}(\|\theta+\bm k\cdot\bm\omega+\sigma\theta_{s}\|_{\T})<\delta_{s}\right\},
	\end{align}
	which proves \eqref{as1} in the case $(\bm C1)_{s+1}$. Thus from \eqref{dqs1+}, we obtain for $\bm k,\bm k'\in P_{s+1}$ with $\bm k\ne \bm k'$,
	\begin{align}\label{dk}
		\|\bm k-\bm k'\|\ge\min\left(100N_{s+1}^3,\left(\frac{\g}{2\delta_s}\right)^{\frac{1}{\tau}}\right)\ge100N_{s+1}^3.
	\end{align}
	In the following, we associate each $\bm k\in P_{s+1}$ with blocks $\Omega_{\bm k}^{s+1}$ and $\tilde{\Omega}_{\bm k}^{s+1}$ so that
	\begin{align*}
		&\lg_{N_{s+1}}(\bm k)\subset\Omega_{\bm k}^{s+1}\subset\lg_{N_{s+1}+50N_s^{5}}(\bm k),\\
		&\lg_{N_{s+1}^3}(\bm k)\subset\tilde{\Omega}_{\bm k}^{s+1}\subset\lg_{N_{s+1}^3+50N_s^{5}}(\bm k),
	\end{align*}
	and
	\begin{align}\label{sb}
		\left\{\begin{array}{l}
			\Omega_{\bm k}^{s+1}\cap\tilde{\Omega}_{\bm k'}^{s'}\ne\emptyset\ (s'<s+1)\Rightarrow\tilde{\Omega}_{\bm k'}^{s'}\subset\Omega_{\bm k}^{s+1},\\
			\tilde{\Omega}_{\bm k}^{s+1}\cap\tilde{\Omega}_{\bm k'}^{s'}\ne\emptyset\ (s'<s+1)\Rightarrow\tilde{\Omega}_{\bm k'}^{s'}\subset\tilde{\Omega}_{\bm k}^{s+1},\\
			\dist(\tilde{\Omega}_{\bm k}^{s+1},\tilde{\Omega}_{\bm k'}^{s+1})>10\tilde{\zeta}_{s+1}\ \text{for}\ \bm k\ne \bm k'\in P_{s+1}.
		\end{array}\right.
	\end{align}
	In addition, the translation of $\tilde{\Omega}_{\bm k}^{s+1}$
	\begin{align}\label{trans1}
		\tilde{\Omega}_{\bm k}^{s+1}-\bm k\subset\Z^d+\frac{1}{2}\sum_{i=0}^s \bm l_i
	\end{align}
	is independent of $\bm k\in P_{s+1}$ and symmetrical about the origin. The details  of proof  \eqref{sb} and \eqref{trans1} can be found in the page 23 of \cite{CSZ24a}. In summary, we have proven $(\bm a)_{s+1}$ and $(\bm b)_{s+1}$ in the case $(\bm C1)_{s}$.
	
	Now we turn to the proof of $(\bm c)_{s+1}$. First, we have that for $\bm k'\in Q_s\ (=P_{s+1})$,
	\begin{align*}
		\tilde{\Omega}_{\bm k'}^s\subset\tilde{\Omega}_{\bm k'}^{s+1}.
	\end{align*}
	For each $\bm k\in P_{s+1}$, we define
	\begin{align*}
		A_{\bm k}^{s+1}=A_{\bm k}^s.
	\end{align*}
	Then $A_{\bm k}^{s+1}\subset\Omega_{\bm k}^s$ and $\# A_{\bm k}^{s+1}=\# A_{\bm k}^s\le 2^s$. It remains to show that $\tilde{\Omega}_{\bm k}^{s+1}\setminus A_{\bm k}^{s+1}$ is $s$-$\good$ and the set $\tilde{\Omega}_{\bm k}^{s+1}-\bm k$ is independent of $\bm k\in P_{s+1}$ and symmetrical about the origin. The details can  be  found in the page 26 of \cite{CSZ24a} as well. This finishes the proof of $(\bm c)_{s+1}$ in the case $(\bm C1)_s$.

	Next, we try to prove $(\bm d)_{s+1}$ and $(\bm f)_{s+1}$ in the case of  $(\bm C1)_s$. For $\bm k\in Q_s^-$, we consider 
	\begin{align*}
		\mathcal{M}_{s+1}(z):=\mathcal{T}_{\tilde{\Omega}_{\bm k}^{s+1}-\bm k}(z)=((v(z+\bm n\cdot\bm\omega)-E)\delta_{\bm n,\bm n'}+\ep \mathcal{W})_{\bm n\in\tilde{\Omega}_{\bm k}^{s+1}-\bm k}
	\end{align*}
	defined in
	\begin{align}\label{zs1}
		\{z\in\C:\ |z-\theta_{s}|<\delta_s^{\frac{18}{19}}\}.
	\end{align}
	If $\bm k'\in P_s$ and $\tilde{\Omega}_{\bm k'}^s\subset(\tilde{\Omega}_{\bm k}^{s+1}\setminus A_{\bm k}^{s+1})$, then $0\ne\|\bm k'-\bm k\|\le3N_{s+1}^3$. Thus
	\begin{align*}
		\|\theta+\bm k'\cdot\bm\omega-\theta_s\|_{\T}&\ge\|(\bm k-\bm k')\cdot\bm\omega\|_{\T}-\|\theta+\bm k\cdot\bm\omega-\theta_s\|_{\T}\\
		&\ge\frac{\g}{(3N_{s+1}^3)^\tau}-\delta_{s}\gtrsim\delta_s^{\frac{1}{10}}.
	\end{align*}
	By \eqref{dqs1+}, we have $\bm k'\notin\tilde{Q}_s^{+}$, and thus
	\begin{align*}
		\|\theta+\bm k'\cdot\bm\omega+\theta_s\|_{\T}>\delta_s^{\frac{2}{3}}.
	\end{align*}
	From $\tilde{\Omega}_{\bm k}^{s+1}\setminus A_{\bm k}^{s+1}$ is $s$-$\good$ (cf. $(\bm c)_{s+1}$) and \eqref{tsg01}, we obtain
	\begin{align}
		\nonumber\|\mathcal{T}_{\tilde{\Omega}_{\bm k}^{s+1}\setminus A_{\bm k}^{s+1}}^{-1}\|_{0}&< \delta_{s}^{-\frac{2}{15}}\times\sup\limits_{\{\bm k'\in P_s:\ \tilde{\Omega}_{\bm k'}^s\subset(\tilde{\Omega}_{\bm k}^{s+1}\setminus A_{\bm k}^{s+1})\}}(\|\theta+\bm k'\cdot\bm\omega-\theta_s\|_{\T}^{-1}\cdot\|\theta+\bm k'\cdot\bm\omega+\theta_s\|_{\T}^{-1})\\
		\label{ts0}&\lesssim\delta_s^{-\frac{9}{10}}.
	\end{align}
	One may restate as
	\begin{align*}
		\|((\mathcal{M}_{s+1}(\theta+\bm k\cdot\bm\omega))_{(\tilde{\Omega}_{\bm k}^{s+1}\setminus A_{\bm k}^{s+1})-\bm k})^{-1}\|_0&\lesssim\delta_s^{-\frac{9}{10}}.
	\end{align*}
	Notice that
	\begin{align}\label{qs-}
		\nonumber\|z-(\theta+\bm k\cdot\bm\omega)\|_{\T}&\le|z-\theta_s|+\|\theta+\bm k\cdot\bm\omega-\theta_s\|_{\T}\\
		&<\delta_s^{\frac{18}{19}}+\delta_s<2\delta_s^{\frac{18}{19}}.
	\end{align}
	Thus by Neumann series argument, we can show
	\begin{align}\label{375}
		\|((\mathcal{M}_{s+1}(z))_{(\tilde{\Omega}_{\bm k}^{s+1}\setminus A_{\bm k}^{s+1})-\bm k})^{-1}\|_0&\lesssim2\delta_s^{-\frac{9}{10}}.
	\end{align}
	Now, we can apply the Schur complement lemma to establish desired estimates. By Lemma \ref{scl}, $(\mathcal{M}_{s+1}(z))^{-1}$ is controlled by the Schur complement of $((\tilde{\Omega}_{\bm k}^{s+1}\setminus A_{\bm k}^{s+1})-\bm k)$
	\begin{align*}
		\mathcal{S}_{s+1}(z)&=(\mathcal{M}_{s+1}(z))_{A_{\bm k}^{s+1}-\bm k}-\mathcal{R}_{A_{\bm k}^{s+1}-\bm k}\mathcal{M}_{s+1}(z)\mathcal{R}_{(\tilde{\Omega}_{\bm k}^{s+1}\setminus A_{\bm k}^{s+1})-\bm k}\\
		&\ \  \times((\mathcal{M}_{s+1}(z))_{(\tilde{\Omega}_{\bm k}^{s+1}\setminus A_{\bm k}^{s+1})-\bm k})^{-1} \mathcal{R}_{(\tilde{\Omega}_{\bm k}^{s+1}\setminus A_{\bm k}^{s+1})-\bm k}\mathcal{M}_{s+1}(z)\mathcal{R}_{A_{\bm k}^{s+1}-\bm k}.
	\end{align*}
	Our next aim is to analyze $\det \mathcal{S}_{s+1}(z)$. Since $A_{\bm k}^{s+1}-\bm k=A_{\bm k}^s-\bm k\subset\Omega_{\bm k}^s-\bm k$ and $\dist(A_{\bm k}^{s},\p \tilde{\Omega}_{\bm k}^s)>\frac{1}{2}\tilde{\zeta}_s$, we obtain 
	\begin{align*}
		&\ \mathcal{R}_{A_{\bm k}^{s+1}-\bm k}\mathcal{M}_{s+1}(z)\mathcal{R}_{(\tilde{\Omega}_{\bm k}^{s+1}\setminus A_{\bm k}^{s+1})-\bm k}\\
		=&\ \mathcal{R}_{A_{\bm k}^{s}-\bm k}\mathcal{M}_{s+1}(z)\mathcal{R}_{(\tilde{\Omega}_{\bm k}^{s+1}\setminus A_{\bm k}^{s})-\bm k}\\
		=&\ \mathcal{R}_{A_{\bm k}^{s}-\bm k}\mathcal{M}_{s+1}(z)\mathcal{R}_{(\tilde{\Omega}_{\bm k}^{s}\setminus A_{\bm k}^{s})-\bm k}+\mathcal{R}_{A_{\bm k}^{s}-\bm k}\mathcal{M}_{s+1}(z)\mathcal{R}_{(\tilde{\Omega}_{\bm k}^{s+1}\setminus \tilde{\Omega}_{\bm k}^{s})-\bm k}\\
		=&\ \mathcal{R}_{A_{\bm k}^{s}-\bm k}\mathcal{M}_{s+1}(z)\mathcal{R}_{(\tilde{\Omega}_{\bm k}^{s}\setminus A_{\bm k}^{s})-\bm k}+O(\tilde{\zeta}_s^{-\alpha_1}).
	\end{align*}
	Similarly, we have
	\begin{align*}
		\mathcal{R}_{(\tilde{\Omega}_{\bm k}^{s+1}\setminus A_{\bm k}^{s+1})-\bm k}\mathcal{M}_{s+1}(z)\mathcal{R}_{A_{\bm k}^{s+1}-\bm k}=\mathcal{R}_{(\tilde{\Omega}_{\bm k}^{s}\setminus A_{\bm k}^{s})-\bm k}\mathcal{M}_{s+1}(z)\mathcal{R}_{A_{\bm k}^{s}-\bm k}+O(\tilde{\zeta}_s^{-\alpha_1}).
	\end{align*}
	Combining \eqref{375} and $\alpha_1>2200\tau$, we can get
	\begin{align*}
		\mathcal{S}_{s+1}(z)&=(\mathcal{M}_{s+1}(z))_{A_{\bm k}^{s}-\bm k}-\mathcal{R}_{A_{\bm k}^{s}-\bm k}\mathcal{M}_{s+1}(z)\mathcal{R}_{(\tilde{\Omega}_{\bm k}^{s}\setminus A_{\bm k}^{s})-\bm k}\\
		&\ \ \times((\mathcal{M}_{s+1}(z))_{(\tilde{\Omega}_{\bm k}^{s+1}\setminus A_{\bm k}^{s+1})-\bm k})^{-1} \mathcal{R}_{(\tilde{\Omega}_{\bm k}^{s}\setminus A_{\bm k}^{s})-\bm k}\mathcal{M}_{s+1}(z)\mathcal{R}_{A_{\bm k}^{s}-\bm k}+O(\delta_s^{6}).
	\end{align*}
	Since $\tilde{\Omega}_{\bm k}^s\setminus A_{\bm k}^s$ is $(s-1)$-$\good$ (cf. $(\bm c)_{s}$), by \eqref{tsg01} and $\eqref{tsg1}$, we have
	\begin{align*}
		\|\mathcal{T}_{\tilde{\Omega}_{\bm k}^s\setminus A_{\bm k}^s}^{-1}\|_0&\le \delta_{s-1}^{-\frac{32}{15}},\\
		\|\mathcal{T}_{\tilde{\Omega}_{\bm k}^s\setminus A_{\bm k}^s}^{-1}\|_{\alpha}&\le \zeta_{s-1}^{\alpha} \delta_{s-1}^{-\frac{14}{3}}.
	\end{align*}
	Equivalently,
	\begin{align*}
		\|((\mathcal{M}_{s+1}(\theta+\bm k\cdot\bm\omega))_{(\tilde{\Omega}_{\bm k}^s\setminus A_{\bm k}^s)-\bm k})^{-1}\|_0&\le  \delta_{s-1}^{-\frac{32}{15}},\\
		\|((\mathcal{M}_{s+1}(\theta+\bm k\cdot\bm\omega))_{(\tilde{\Omega}_{\bm k}^s\setminus A_{\bm k}^s)-\bm k})^{-1}\|_{\alpha}&\le\zeta_{s-1}^{\alpha} \delta_{s-1}^{-\frac{14}{3}}.
	\end{align*}
	In the set defined by \eqref{zs1}, we claim that
	\begin{align}
		\label{369}\|((\mathcal{M}_{s+1}(z))_{(\tilde{\Omega}_{\bm k}^s\setminus A_{\bm k}^s)-\bm k})^{-1}\|_0&\lesssim \delta_{s-1}^{-\frac{32}{15}},\\
		\label{370}\|((\mathcal{M}_{s+1}(z))_{(\tilde{\Omega}_{\bm k}^s\setminus A_{\bm k}^s)-\bm k})^{-1}\|_{\alpha}&\lesssim \zeta_{s-1}^{\alpha} \delta_{s-1}^{-\frac{14}{3}}.
	\end{align}
	\begin{proof}[Proof of the Claim]
		Define
		\begin{align*}
			\mathcal{T}_1=(\mathcal{M}_{s+1}(\theta+\bm k\cdot\bm\omega))_{(\tilde{\Omega}_{\bm k}^s \setminus A_{\bm k}^s)-\bm k},\ \mathcal{T}_2=(\mathcal{M}_{s+1}(z))_{(\tilde{\Omega}_{\bm k}^s \setminus A_{\bm k}^s)-\bm k}.
		\end{align*}
		Then $\mathcal{D}_1=\mathcal{T}_1-\mathcal{T}_2$ is diagonal so that $\|\mathcal{D}_1\|_0=\|\mathcal{D}_1\|_{\alpha}\lesssim\delta_s^{\frac{18}{19}}$ by \eqref{vdefn} and \eqref{qs-}. Using the Neumann series expansion yields
		\begin{align*}
			\mathcal{T}_2^{-1}=(\mathcal{I}_{(\tilde{\Omega}_{\bm k}^s \setminus A_{\bm k}^s)-\bm k}-\mathcal{T}_1^{-1}\mathcal{D}_1)^{-1}\mathcal{T}_1^{-1}=\sum_{i=0}^{\infty}(\mathcal{T}_1^{-1}\mathcal{D}_1)^i \mathcal{T}_1^{-1}.
		\end{align*}
		By \eqref{tame} and \eqref{369}, we have
		\begin{align*}
			\|\mathcal{T}_2^{-1}\|_\alpha&\le \|\mathcal{T}_1^{-1}\|_\alpha\sum_{i=0}^{\infty}(i+1)K(i+1,\alpha)(\|\mathcal{T}_1^{-1}\|_0\|\mathcal{D}_1\|_0)^i\\
			&\le\|\mathcal{T}_1^{-1}\|_\alpha\sum_{i=0}^{\infty}\frac{(i+1)K(i+1,\alpha)}{2^i}\lesssim \|\mathcal{T}_1^{-1}\|_\alpha,
		\end{align*}
		where $K(n,\alpha)$ is defined in \eqref{kns}.
	\end{proof}
	
	Next, we will use the resolvent identity and the decay  of $\mathcal W$ to estimate the difference of $\mathcal{S}_s$ and $\mathcal{S}_{s+1}$.
	For the convenience, we let $X=(\tilde{\Omega}_{\bm k}^s\setminus A_{\bm k}^s)-\bm k$, $Z_1=\lg_{\frac{\tilde{\zeta}_s}{4}}\cap X$, $Z_2=\lg_{\frac{\tilde{\zeta}_s}{8}}\cap X$ and $Y=(\tilde{\Omega}_{\bm k}^{s+1}\setminus A_{\bm k}^{s+1})-\bm k$. Let $\bm m\in X$. By the resolvent identity, we have for any $\bm n\in Y$,
	\begin{align*}
		&((\mathcal{M}_{s+1}(z))_Y)^{-1}(\bm m,\bm n)-\chi_X(\bm n)((\mathcal{M}_{s+1}(z))_X)^{-1}(\bm m,\bm n)\\
		=&-\ep\sum_{\bm l\in X \atop \bm l'\in Y\setminus X}((\mathcal{M}_{s+1}(z))_X)^{-1}(\bm m,\bm l)\mathcal{W}(\bm l,\bm l')((\mathcal{M}_{s+1}(z))_Y)^{-1}(\bm l',\bm n).
	\end{align*}
	If $\bm m\in Z_2$, since \eqref{oddgf}, $\dist(Z_1,Y\setminus X)\ge\frac{\tilde{\zeta}_s}{8}$ and $\dist(Z_2,X\setminus Z_1)\ge\frac{\tilde{\zeta}_s}{8}$, we can get
	\begin{align*}
		&|((\mathcal{M}_{s+1}(z))_Y)(\bm m,\bm n)-\chi_X(\bm n)((\mathcal{M}_{s+1}(z))_X)^{-1}(\bm m,\bm n)|\\
		\le&\sum_{\bm l\in Z_1 \atop \bm l'\in Y\setminus X}|((\mathcal{M}_{s+1}(z))_X)^{-1}(\bm m,\bm l)|\cdot|\mathcal{W}(\bm l,\bm l')|\cdot|((\mathcal{M}_{s+1}(z))_Y)^{-1}(\bm l',\bm n)|\\
		&\ \  +\sum_{\bm l\in X\setminus Z_1 \atop\bm l'\in Y\setminus X}|((\mathcal{M}_{s+1}(z))_X)^{-1}(\bm m,\bm l)|\cdot|\mathcal{W}(\bm l,\bm l')|\cdot|((\mathcal{M}_{s+1}(z))_Y)^{-1}(\bm l',\bm n)|\\
		\lesssim&\tilde{\zeta}_s^{-\alpha_1}\|((\mathcal{M}_{s+1}(z))_Y)^{-1}\|_0\left(\|\mathcal{W}\|_{\alpha_1}\|((\mathcal{M}_{s+1}(z))_X)^{-1}\|_0+\|((\mathcal{M}_{s+1}(z))_X)^{-1}\|_{\alpha_1}\|\mathcal{W}\|_0\right).
	\end{align*}
	If $\bm m\in X\setminus Z_2$, we obtain
	\begin{align*}
		&|((\mathcal{M}_{s+1}(z))_Y)^{-1}(\bm m,\bm n)-\chi_X(\bm n)((\mathcal{M}_{s+1}(z))_X)^{-1}(\bm m,\bm n)|\\
		\le&\sum_{\bm l\in X \atop \bm l'\in Y\setminus X}|((\mathcal{M}_{s+1}(z))_X)^{-1}(\bm m,\bm l)|\cdot|\mathcal{W}(\bm l,\bm l')|\cdot|((\mathcal{M}_{s+1}(z))_Y)^{-1}(\bm l',\bm n)|\\
		\le&\|((\mathcal{M}_{s+1}(z))_X)^{-1}\|_0\|\mathcal{W}\|_0\|((\mathcal{M}_{s+1}(z))_Y)^{-1}\|_0.
	\end{align*}
	For $\bm i\in A_{\bm k}^s-\bm k$, $\bm n\in Y$, since \eqref{oddgf}, $\dist(A_{\bm k}^s-\bm k,X\setminus Z_2)\ge\frac{\tilde{\zeta}_s}{16}$ and \eqref{375}--\eqref{370},  we have 
	\begin{align*}
		&|(\mathcal{R}_{A_{\bm k}^s-\bm k}\mathcal{M}_{s+1}(z)\mathcal{R}_X ((\mathcal{M}_{s+1}(z))_Y)^{-1})(\bm i,\bm n)-(\mathcal{R}_{A_{\bm k}^s-\bm k}\mathcal{M}_{s+1}(z)\mathcal{R}_X ((\mathcal{M}_{s+1}(z))_X)^{-1}\mathcal{R}_X)(\bm i,\bm n)|\\
		\lesssim&\sum_{\bm m\in Z_2}|\mathcal{W}(\bm i,\bm m)|\cdot|((\mathcal{M}_{s+1}(z))_Y)^{-1}(\bm m,\bm n)-\chi_X(\bm n)((\mathcal{M}_{s+1}(z))_X)^{-1}(\bm m,\bm n)|\\
		&\ \ +\sum_{\bm m\in X\setminus Z_2}|\mathcal{W}(\bm i,\bm m)|\cdot|((\mathcal{M}_{s+1}(z))_Y)^{-1}(\bm m,\bm n)-\chi_X(\bm n)((\mathcal{M}_{s+1}(z))_X)^{-1}(\bm m,\bm n)|\\
		\lesssim&\|\mathcal{W}\|_0\tilde{\zeta}_s^{-\alpha_1}\|((\mathcal{M}_{s+1}(z))_Y)^{-1}\|_0\left(\|\mathcal{W}\|_{\alpha_1}\|((\mathcal{M}_{s+1}(z))_X)^{-1}\|_0+\|((\mathcal{M}_{s+1}(z))_X)^{-1}\|_{\alpha_1}\|\mathcal{W}\|_0\right)\\
		&\ \ +\|\mathcal{W}\|_0\|\mathcal{W}\|_{\alpha_1}\tilde{\zeta}_s^{-\alpha_1}\|((\mathcal{M}_{s+1}(z))_X)^{-1}\|_0\|((\mathcal{M}_{s+1}(z))_Y)^{-1}\|_0\\
		\lesssim&N_s^{-3\alpha_1}N_{s-1}^{3\alpha_1}\delta_s^{-\frac{19}{18}}\lesssim\delta_s^{\frac{29\alpha_1}{9000\tau}-\frac{19}{18}}<\delta_s^{4}.
	\end{align*}
	It then follows that
	\begin{align*}
		\mathcal{R}_{A_{\bm k}^s-\bm k}\mathcal{M}_{s+1}(z)\mathcal{R}_X ((\mathcal{M}_{s+1}(z))_Y)^{-1}=\mathcal{R}_{A_{\bm k}^s-\bm k}\mathcal{M}_{s+1}(z)\mathcal{R}_X ((\mathcal{M}_{s+1}(z))_X)^{-1}\mathcal{R}_X+O(\delta_s^{4}).
	\end{align*}
	As a result,
	\begin{align*}
		&\ \mathcal{R}_{A_{\bm k}^s-\bm k}\mathcal{M}_{s+1}(z)\mathcal{R}_X ((\mathcal{M}_{s+1}(z))_Y)^{-1}\mathcal{R}_X \mathcal{M}_{s+1}(z)\mathcal{R}_{A_{\bm k}^s-\bm k}\\
		=&\ \mathcal{R}_{A_{\bm k}^s-\bm k}\mathcal{M}_{s+1}(z)\mathcal{R}_X ((\mathcal{M}_{s+1}(z))_X)^{-1}\mathcal{R}_X \mathcal{M}_{s+1}(z)\mathcal{R}_{A_{\bm k}^s-\bm k}+O(\delta_s^{4})\\
		=&\ \mathcal{R}_{A_{\bm k}^s-\bm k}\mathcal{M}_{s}(z)\mathcal{R}_X ((\mathcal{M}_{s}(z))_X)^{-1}\mathcal{R}_X \mathcal{M}_{s}(z)\mathcal{R}_{A_{\bm k}^s-\bm k}+O(\delta_s^{4})
	\end{align*}
	and
	\begin{align*}
		\mathcal{S}_{s+1}(z)&=\mathcal{M}_{s}(z)_{A_{\bm k}^{s}-\bm k}-\mathcal{R}_{A_{\bm k}^{s}-\bm k}\mathcal{M}_{s}(z)\mathcal{R}_{X}((\mathcal{M}_{s}(z))_{X})^{-1} \mathcal{R}_{X}\mathcal{M}_{s}(z)\mathcal{R}_{A_{\bm k}^{s}-\bm k}+O(\delta_s^{4})\\
		&=\mathcal{S}_s(z)+O(\delta_s^{4}),
	\end{align*}
	which implies \eqref{ss} for the $(s+1)$-th step. Recalling \eqref{zs}, \eqref{detss} and \eqref{zs1}, we have
	\begin{align*}
		|\det \mathcal{S}_s(z)|\gtrsim\delta_{s}^{\frac{2}{75}}\|z-\theta_s\|_{\T}\cdot\|z+\theta_s\|_{\T}.
	\end{align*}
	By Lemma \ref{det1}, $\#(A_{\bm k}^s-\bm k)\le 2^s$ and \eqref{ss}, we obtain
	\begin{align*}
		\det \mathcal{S}_{s+1}(z)&=\det \mathcal{S}_s(z)+O((2^s)^2 (4|v|_R)^{2^s}\delta_s^{4})\\
		&=\det \mathcal{S}_s(z)+O(\delta_s^{\frac{7}{2}}). 
	\end{align*}
	Notice that
	\begin{align*}
		\|z+\theta_s\|_{\T}&\ge\|\theta+\bm k\cdot\bm\omega+\theta_s\|_{\T}-\|z-\theta_s\|_{\T}-\|\theta+\bm k\cdot\bm\omega-\theta_s\|_{\T}\\
		&>\delta_s^{\frac{2}{3}}-\delta_{s}^{\frac{18}{19}}-\delta_s\\
		&>\frac{1}{2}\delta_s^{\frac{2}{3}}.
	\end{align*}
	Then we have
	\begin{align*}
		|\det \mathcal{S}_{s+1}(z)|\gtrsim\delta_{s+1}^{\frac{2}{75}} (z-\theta_s+r_{s+1}(z)),
	\end{align*}
	where $r_{s+1}(z)$ is an analytic function defined in the set of \eqref{zs1} with $|r_{s+1}(z)|\lesssim\delta_s^{\frac{7}{2}-\frac{2}{3}-\frac{2}{75}}\ll\delta_s^{\frac{3}{2}}\ll\delta_s^{\frac{18}{19}}$. Finally, by the Rouch\'e's theorem, the equation
	\begin{align*}
		(z-\theta_s)+r_{s+1}(z)=0
	\end{align*}
	has a unique root $\theta_{s+1}$ in the set of \eqref{zs1} satisfying
	\begin{align}\label{ts+1-ts}
		|\theta_{s+1}-\theta_s|=|r_{s+1}(\theta_{s+1})|\ll\delta_s^{\frac{3}{2}},\ |(z-\theta_s)+r_{s+1}(z)|\sim|z-\theta_{s+1}|.
	\end{align}
	Moreover, $\theta_{s+1}$ is the unique root of $\det \mathcal{M}_{s+1}(z)=0$ in the set of \eqref{zs1}. Since $\|z+\theta_s\|_{\T}>\frac{1}{2}\delta_s^{\frac{2}{3}}$ and $|\theta_{s+1}-\theta_s|\ll\delta_s^{\frac{3}{2}}$, we have
	\begin{align*}
		\|z+\theta_s\|_{\T}\sim\|z+\theta_{s+1}\|_{\T}.
	\end{align*}
	Thus, if $z$ belongs to the set of \eqref{zs1}, we get
	\begin{align}\label{detss+11}
		|\det \mathcal{S}_{s+1}(z)|\gtrsim\delta_{s+1}^{\frac{2}{75}}\|z-\theta_{s+1}\|_{\T}\cdot\|z+\theta_{s+1}\|_{\T}.
	\end{align}
	Since $\delta_{s+1}\sim\delta_s^{30}$, we obtain $\delta_{s+1}^{\frac{1}{2}}\ll\frac{1}{2}\delta_s^{\frac{18}{19}}$. Recalling \eqref{zs1} and \eqref{ts+1-ts}, we see that \eqref{detss+11} remains valid for $z$ satisfying
	\begin{align*}
		\|z-\theta_{s+1}\|_{\T}<\delta_{s+1}^{\frac{1}{2}}.
	\end{align*}
	Now, for $\bm k\in Q_s^+$, we consider $\mathcal{M}_{s+1}(z)$ in the set
	\begin{align}\label{zs2}
		\{z\in\C:\ |z+\theta_s|<\delta_{s}^{\frac{18}{19}}\}.
	\end{align}
	The similar argument shows that $\det \mathcal{M}_{s+1}(z)=0$ has a unique root $\theta_{s+1}^{'}$ in the set of \eqref{zs2}. We will show $\theta_{s+1}+\theta_{s+1}^{'}=0$. In fact, by Lemma \ref{ef}, $\det \mathcal{M}_{s+1}(z)$ is an even function of $z$. Then the uniqueness of the root implies $\theta_{s+1}^{'}=-\theta_{s+1}$. Thus for $z$ being in the set of \eqref{zs2}, we also have \eqref{detss+11}. In conclusion, \eqref{detss+11} is established for $z$ being in
	\begin{align*}
		\{z\in\C:\min_{\sigma=\pm1}\|z+\sigma\theta_{s+1}\|_{\T}<\delta_{s+1}^{\frac{1}{2}}\},
	\end{align*}
	which proves \eqref{detss} for the $(s+1)$-th step. Combining \eqref{qspm}--\eqref{ps+1} and the following
	\begin{align*}
		\|\theta+\bm k\cdot\bm\omega\pm\theta_{s+1}\|_{\T}<10\delta_{s+1}^{\frac{2}{3}},\ |\theta_{s+1}-\theta_s|<\delta_s^{\frac{3}{2}}\Rightarrow\|\theta+\bm k\cdot\bm\omega\pm\theta_s\|_{\T}<\delta_s,
	\end{align*}
	we get
	\begin{align*}
		\left\{\bm k\in\Z^d+\frac{1}{2}\sum_{i=0}^{s}\bm l_i:\ \min_{\sigma=\pm1}\|\theta+\bm k\cdot\bm\omega+\sigma\theta_{s+1}\|_{\T}<10\delta_{s+1}^{\frac{2}{3}}\right\}\subset P_{s+1},
	\end{align*}
	which proves \eqref{fs} at the $(s+1)$-th step. Finally, we want to estimate $\mathcal{T}_{\tilde{\Omega}_{\bm k}^{s+1}}^{-1}$. For $\bm k\in P_{s+1}$, by \eqref{ps+11}, we obtain
	\begin{align*}
		\theta+\bm k\cdot\bm\omega\in\{z\in\C:\ \min_{\sigma=\pm1}\|z+\sigma\theta_s\|_{\T}<\delta_s^{\frac{18}{19}}\},
	\end{align*}
	which together with \eqref{detss+11} implies
	\begin{align*}
		&\ |\det(\mathcal{T}_{A_{\bm k}^{s+1}}-\mathcal{R}_{A_{\bm k}^{s+1}}\mathcal{T}\mathcal{R}_{\tilde{\Omega}_{\bm k}^{s+1}\setminus A_{\bm k}^{s+1}}\mathcal{T}_{\tilde{\Omega}_{\bm k}^{s+1}\setminus A_{\bm k}^{s+1}}^{-1}\mathcal{R}_{\tilde{\Omega}_{\bm k}^{s+1}\setminus A_{\bm k}^{s+1}}\mathcal{T}\mathcal{R}_{A_{\bm k}^{s+1}})|\\
		=&\ |\det \mathcal{S}_{s+1}(\theta+\bm k\cdot\bm\omega)|\\
		\gtrsim&\ \delta_{s+1}^{\frac{2}{75}}\|\theta+\bm k\cdot\bm\omega-\theta_{s+1}\|_{\T}\cdot\|\theta+\bm k\cdot\bm\omega+\theta_{s+1}\|_{\T}.
	\end{align*}
	By \eqref{ss}, Cramer's rule and Lemma \ref{chi}, one has
	\begin{align*}
		&\|(\mathcal{T}_{A_{\bm k}^{s+1}}-\mathcal{R}_{A_{\bm k}^{s+1}}\mathcal{T}\mathcal{R}_{\tilde{\Omega}_{\bm k}^{s+1}\setminus A_{\bm k}^{s+1}}\mathcal{T}_{\tilde{\Omega}_{\bm k}^{s+1}\setminus A_{\bm k}^{s+1}}^{-1}\mathcal{R}_{\tilde{\Omega}_{\bm k}^{s+1}\setminus A_{\bm k}^{s+1}}\mathcal{T}\mathcal{R}_{A_{\bm k}^{s+1}})^{-1}\|_0\\
		=&|\det \mathcal{S}_{s+1}(\theta+\bm k\cdot\bm\omega)|^{-1}\|(\mathcal{T}_{A_{\bm k}^{s+1}}-\mathcal{R}_{A_{\bm k}^{s+1}}\mathcal{T}\mathcal{R}_{\tilde{\Omega}_{\bm k}^{s+1}\setminus A_{\bm k}^{s+1}}\mathcal{T}_{\tilde{\Omega}_{\bm k}^{s+1}\setminus A_{\bm k}^{s+1}}^{-1}\mathcal{R}_{\tilde{\Omega}_{\bm k}^{s+1}\setminus A_{\bm k}^{s+1}}\mathcal{T}\mathcal{R}_{A_{\bm k}^{s+1}})^{*}\|_0\\
		\lesssim&(2^s)^2 (4|v|_R)^{2^s}\delta_{s+1}^{-\frac{2}{75}}\|\theta+\bm k\cdot\bm\omega-\theta_{s+1}\|_{\T}^{-1}\cdot\|\theta+\bm k\cdot\bm\omega+\theta_{s+1}\|_{\T}^{-1}.
	\end{align*}
	From Lemma \ref{scl} and \eqref{ts0}, we get
	\begin{align}
		\nonumber\|\mathcal{T}_{\tilde{\Omega}_{\bm k}^{s+1}}^{-1}\|_0&<4(1+\|\mathcal{T}_{\tilde{\Omega}_{\bm k}^{s+1}\setminus A_{\bm k}^{s+1}}\|_0)^2\\
		\nonumber&\times(1+\|(\mathcal{T}_{A_{\bm k}^{s+1}}-\mathcal{R}_{A_{\bm k}^{s+1}}\mathcal{T}\mathcal{R}_{\tilde{\Omega}_{\bm k}^{s+1}\setminus A_{\bm k}^{s+1}}\mathcal{T}_{\tilde{\Omega}_{\bm k}^{s+1}\setminus A_{\bm k}^{s+1}}^{-1}\mathcal{R}_{\tilde{\Omega}_{\bm k}^{s+1}\setminus A_{\bm k}^{s+1}}\mathcal{T}\mathcal{R}_{A_{\bm k}^{s+1}})^{-1}\|_0)\\
		\label{tos+1-11}&<\delta_{s+1}^{-\frac{1}{15}}\|\theta+\bm k\cdot\bm\omega-\theta_{s+1}\|_{\T}^{-1}\cdot\|\theta+\bm k\cdot\bm\omega+\theta_{s+1}\|^{-1}_{\T}.
	\end{align}
	
	\begin{itemize}
		\item[\textbf{Case 2.}] The case $(\bm C2)_s$ occurs, i.e., 
	\end{itemize}
	\begin{align*}
		\dist(\tilde{Q}_s^-,Q_s^+)\le100N_{s+1}^3.
	\end{align*}
	Then there exist $\bm i_s\in Q_{s}^{+}$ and $\bm j_s\in \tilde{Q}_s^-$ with $\|\bm i_s-\bm j_s\|\le100 N_{s+1}^3$ such that
	\begin{align*}
		\|\theta+\bm i_s\cdot\bm\omega+\theta_s\|_{\T}<\delta_s,\ \|\theta+\bm j_s\cdot\bm\omega-\theta_s\|_{\T}<\delta_s^{\frac{2}{3}}.
	\end{align*}
	Define
	\begin{align*}
		\bm l_s=\bm i_s-\bm j_s.
	\end{align*}
	Using \eqref{as1} and \eqref{as2} yields
	\begin{align*}
		Q_s^+,\tilde{Q}_s^-\subset P_s\subset \Z^d+\frac{1}{2}\sum_{i=0}^{s-1}\bm l_i.
	\end{align*}
	Thus $\bm i_s\equiv \bm j_s\ (\text{mod}\ \Z^d)$ and $\bm l_s\in\Z^d$. Define
	\begin{align}\label{os+12}
		O_{s+1}=Q_s^-\cup(Q_s^+-\bm l_s).
	\end{align}
	For every $\bm o\in O_{s+1}$, define its mirror point as 
	\begin{align*}
		\bm o^*=\bm o+\bm l_s.
	\end{align*}
	Then we have
	\begin{align*}
		O_{s+1}\subset\{\bm o\in\Z^d+\frac{1}{2}\sum_{i=0}^{s-1}\bm l_i:\ \|\theta+\bm o\cdot\bm\omega-\theta_s\|_{\T}<2\delta_s^{\frac{2}{3}}\}
	\end{align*}
	and
	\begin{align*}
		O_{s+1}+\bm l_s\subset\{\bm o^*\in\Z^d+\frac{1}{2}\sum_{i=0}^{s-1}\bm l_i:\ \|\theta+\bm o^*\cdot\bm\omega+\theta_s\|_{\T}<2\delta_s^{\frac{2}{3}}\}.
	\end{align*}
	Then by \eqref{fs}, we obtain
	\begin{align}\label{os+13}
		O_{s+1}\cup(O_{s+1}+\bm l_s)\subset P_s.
	\end{align}
	Next, define
	\begin{align}\label{ps+12}
		P_{s+1}=\left\{\frac{1}{2}(\bm o+\bm o^*):\ \bm o\in O_{s+1}\right\}=\left\{\bm o+\frac{\bm l_s}{2}:\ \bm o\in O_{s+1}\right\}.
	\end{align}
	Notice that
	\begin{align*}
		\min&\left(\left\|\frac{\bm l_s}{2}\cdot\bm\omega+\theta_s\right\|_{\T},\left\|\frac{\bm l_s}{2}\cdot\bm\omega+\theta_s-\frac{1}{2}\right\|_{\T}\right)\\
		&=\frac{1}{2}\|\bm l_s\cdot\bm\omega+2\theta_s\|_{\T}\\
		&\le\frac{1}{2}(\|\theta+\bm i_s\cdot\bm\omega+\theta_s\|_{\T}+\|\theta+\bm j_s\cdot\bm\omega-\theta_s\|_{\T})<\delta_s^{\frac{2}{3}}.
	\end{align*}
	Since $\delta_s\ll1$, only one of
	\begin{align*}
		\left\|\frac{\bm l_s}{2}\cdot\bm\omega+\theta_s\right\|_{\T}<\delta_0^{\frac{2}{3}}\ \text{and}\ \left\|\frac{\bm l_s}{2}\cdot\bm\omega+\theta_s-\frac{1}{2}\right\|_{\T}<\delta_0^{\frac{2}{3}}
	\end{align*}
	occurs. First, we consider the case of
	\begin{align}\label{ls1}
		\left\|\frac{\bm l_s}{2}\cdot\bm\omega+\theta_s\right\|_{\T}<\delta_s^{\frac{2}{3}}.
	\end{align}
	Let $\bm k\in P_{s+1}$. Since $\bm k=\bm o+\frac{\bm l_s}{2}$ (for some $\bm o\in O_{s+1}$), we have
	\begin{align}\label{t+kos}
		\|\theta+\bm k\cdot\bm\omega\|_{\T}\le\|\theta+\bm o\cdot\bm\omega-\theta_s\|_{\T}+\left\|\frac{\bm l_s}{2}\cdot\bm\omega+\theta_s\right\|_{\T}<3\delta_s^{\frac{2}{3}},
	\end{align}
	which implies
	\begin{align}\label{ps+1s1}
		P_{s+1}\subset\{\bm k\in\Z^d+\frac{1}{2}\sum_{i=0}^s \bm l_i:\ \|\theta+\bm k\cdot\bm \omega\|_{\T}<3\delta_s^{\frac{2}{3}}\}.
	\end{align}
	Moreover, if $\bm k\ne \bm k'\in P_{s+1}$, we obtain
	\begin{align*}
		\|\bm k-\bm k'\|\ge\left(\frac{\g}{6\delta_s^{\frac{2}{3}}}\right)^{\frac{1}{\tau}}\sim N_{s+1}^{20}\gg 100N_{s+1}^{5}.
	\end{align*}
	Similar to the proof that appears in {\bf Case 1}  (i.e., the $(\bm C1)_s$ case), we can associate each $\bm k\in P_{s+1}$ with the blocks $\Omega_{\bm k}^{s+1}$ and $\tilde{\Omega}_{\bm k}^{s+1}$ satisfying
	\begin{align*}
		&\lg_{100N_{s+1}^3}(\bm k)\subset \Omega_{\bm k}^{s+1}\subset\lg_{100N_{s+1}^3+50N_s^{5}}(\bm k),\\
		&\lg_{N_{s+1}^{5}}(\bm k)\subset \tilde{\Omega}_{\bm k}^{s+1}\subset\lg_{N_{s+1}^{5}+50N_s^{5}}(\bm k)
	\end{align*}
	and
	\begin{align}\label{sb3}
		\left\{\begin{array}{l}
			\Omega_{\bm k}^{s+1}\cap\tilde{\Omega}_{\bm k'}^{s'}\ne\emptyset\ (s'<s+1)\Rightarrow\tilde{\Omega}_{\bm k'}^{s'}\subset\Omega_{\bm k}^{s+1},\\
			\tilde{\Omega}_{\bm k}^{s+1}\cap\tilde{\Omega}_{\bm k'}^{s'}\ne\emptyset\ (s'<s+1)\Rightarrow\tilde{\Omega}_{\bm k'}^{s'}\subset\tilde{\Omega}_{\bm k}^{s+1},\\
			\dist(\tilde{\Omega}_{\bm k}^{s+1},\tilde{\Omega}_{\bm k'}^{s+1})>10\tilde{\zeta}_{s+1}\ \text{for}\ \bm k\ne \bm k'\in P_{s+1}.
		\end{array}\right.
	\end{align}
	In addition, the translation
	\begin{align*}
		\tilde{\Omega}_{\bm k}^{s+1}-\bm k\subset\Z^d+\frac{1}{2}\sum_{i=0}^s \bm l_i
	\end{align*}
	is independent of $\bm k\in P_{s+1}$ and symmetrical about the origin. In summary, we have proven $(\bm a)_{s+1}$ and $(\bm b)_{s+1}$ in the case of $(\bm C2)_{s}$.
	
	For each $\bm k\in P_{s+1}$, we have $\bm o,\bm o^*\in P_s$ by \eqref{os+13}. Define
	\begin{align*}
		A_{\bm k}^{s+1}=A_{\bm o}^s\cup A_{\bm o^*}^s,
	\end{align*}
	where $\bm o\in O_{s+1}$ and $\bm k=\frac{\bm o+\bm o^*}{2}$ (see \eqref{ps+12}). Then
	\begin{align*}
		A_{\bm k}^{s+1}&\subset \Omega_{\bm o}^s\cup \Omega_{\bm o^*}^s\subset \Omega_{\bm k}^{s+1},\\
		\# A_{\bm k}^{s+1}&=\# A_{\bm o}^s+\# A_{\bm o^*}^s\le 2^{s+1}.
	\end{align*}
	The proof of  $\tilde{\Omega}_{\bm k}^{s+1}\setminus A_{\bm k}^{s+1}$ is $s$-$\good$ can be found at page 32 of \cite{CSZ24a}. This finishes the proof of $(\bm c)_{s+1}$ in the case of $(\bm C2)_s$.
	
	Consider
	\begin{align*}
		\mathcal{M}_{s+1}(z):=\mathcal{T}_{\tilde{\Omega}_{\bm k}^{s+1}-\bm k}(z)=((v(z+\bm n\cdot\bm\omega)-E)\delta_{\bm n,\bm n'}+\ep \mathcal{W})_{\bm n\in\tilde{\Omega}_{\bm k}^{s+1}-\bm k}
	\end{align*}
	in the set
	\begin{align}\label{zs3}
		\{z\in\C:\ |z|<\delta_{s}^{\frac{5}{8}}\}.
	\end{align}
	If $\bm k'\in P_s$ and $\tilde{\Omega}_{\bm k'}^{s}\subset(\tilde{\Omega}_{\bm k}^{s+1}\setminus A_{\bm k}^{s+1})$, then $\bm k'\ne \bm o,\bm o^*$ and $\|\bm k'-\bm o\|,\|\bm k'-\bm o^*\|\le 4N_{s+1}^{5}$. Thus
	\begin{align*}
		\|\theta+\bm k'\cdot\bm\omega-\theta_s\|_{\T}&\ge\|(\bm k'-\bm o)\cdot\bm\omega\|_{\T}-\|\theta+\bm o\cdot\bm\omega-\theta_s\|_{\T}\\
		&\ge\frac{\g}{(4N_{s+1}^{5})^\tau}-2\delta_s^{\frac{2}{3}}\gtrsim\delta_s^{\frac{1}{6}}
	\end{align*}
	and
	\begin{align*}
		\|\theta+\bm k'\cdot\bm\omega+\theta_s\|_{\T}&\ge\|(\bm k'-\bm o^*)\cdot\bm\omega\|_{\T}-\|\theta+\bm o^*\cdot\bm\omega+\theta_s\|_{\T}\\
		&\ge\frac{\g}{(4N_{s+1}^{5})^\tau}-2\delta_s^{\frac{2}{3}}\gtrsim\delta_s^{\frac{1}{6}}.
	\end{align*}
	By \eqref{tsg01} and $\tilde{\Omega}_{\bm k}^{s+1}\setminus A_{\bm k}^{s+1}$ is $s$-$\good$ (cf. $(\bm c)_{s+1}$), we have
	\begin{align}
		\nonumber\|\mathcal{T}_{\tilde{\Omega}_{\bm k}^{s+1}\setminus A_{\bm k}^{s+1}}^{-1}\|_{0}&\le\delta_{s}^{-\frac{2}{15}}\times\sup\limits_{\{\bm k'\in P_s:\ \tilde{\Omega}_{\bm k'}^s\subset(\tilde{\Omega}_{\bm k}^{s+1}\setminus A_{\bm k}^{s+1})\}}(\|\theta+\bm k'\cdot\bm\omega-\theta_s\|_{\T}^{-1}\cdot\|\theta+\bm k'\cdot\bm\omega+\theta_s\|_{\T}^{-1})\\
		\label{tosa2}&<\delta_s^{-\frac{7}{15}}.
	\end{align}
	One may restate \eqref{tosa2} as
	\begin{align*}
		\|((\mathcal{M}_{s+1}(\theta+\bm k\cdot\bm\omega))_{(\tilde{\Omega}_{\bm k}^{s+1}\setminus A_{\bm k}^{s+1})-\bm k})^{-1}\|_0<\delta_s^{-\frac{7}{15}}.
	\end{align*}
	Since
	\begin{align}\label{qs-2}
		\nonumber\|z-(\theta+\bm k\cdot\bm\omega)\|_{\T}&\le|z|+\|\theta+\bm k\cdot\bm\omega\|_{\T}\\
		&<\delta_s^{\frac{5}{8}}+3\delta_s^{\frac{2}{3}}<2\delta_s^{\frac{5}{8}},
	\end{align}
	using Neumann series argument  shows 
	\begin{align}\label{ms+1-1}
		\|(\mathcal{M}_{s+1}(z)_{(\tilde{\Omega}_{\bm k}^{s+1}\setminus A_{\bm k}^{s+1})-\bm k})^{-1}\|_0\lesssim\delta_s^{-\frac{7}{15}}.
	\end{align}
	Thus by Lemma \ref{scl}, $(\mathcal{M}_{s+1}(z))^{-1}$ is controlled by the Schur complement of $((\tilde{\Omega}_{\bm k}^{s+1}\setminus A_{\bm k}^{s+1})-\bm k)$, i.e.,
	\begin{align*}
		\mathcal{S}_{s+1}(z)&=(\mathcal{M}_{s+1}(z))_{A_{\bm k}^{s+1}-\bm k}-\mathcal{R}_{A_{\bm k}^{s+1}-\bm k}\mathcal{M}_{s+1}(z)\mathcal{R}_{(\tilde{\Omega}_{\bm k}^{s+1}\setminus A_{\bm k}^{s+1})-\bm k}\\
		&\ \ \times((\mathcal{M}_{s+1}(z))_{(\tilde{\Omega}_{\bm k}^{s+1}\setminus A_{\bm k}^{s+1})-\bm k})^{-1} \mathcal{R}_{(\tilde{\Omega}_{\bm k}^{s+1}\setminus A_{\bm k}^{s+1})-\bm k}\mathcal{M}_{s+1}(z)\mathcal{R}_{A_{\bm k}^{s+1}-\bm k}.
	\end{align*}
	Our next aim is to analyze $\det \mathcal{S}_{s+1}(z)$. Since
	\begin{align*}
		A_{\bm k}^{s+1}-\bm k&=(A_{\bm o}^s-\bm k)\cup(A_{\bm o^*}^s-\bm k),\\
		A_{\bm o}^s-\bm k&\subset\Omega_{\bm o}^s-\bm k,\ A_{\bm o^*}^s-\bm k\subset \Omega_{\bm o^*}^s-\bm k
	\end{align*}
	and
	\begin{align*}
		\dist(\Omega_{\bm o}^s-\bm k,\Omega_{\bm o^*}^s-\bm k)>10\tilde{\zeta}_s,
	\end{align*}
	we have
	\begin{align*}
		(\mathcal{M}_{s+1}(z))_{A_{\bm k}^{s+1}-\bm k}=((\mathcal{M}_{s+1}(z))_{A_{\bm o}^{s}-\bm k})\oplus ((\mathcal{M}_{s+1}(z))_{A_{\bm o^*}^{s}-\bm k}).
	\end{align*}
	From  $\dist(A_{\bm o}^{s},\p \tilde{\Omega}_{\bm o}^s)>\frac{1}{2}\tilde{\zeta}_s$ and $\dist(A_{\bm o^*}^{s},\p \tilde{\Omega}_{\bm o^*}^s)>\frac{1}{2}\tilde{\zeta}_s$, we have
	\begin{align*}
		\mathcal{R}_{A_{\bm o}^{s}-\bm k}\mathcal{M}_{s+1}(z)\mathcal{R}_{(\tilde{\Omega}_{\bm k}^{s+1}\setminus A_{\bm k}^{s+1})-\bm k}&=\mathcal{R}_{A_{\bm o}^{s}-\bm k}\mathcal{M}_{s+1}(z)\mathcal{R}_{(\tilde{\Omega}_{\bm o}^{s}\setminus A_{\bm o}^{s})-\bm k}+O(\delta_s^6),\\
		\mathcal{R}_{A_{\bm o^*}^{s}-\bm k}\mathcal{M}_{s+1}(z)\mathcal{R}_{(\tilde{\Omega}_{\bm k}^{s+1}\setminus A_{\bm k}^{s+1})-\bm k}&=\mathcal{R}_{A_{\bm o^*}^{s}-\bm k}\mathcal{M}_{s+1}(z)\mathcal{R}_{(\tilde{\Omega}_{\bm o^*}^{s}\setminus A_{\bm o}^{s})-\bm k}+O(\delta_s^6).
	\end{align*}
	For the convenience, we define
	\begin{align*}
		&X=(\tilde{\Omega}_{\bm o}^s\setminus A_{\bm o}^s)-\bm k,\ X^*=(\tilde{\Omega}_{\bm o^*}^s\setminus A_{\bm o^*}^s)-k,\ Y=(\tilde{\Omega}_{\bm k}^{s+1}\setminus A_{\bm k}^{s+1})-\bm k,\\
		&Z_1=\lg_{\frac{\tilde{\zeta}_s}{4}}\cap X,\ Z_2=\lg_{\frac{\tilde{\zeta}_s}{8}}\cap X,\ Z_1^*=\lg_{\frac{\tilde{\zeta}_s}{4}}\cap X^*,\ Z_2^*=\lg_{\frac{\tilde{\zeta}_s}{8}}\cap X^*.
	\end{align*}
	Then direct computation yields
	\begin{align*}
		&\mathcal{S}_{s+1}(z)\\
		=&((\mathcal{M}_{s+1}(z))_{A_{\bm o}^{s}-\bm k})\oplus ((\mathcal{M}_{s+1}(z))_{A_{\bm o^*}^{s}-\bm k})-(\mathcal{R}_{A_{\bm o}^{s}-\bm k}\oplus \mathcal{R}_{A_{\bm o*}^{s}-\bm k})\mathcal{M}_{s+1}(z)\mathcal{R}_{(\tilde{\Omega}_{\bm k}^{s+1}\setminus A_{\bm k}^{s+1})-\bm k}\\
		&\ \ \times((\mathcal{M}_{s+1}(z))_{(\tilde{\Omega}_{\bm k}^{s+1}\setminus A_{\bm k}^{s+1})-\bm k})^{-1} \mathcal{R}_{(\tilde{\Omega}_{\bm k}^{s+1}\setminus A_{\bm k}^{s+1})-\bm k}\mathcal{M}_{s+1}(z)(\mathcal{R}_{A_{\bm o}^{s}-\bm k}\oplus \mathcal{R}_{A_{\bm o*}^{s}-\bm k}).
	\end{align*}
	Since $\tilde{\Omega}_{\bm o}^s\setminus A_{\bm o}^s$ is $(s-1)$-$\good$, by \eqref{tsg01} and $\eqref{tsg1}$, we get
	\begin{align*}
		\|\mathcal{T}_{\tilde{\Omega}_{\bm o}^s\setminus A_{\bm o}^s}^{-1}\|_0&<\delta_{s-1}^{-\frac{32}{15}},\\
		\|\mathcal{T}_{\tilde{\Omega}_{\bm o}^s\setminus A_{\bm o}^s}^{-1}\|_{\alpha}&< \zeta_{s-1}^{\alpha}\delta_{s-1}^{-\frac{14}{3}}.
	\end{align*}
	Equivalently,
	\begin{align*}
		\|((\mathcal{M}_{s+1}(\theta+\bm k\cdot\bm\omega))_{X})^{-1}\|_0&<\delta_{s-1}^{-\frac{32}{15}},\\
		\|((\mathcal{M}_{s+1}(\theta+\bm k\cdot\bm\omega))_{X})^{-1}\|_{\alpha}&< \zeta_{s-1}^{\alpha}\delta_{s-1}^{-\frac{14}{3}}.
	\end{align*}
	From \eqref{qs-2}, we deduce by the same argument as \eqref{369} and \eqref{370} that
	\begin{align}
		\label{392}\|((\mathcal{M}_{s+1}(z))_{X})^{-1}\|_0&\lesssim\delta_{s-1}^{-\frac{32}{15}},\\
		\label{393}\|((\mathcal{M}_{s+1}(z))_{X})^{-1}\|_{\alpha}&\lesssim\zeta_{s-1}^{\alpha}\delta_{s-1}^{-\frac{14}{3}}.
	\end{align}
	Let $\bm m\in X$. By the resolvent identity, we have for any $\bm n\in Y$,
	\begin{align*}
		&((\mathcal{M}_{s+1}(z))_Y)^{-1}(\bm m,\bm n)-\chi_X(\bm n)((\mathcal{M}_{s+1}(z))_X)^{-1}(\bm m,\bm n)\\
		=&-\ep\sum_{\bm l\in X \atop \bm l'\in Y\setminus X}((\mathcal{M}_{s+1}(z))_X)^{-1}(\bm m,\bm l)\mathcal{W}(\bm l,\bm l')((\mathcal{M}_{s+1}(z))_Y)^{-1}(\bm l',\bm n).
	\end{align*}
	If $\bm m\in Z_2$, since \eqref{oddgf}, $\dist(Z_1,Y\setminus X)\ge\frac{\tilde{\zeta}_s}{8}$ and $\dist(Z_2,X\setminus Z_1)\ge\frac{\tilde{\zeta}_s}{8}$, we  get
	\begin{align*}
		&|((\mathcal{M}_{s+1}(z))_Y)(\bm m,\bm n)-\chi_X(\bm n)((\mathcal{M}_{s+1}(z))_X)^{-1}(\bm m,\bm n)|\\
		\le&\sum_{\bm l\in Z_1 \atop \bm l'\in Y\setminus X}|((\mathcal{M}_{s+1}(z))_X)^{-1}(\bm m,\bm l)|\cdot|\mathcal{W}(\bm l,\bm l')|\cdot|((\mathcal{M}_{s+1}(z))_Y)^{-1}(\bm l',\bm n)|\\
		&\ \ +\sum_{\bm l\in X\setminus Z_1 \atop \bm l'\in Y\setminus X}|((\mathcal{M}_{s+1}(z))_X)^{-1}(\bm m,\bm l)|\cdot|\mathcal{W}(\bm l,\bm l')|\cdot|((\mathcal{M}_{s+1}(z))_Y)^{-1}(\bm l',\bm n)|\\
		\lesssim&\tilde{\zeta}_s^{-\alpha_1}\|((\mathcal{M}_{s+1}(z))_Y)^{-1}\|_0\left(\|\mathcal{W}\|_{\alpha_1}\|((\mathcal{M}_{s+1}(z))_X)^{-1}\|_0+\|((\mathcal{M}_{s+1}(z))_X)^{-1}\|_{\alpha_1}\|\mathcal{W}\|_0\right).
	\end{align*}
	If $\bm m\in X\setminus Z_2$, we obtain
	\begin{align*}
		&|((\mathcal{M}_{s+1}(z))_Y)^{-1}(\bm m,\bm n)-\chi_X(\bm n)((\mathcal{M}_{s+1}(z))_X)^{-1}(\bm m,\bm n)|\\
		\le&\sum_{\bm l\in X \atop \bm l'\in Y\setminus X}|((\mathcal{M}_{s+1}(z))_X)^{-1}(\bm m,\bm l)|\cdot|\mathcal{W}(\bm l,\bm l')|\cdot|((\mathcal{M}_{s+1}(z))_Y)^{-1}(\bm l',\bm n)|\\
		\le&\|((\mathcal{M}_{s+1}(z))_X)^{-1}\|_0\|\mathcal{W}\|_0\|((\mathcal{M}_{s+1}(z))_Y)^{-1}\|_0.
	\end{align*}
	For $\bm i\in A_{\bm o}^s-\bm k$, $\bm n\in Y$, since \eqref{oddgf}, $\dist(A_{\bm o}^s-\bm k,X\setminus Z_2)\ge\frac{\tilde{\zeta}_s}{16}$ and \eqref{ms+1-1}--\eqref{393}, we have 
	\begin{align*}
		&|\mathcal{R}_{A_{\bm o}^s-\bm k}\mathcal{M}_{s+1}(z)\mathcal{R}_X ((\mathcal{M}_{s+1}(z))_Y)^{-1}(\bm i,\bm n)-\mathcal{R}_{A_{\bm o}^s-\bm k}\mathcal{M}_{s+1}(z)\mathcal{R}_X ((\mathcal{M}_{s+1}(z))_X)^{-1}\mathcal{R}_X(\bm i,\bm n)|\\
		\le&\sum_{\bm m\in Z_2}|\mathcal{W}(\bm i,\bm m)|\cdot|((\mathcal{M}_{s+1}(z))_Y)^{-1}(\bm m,\bm n)-\chi_X(\bm n)((\mathcal{M}_{s+1}(z))_X)^{-1}(\bm m,\bm n)|\\
		&\ \ +\sum_{\bm m\in X\setminus Z_2}|\mathcal{W}(\bm i,\bm m)|\cdot|((\mathcal{M}_{s+1}(z))_Y)^{-1}(\bm m,\bm n)-\chi_X(\bm n)((\mathcal{M}_{s+1}(z))_X)^{-1}(\bm m,\bm n)|\\
		\lesssim&\|\mathcal{W}\|_0\tilde{\zeta}_s^{-\alpha_1}\|((\mathcal{M}_{s+1}(z))_Y)^{-1}\|_0\left(\|\mathcal{W}\|_{\alpha_1}\|((\mathcal{M}_{s+1}(z))_X)^{-1}\|_0+\|((\mathcal{M}_{s+1}(z))_X)^{-1}\|_{\alpha_1}\|\mathcal{W}\|_0\right)\\
		&\ \ +\|\mathcal{W}\|_0\|\mathcal{W}\|_{\alpha_1}\tilde{\zeta}_s^{-\alpha_1}\|((\mathcal{M}_{s+1}(z))_X)^{-1}\|_0\|((\mathcal{M}_{s+1}(z))_Y)^{-1}\|_0\\
		\lesssim&N_s^{-3\alpha_1}N_{s-1}^{3\alpha_1}\delta_s^{-\frac{28}{45}}\lesssim\delta_s^{\frac{29\alpha_1}{9000\tau}-\frac{28}{45}}<\delta_s^{4}.
	\end{align*}
	It then follows that
	\begin{align*}
		\mathcal{R}_{A_{\bm o}^s-\bm k}\mathcal{M}_{s+1}(z)\mathcal{R}_X ((\mathcal{M}_{s+1}(z))_Y)^{-1}=\mathcal{R}_{A_{\bm o}^s-\bm k}\mathcal{M}_{s+1}(z)\mathcal{R}_X ((\mathcal{M}_{s+1}(z))_X)^{-1}\mathcal{R}_X+O(\delta_s^4).
	\end{align*}
	Similarly,
	\begin{align*}
		\mathcal{R}_{A_{\bm o^*}^s-\bm k}\mathcal{M}_{s+1}(z)\mathcal{R}_{X^*} ((\mathcal{M}_{s+1}(z))_Y)^{-1}=\mathcal{R}_{A_{\bm o^*}^s-\bm k}\mathcal{M}_{s+1}(z)\mathcal{R}_{X^*} ((\mathcal{M}_{s+1}(z))_{X^*})^{-1}\mathcal{R}_{X^*}+O(\delta_s^4).
	\end{align*}
	As a result,
	\begin{align*}
		\mathcal{S}_{s+1}(z)&=\left((\mathcal{M}_{s+1}(z))_{A_{\bm o}^{s}-\bm k}-\mathcal{R}_{A_{\bm o}^{s}-\bm k}\mathcal{M}_{s+1}(z)\mathcal{R}_X((\mathcal{M}_{s+1}(z))_{X})^{-1}\mathcal{R}_X \mathcal{M}_{s+1}(z)\mathcal{R}_{A_{\bm o}^{s}-\bm k}\right)\\
		&\ \ \oplus\left((\mathcal{M}_{s+1}(z))_{A_{\bm o^*}^{s}-\bm k}-\mathcal{R}_{A_{\bm o^*}^{s}-\bm k}\mathcal{M}_{s+1}(z)\mathcal{R}_{X^*}((\mathcal{M}_{s+1}(z))_{X^*})^{-1}\mathcal{R}_{X^*} \mathcal{M}_{s+1}(z)\mathcal{R}_{A_{\bm o^*}^{s}-\bm k}\right)\\
		&\ \ +O(\delta_s^{4})\\
		&=\mathcal{S}_s\left(z-\frac{\bm l_s}{2}\cdot\bm\omega\right)\oplus \mathcal{S}_s\left(z+\frac{\bm l_s}{2}\cdot\bm\omega\right)+O(\delta_s^{4}).
	\end{align*}
	From \eqref{ls1} and \eqref{zs3}, we have
	\begin{align*}
		\left\|z-\frac{\bm l_s}{2}\cdot\bm\omega-\theta_s\right\|_{\T}\le|z|+\left\|\frac{\bm l_s}{2}\cdot\bm\omega+\theta_s\right\|_{\T}<\delta_s^{\frac{5}{8}}+\delta_s^{\frac{2}{3}}<2\delta_s^{\frac{5}{8}}
	\end{align*}
	and
	\begin{align*}
		\left\|z+\frac{\bm l_s}{2}\cdot\bm\omega+\theta_s\right\|_{\T}|\le|z|+\left\|\frac{\bm l_s}{2}\cdot\bm\omega+\theta_s\right\|_{\T}<\delta_s^{\frac{5}{8}}+\delta_s^{\frac{2}{3}}<2\delta_s^{\frac{5}{8}}.
	\end{align*}
	Thus, both $z-\frac{\bm l_s}{2}\cdot\bm\omega$ and $z+\frac{\bm l_s}{2}\cdot\bm\omega$ belong to the set of \eqref{zs}, which together with \eqref{detss} implies
	\begin{align}
		\label{detss-}\left|\det \mathcal{S}_s\left(z-\frac{\bm l_s}{2}\cdot\bm\omega\right)\right|&\gtrsim\delta_{s}^{\frac{2}{75}}	\left\|\left(z-\frac{\bm l_s}{2}\cdot\bm\omega\right)-\theta_s\right\|_{\T}\cdot\left\|\left(z-\frac{\bm l_s}{2}\cdot\bm\omega\right)+\theta_s\right\|_{\T},\\
		\label{detss+}\left|\det \mathcal{S}_s\left(z+\frac{\bm l_s}{2}\cdot\bm\omega\right)\right|&\gtrsim\delta_{s}^{\frac{2}{75}}	\left\|\left(z+\frac{\bm l_s}{2}\cdot\bm\omega\right)-\theta_s\right\|_{\T}\cdot\left\|\left(z+\frac{\bm l_s}{2}\cdot\bm\omega\right)+\theta_s\right\|_{\T}.
	\end{align}
	Moreover, since $\#(A_{\bm k}^{s+1}-\bm k)\le 2^{s+1}$, \eqref{ss} and Lemma \ref{det1},  we have
	\begin{align}
		\nonumber\|\mathcal{S}_{s+1}(z)\|_0&\le \left\|\mathcal{S}_s\left(z-\frac{\bm l_s}{2}\cdot\bm\omega\right)\oplus \mathcal{S}_s\left(z+\frac{\bm l_s}{2}\cdot\bm\omega\right)\right\|_0+O(\delta_s^3)\\
		\label{Ss+10}&\le2|v|_R+\sum_{i=0}^{s}\delta_s<4|v|_R,
	\end{align}
	and
	\begin{align}
		\nonumber\det \mathcal{S}_{s+1}(z)&=\det \mathcal{S}_s\left(z-\frac{\bm l_s}{2}\cdot\bm\omega\right)\cdot\det \mathcal{S}_s\left(z+\frac{\bm l_s}{2}\cdot\bm\omega\right)+O((2^{s+1})^2(4|v|_R)^{2^{s+1}}\delta_s^4)\\
		\label{dets2}&=\det \mathcal{S}_s\left(z-\frac{\bm l_s}{2}\cdot\bm\omega\right)\cdot\det \mathcal{S}_s\left(z+\frac{\bm l_s}{2}\cdot\bm\omega\right)+O(\delta_s^{\frac{7}{2}}).
	\end{align}
	Notice that
	\begin{align}
		\nonumber\left\|z+\frac{\bm l_s}{2}\cdot\bm\omega-\theta_s\right\|_{\T}&\ge\|\bm l_s\cdot\bm\omega\|_{\T}-\left\|z-\frac{\bm l_s}{2}\cdot\bm\omega-\theta_0\right\|_{\T}\\
		\label{398}&>\frac{\g}{(100N_{s+1}^3)^\tau}-2\delta_s^{\frac{5}{8}}\gtrsim\delta_s^{\frac{1}{10}},
	\end{align}
	and
	\begin{align}
		\nonumber\left\|z-\frac{\bm l_s}{2}\cdot\bm\omega+\theta_s\right\|_{\T}&\ge\|\bm l_s\cdot\bm\omega\|_{\T}-\left\|z+\frac{\bm l_s}{2}\cdot\bm\omega+\theta_0\right\|_{\T}\\
		\label{399}&>\frac{\g}{(100N_{s+1}^3)^\tau}-2\delta_s^{\frac{5}{8}}\gtrsim\delta_s^{\frac{1}{10}}.
	\end{align}
	Let $z_{s+1}$ satisfy
	\begin{align}\label{zs+1}
		z_{s+1}\equiv\frac{\bm l_s}{2}\cdot\bm\omega+\theta_s\ (\text{mod}\ \Z),\ |z_{s+1}|=\left\|\frac{\bm l_s}{2}\cdot\bm\omega+\theta_s\right\|_{\T}<\delta_s^{\frac{2}{3}}.
	\end{align}
	From \eqref{detss-}--\eqref{399}, we get
	\begin{align*}
		|\det \mathcal{S}_{s+1}(z)|\gtrsim \delta_{s+1}^{\frac{2}{75}}((z-z_{s+1})\cdot(z+z_{s+1})+r_{s+1}(z)),
	\end{align*}
	where $r_{s+1}(z)$ is an analytic function in the set of \eqref{zs3} with 
	\begin{align}\label{rs+12}
		|r_{s+1}(z)|\lesssim\delta_s^{\frac{7}{2}-\frac{4}{75}-\frac{1}{5}}\ll\delta_s^3\ll\delta_s^{\frac{5}{8}}.
	\end{align}
	By the Rouch\'e's  theorem, the equation
	\begin{align*}
		(z-z_{s+1})\cdot(z+z_{s+1})+r_{s+1}(z)=0
	\end{align*}
	has exactly two roots $\theta_{s+1}$ and $\theta_{s+1}^{'}$ in the set of \eqref{zs3}, which are perturbations of $\pm z_{s+1}$. Notice that
	\begin{align*}
		\{|z|<\delta_s^{\frac{5}{8}}:\ \det \mathcal{M}_{s+1}(z)=0\}=\{|z|<\delta_s^{\frac{5}{8}}:\  \det \mathcal{S}_{s+1}(z)=0\}
	\end{align*}
	and $\det \mathcal{M}_{s+1}(z)$ is an even function of $z$. Thus
	\begin{align*}
		\theta_{s+1}^{'}=-\theta_{s+1}.
	\end{align*}
	If
	\begin{align*}
		|z_{s+1}-\theta_{s+1}|>|r_{s+1}(\theta_{s+1})|^{\frac{1}{2}}\ \text{and}\ |z_{s+1}+\theta_{s+1}|>|r_{s+1}(\theta_{s+1})|^{\frac{1}{2}},
	\end{align*}
	then
	\begin{align*}
		|r_{s+1}(\theta_{s+1})|>|z_{s+1}-\theta_{s+1}|\cdot|z_{s+1}+\theta_{s+1}|>|r_{s+1}(\theta_{s+1})|,
	\end{align*}
	which is a contradiction. Without loss of generality, we assume that
	\begin{align}\label{tzs+1}
		|\theta_{s+1}-z_{s+1}|\le|r_{s+1}(\theta_{s+1})|^{\frac{1}{2}}<\delta_s^{\frac{3}{2}}.
	\end{align}
	Moreover, since \eqref{zs+1} and \eqref{rs+12}, we get for $|z|=\delta_{s}^{\frac{5}{8}}$, 
	\begin{align*}
		\frac{|r_{s+1}(z)-r_{s+1}(\theta_{s+1})|}{|(z-z_{s+1})\cdot(z+z_{s+1})+r_{s+1}(\theta_{s+1})|}\le 2\delta_s^{\frac{7}{4}},
	\end{align*}
	which combined with $\theta_{s+1}^2-z_{s+1}^2+r_{s+1}(\theta_{s+1})=0$ shows
	\begin{align*}
		&\ \frac{|(z-z_{s+1})\cdot(z+z_{s+1})+r_{s+1}(z)|}{|(z-\theta_{s+1})\cdot(z+\theta_{s+1})|}\\
		=&\ \frac{|(z-z_{s+1})\cdot(z+z_{s+1})+r_{s+1}(z)|}{|(z-z_{s+1})\cdot(z+z_{s+1})+r_{s+1}(\theta_{s+1})|}\\
		\in&\ \left[1-2\delta_s^{\frac{7}{4}},1+2\delta_s^{\frac{7}{4}}\right].
	\end{align*}
	By the maximum modulus principle, we have
	\begin{align*}
		|(z-z_{s+1})(z+z_{s+1})+r_{s+1}(z)|\sim|(z-\theta_{s+1})(z+\theta_{s+1})|.
	\end{align*}
	Thus for $z$ being in the set of \eqref{zs3}, we have
	\begin{align}\label{detss+12}
		|\det \mathcal{S}_{s+1}(z)|\gtrsim\delta_{s+1}^{\frac{2}{75}}\|z-\theta_{s+1}\|_{\T}\cdot\|z+\theta_{s+1}\|_{\T}.
	\end{align}
	Given $\delta_{s+1}^{\frac{1}{2}}<\frac{1}{2}\delta_{s}^{\frac{5}{8}}$, the combination of \eqref{zs+1} and \eqref{tzs+1} implies 
	\begin{align*}
		\{z\in\C\ :\ \min_{\sigma=\pm1}|z+\sigma\theta_{s+1}|<\delta_{s+1}^{\frac{1}{2}}\}\subset\{z\in\C\ :\ |z|<\delta_s^{\frac{5}{8}}\}.
	\end{align*}
	Hence, \eqref{detss+12} also holds true for $z$ being in
	\begin{align*}
		\{z\in\C:\ \|z\pm\theta_{s+1}\|_{\T}<\delta_{s+1}^{\frac{1}{2}}\},
	\end{align*}
	which proves \eqref{detss} for the $(s+1)$-th step. Notice that
	\begin{align*}
		\|\theta+\bm k\cdot\bm\omega+\theta_{s+1}\|_{\T}<10\delta_{s+1}^{\frac{2}{3}},\ |\theta_{s+1}-z_{s+1}|<\delta_s^{\frac{3}{2}}\Rightarrow\left\|\theta+\left(\bm k+\frac{\bm l_s}{2}\right)\cdot\bm\omega+\theta_s\right\|_{\T}<\delta_s.
	\end{align*}
	Thus if 
	\begin{align*}
		\bm k\in\Z^d+\frac{1}{2}\sum_{i=0}^{s}\bm l_i\ \text{and}\ \|\theta+\bm k\cdot\bm\omega+\theta_{s+1}\|_{\T}<10\delta_{s+1}^{\frac{2}{3}},
	\end{align*}
	then
	\begin{align*}
		\bm k+\frac{\bm l_s}{2}\in\Z^d+\frac{1}{2}\sum_{i=0}^{s-1}\bm l_i\ \text{and}\ \left\|\theta+\left(\bm k+\frac{\bm l_s}{2}\right)\cdot\bm\omega+\theta_s\right\|_{\T}<\delta_s.
	\end{align*}
	Therefore, by \eqref{qspm}, we have $\bm k+\frac{\bm l_s}{2}\in Q_s^+$. Recalling \eqref{os+12} and \eqref{ps+12}, we have $\bm k\in P_{s+1}$. Thus
	\begin{align*}
		\left\{\bm k\in\Z^d+\frac{1}{2}\sum_{i=0}^{s}\bm l_i:\ \|\theta+\bm k\cdot\bm\omega+\theta_s\|_{\T}<10\delta_{s+1}^{\frac{2}{3}}\right\}\subset P_{s+1}.
	\end{align*}
	Similarly,
	\begin{align*}
		\left\{\bm k\in\Z^d+\frac{1}{2}\sum_{i=0}^{s}\bm l_i:\ \|\theta+\bm k\cdot\bm\omega-\theta_s\|_{\T}<10\delta_{s+1}^{\frac{2}{3}}\right\}\subset P_{s+1}.
	\end{align*}
	Hence, we prove $\eqref{fs}$ for the $(s+1)$-th step.
	
	Finally, we estimate $\mathcal{T}_{\tilde{\Omega}_{\bm k}^{s+1}}^{-1}$. For $\bm k\in P_{s+1}$, by \eqref{t+kos}, we have
	\begin{align*}
		\theta+\bm k\cdot\bm\omega\in\{z\in\C:\ \|z\|_{\T}<\delta_{s}^{\frac{5}{8}}\}.
	\end{align*}
	Thus from \eqref{detss+12}, we obtain
	\begin{align*}
		&|\det(\mathcal{T}_{A_{\bm k}^{s+1}}-\mathcal{R}_{A_{\bm k}^{s+1}}\mathcal{T}\mathcal{R}_{\tilde{\Omega}_{\bm k}^{s+1}\setminus A_{\bm k}^{s+1}}\mathcal{T}_{\tilde{\Omega}_{\bm k}^{s+1}\setminus A_{\bm k}^{s+1}}^{-1}\mathcal{R}_{\tilde{\Omega}_{\bm k}^{s+1}\setminus A_{\bm k}^{s+1}}\mathcal{T}\mathcal{R}_{A_{\bm k}^{s+1}})|\\
		=&|\det \mathcal{S}_{s+1}(\theta+\bm k\cdot\bm\omega)|\\
		\gtrsim&\delta_{s+1}^{\frac{2}{75}}\|\theta+\bm k\cdot\bm\omega-\theta_{s+1}\|_{\T}\cdot\|\theta+\bm k\cdot\bm\omega+\theta_{s+1}\|_{\T}.
	\end{align*}
	By \eqref{ss}, Cramer's rule and Lemma \ref{chi}, one has
	\begin{align*}
		&\|(\mathcal{T}_{A_{\bm k}^{s+1}}-\mathcal{R}_{A_{\bm k}^{s+1}}\mathcal{T}\mathcal{R}_{\tilde{\Omega}_{\bm k}^{s+1}\setminus A_{\bm k}^{s+1}}\mathcal{T}_{\tilde{\Omega}_{\bm k}^{s+1}\setminus A_{\bm k}^{s+1}}^{-1}\mathcal{R}_{\tilde{\Omega}_{\bm k}^{s+1}\setminus A_{\bm k}^{s+1}}\mathcal{T}\mathcal{R}_{A_{\bm k}^{s+1}})^{-1}\|_0\\
		=&|\det \mathcal{S}_{s+1}(\theta+\bm k\cdot\bm\omega)|^{-1}\|(\mathcal{T}_{A_{\bm k}^{s+1}}-\mathcal{R}_{A_{\bm k}^{s+1}}\mathcal{T}\mathcal{R}_{\tilde{\Omega}_{\bm k}^{s+1}\setminus A_{\bm k}^{s+1}}\mathcal{T}_{\tilde{\Omega}_{\bm k}^{s+1}\setminus A_{\bm k}^{s+1}}^{-1}\mathcal{R}_{\tilde{\Omega}_{\bm k}^{s+1}\setminus A_{\bm k}^{s+1}}\mathcal{T}\mathcal{R}_{A_{\bm k}^{s+1}})^{*}\|_0\\
		\lesssim&(2^{s+1})^2 (4|v|_R)^{2^{s+1}}\delta_{s+1}^{-\frac{2}{75}}\|\theta+\bm k\cdot\bm\omega-\theta_{s+1}\|_{\T}^{-1}\cdot\|\theta+\bm k\cdot\bm\omega+\theta_{s+1}\|_{\T}^{-1}.
	\end{align*}
	Recalling \eqref{tosa2}, we get since Lemma \ref{scl}
	\begin{align}
		\nonumber\|\mathcal{T}_{\tilde{\Omega}_{\bm k}^{s+1}}^{-1}\|_0&<4(1+\|\mathcal{T}_{\tilde{\Omega}_{\bm k}^{s+1}\setminus A_{\bm k}^{s+1}}\|_0)^2\\
		\nonumber&\ \ \times(1+\|(\mathcal{T}_{A_{\bm k}^{s+1}}-\mathcal{R}_{A_{\bm k}^{s+1}}\mathcal{T}\mathcal{R}_{\tilde{\Omega}_{\bm k}^{s+1}\setminus A_{\bm k}^{s+1}}\mathcal{T}_{\tilde{\Omega}_{\bm k}^{s+1}\setminus A_{\bm k}^{s+1}}^{-1}\mathcal{R}_{\tilde{\Omega}_{\bm k}^{s+1}\setminus A_{\bm k}^{s+1}}\mathcal{T}\mathcal{R}_{A_{\bm k}^{s+1}})^{-1}\|_0)\\
		\label{tos+1-12}&<\delta_{s+1}^{-\frac{1}{15}}\|\theta+\bm k\cdot\bm\omega-\theta_{s+1}\|_{\T}^{-1}\cdot\|\theta+\bm k\cdot\bm\omega+\theta_{s+1}\|_{\T}^{-1}.
	\end{align}
	For the case of 
	\begin{align}\label{ls2}
		\left\|\frac{\bm l_s}{2}\cdot\bm\omega+\theta_0-\frac{1}{2}\right\|_{\T}<\delta_s^{\frac{2}{3}}, 
	\end{align}
	we have
	\begin{align}\label{ps+1s2}
		P_{s+1}\subset\{\bm k\in\Z^d+\frac{1}{2}\sum_{i=0}^s \bm l_i:\ \left\|\theta+\bm k\cdot\bm\omega-\frac{1}{2}\right\|_{\T}<3\delta_s^{\frac{2}{3}}\}.
	\end{align}
	Now we consider $\mathcal{M}_{s+1}(z)$ in the set
	\begin{align}\label{zs4}
		\{z\in\C:\ \left|z-\frac{1}{2}\right|<\delta_{s}^{\frac{5}{8}}\}.
	\end{align}
	The similar argument shows that $\det\mathcal{M}_{s+1}(z)=0$ has two roots $\theta_{s+1}$ and $1-\theta_{s+1}$ in the set of \eqref{zs4} such that \eqref{Ss+10}--\eqref{tos+1-12} hold true for $z$ being in the set of \eqref{zs4}. Hence if \eqref{ls2}, then \eqref{Ss+10}--\eqref{tos+1-12} hold true for $z$ being in
	\begin{align*}
		\{z\in\C:\ \left\|z-\frac{1}{2}\right\|_{\T}<\delta_{s}^{\frac{5}{8}}\}.
	\end{align*}
	By \eqref{ps+1s2}, for $\bm k\in P_{s+1}$, we also have
	\begin{align*}
		\|\mathcal{T}_{\tilde{\Omega}_{\bm k}^{s+1}}^{-1}\|_0<\delta_{s+1}^{-\frac{1}{15}}\|\theta+\bm k\cdot\bm\omega-\theta_{s+1}\|_{\T}^{-1}\cdot\|\theta+\bm k\cdot\bm\omega+\theta_{s+1}\|_{\T}^{-1}.
	\end{align*}
	
	Therefore, we have established desired estimates of $\|\mathcal{T}_{\tilde{\Omega}_{\bm k}^{s+1}}^{-1}\|_0$
	for both cases of $(\bm C1)_{s}$ and $(\bm C2)_{s}$.
	
	\  \\
	
	\begin{itemize}
		\item[\textbf{Step2}]: \textbf{Estimates of $\|\mathcal{T}_{\tilde{\Omega}_{\bm k}^{s+1}}^{-1}\|_{\alpha}$ for $\alpha\in(0,\alpha_1]$.} 
	\end{itemize}
	The main result of this step is Theorem \ref{psqs}. Recalling
	\begin{align*}
		\left(\frac{\g}{\delta_{s+1}}\right)=\left(\frac{\g}{\delta_s}\right)^{30}
	\end{align*}
	and
	\begin{align*}
		Q_{s+1}^{\pm}=\{\bm k\in P_{s+1}:\ \|\theta+\bm k\cdot\bm\omega\pm\theta_{s+1}\|_{\T}<\delta_{s+1}\},\ Q_{s+1}=Q_{s+1}^+\cup Q_{s+1}^-,
	\end{align*}
	we have 
	\begin{thm}\label{psqs}
		For $\bm k\in P_{s+1}\setminus Q_{s+1}$ and $\alpha\in(0,\alpha_1]$,  one has 
		\begin{align*}
			\|T_{\tilde{\Omega}_{\bm k}^{s+1}}^{-1}\|_\alpha< \zeta_{s+1}^{\alpha}\delta_{s+1}^{-\frac{7}{3}}.
		\end{align*}
	\end{thm}
	\begin{rem}\label{remt}
		Actually, for $\alpha\in(0,\alpha_1]$,  we can prove
		\begin{align*}
			\|\mathcal{T}_{\Omega_{\bm k}^{s+1}}^{-1}\|_{\alpha}\lesssim\zeta_{s+1}^{\alpha}\delta_{s+1}^{-\frac{1}{3}}A^{-2}
		\end{align*}
		provided $\min\limits_{\sigma=\pm1}\|\theta+\bm k\cdot\bm\omega+\sigma\theta_{s+1}\|_{\T}\ge A>0$ and $\zeta_{s+1}^{-\alpha_1}\zeta_s^{\alpha_1}\delta_{s+1}^{-\frac{1}{3}}A^{-2}\ll1$.
	\end{rem}
	
	Similar to the proof of Theorem \ref{psq1},   the proof of Theorem \ref{psqs}  also builds on  several lemmas. 
	
	Let $G=\tilde{\Omega}_{\bm k}^{s+1}\setminus\Omega_{\bm k}^{s+1}$ which is $s$-$\good$ (cf.  Subsection 3.3 of \cite{CSZ24a})  and $B=\Omega_{\bm k}^{s+1}$. Thus $X:=\Omega_{\bm k}^{s+1}=B\cup G$, $G\cap A_{\bm k}^{s+1}=\emptyset$ and $A_{\bm k}^{s+1}\subset B$.
	
	We have 
	
	\begin{lem}\label{mn2}
		Let $\ep_0=\ep_0(\alpha_1,\alpha_0,d,\tau,\g,v,R,\phi)\ll1$. Then there exist $\mathcal{M}\in\mathbf{M}_X^G$ and $\mathcal{N}\in\mathbf{M}_B^G$ satisfying
		\begin{align}\label{eq4}
			\mathcal{M}\mathcal{T}_X=\mathcal{R}_G+\mathcal{N}.
		\end{align}
		with the following estimates:  
		\begin{align*}
			\|\mathcal{N}\|_0\lesssim 1,\ \|\mathcal{M}\|_0\lesssim \tilde{\zeta}_s^{\alpha_0}\delta_s^{-\frac{31}{15}},
		\end{align*}
		and for $\alpha\in(0,\alpha_1]$,
		\begin{align}\label{emn1s}
			\|\mathcal{N}\|_\alpha\lesssim\zeta_s^{\alpha}\tilde{\zeta}_s^{\alpha_0}\delta_s^{-\frac{7}{3}},\ \|\mathcal{M}\|_\alpha\lesssim\zeta_s^{\alpha}\tilde{\zeta}_s^{2\alpha_0}\delta_s^{-\frac{22}{5}}.
		\end{align}
		
	\end{lem}

	\begin{proof}
		Define
		\begin{align}\label{tpt1}
			\tilde{P}_t=\{\bm j\in P_t:\ \exists \bm j'\in Q_{t-1}\ \text{s.t.}\ \tilde{\Omega}_{\bm j'}^{t-1}\subset X,\tilde{\Omega}_{\bm j'}^{t-1}\subset\Omega_{\bm j}^t\},\ 1\le t\le s+1.
		\end{align}
		From \eqref{311}, \eqref{a1s} and \eqref{tpt1}, it follows that for $\bm j'\in \tilde{P}_t\cap Q_t\ (1\le t\le s)$, there exists a $\bm j\in\tilde{P}_{t+1}$ such that
		\begin{align*}
			\tilde{\Omega}_{\bm j'}^t\subset\Omega_{\bm j}^{t+1}
		\end{align*}
		and
		\begin{align*}
			\tilde{P}_{s+1}\cap Q_{s+1}=\emptyset.
		\end{align*}
		Moreover,
		\begin{align*}
			\bigcup_{1\le t\le s+1}\bigcup_{\bm j\in\tilde{P}_t}\tilde{\Omega}_{\bm j}^t\subset X.
		\end{align*}
		Hence for any $\bm l\in G$, if 
		\begin{align*}
			\bm l\in\bigcup_{\bm j\in\tilde{P}_1}\Omega_{\bm j}^1,
		\end{align*}
		then there exists a $t\in[1,s]$ such that
		\begin{align*}
			\bm l\in\bigcup_{\bm j\in\tilde{P}_t\setminus Q_t}\Omega_{\bm j}^t.
		\end{align*}
		For every $\bm l\in G$, define %its block in $X$:
		\begin{align*}
			U_{\bm l}=\left\{\begin{array}{ll}
				\lg_{\frac{1}{2}N_1}(\bm l)\cap X, &\text{if}\ \bm l\notin\bigcup_{\bm j\in \tilde{P}_1}\Omega_{\bm j}^1,\\
				\tilde{\Omega}_{\bm j}^t, &\text{if}\ \bm l\in\Omega_{\bm j}^t\ \text{for some}\ \bm j\in\tilde{P}_t\setminus Q_t.
			\end{array}\right.
		\end{align*}
		Let  $\mathcal{Q}_{\bm l}=\ep \mathcal{T}_{U_{\bm l}}^{-1}\mathcal{W}_{X\setminus U_{\bm l}}^{U_{\bm l}}\in\mathbf{M}_{X\setminus U_{\bm l}}^{U_{\bm l}}$. 
		We now vary $\bm l\in G$ and define 
		\begin{align*}
			\mathcal{K}(\bm l,\bm l')=\left\{\begin{array}{ll}
				0, & \text{for}\ \bm l'\in U_{\bm l}\\
				\mathcal{Q}_{\bm l}(\bm l,\bm l'), &\text{for}\ \bm l'\in X\setminus U_{\bm l},
			\end{array}\right.
		\end{align*}
		and
		\begin{align*}
			\mathcal{L}(\bm l,\bm l')=\left\{\begin{array}{ll}
				\mathcal{T}_{U_{\bm l}}^{-1}(\bm l,\bm l'), & \text{for}\ \bm l'\in U_{\bm l}\\
				0, &\text{for}\ \bm l'\in X\setminus U_{\bm l}.
			\end{array}\right.
		\end{align*}
		Then we have 
		\begin{align}\label{eql1}
			\mathcal{L}\mathcal{T}_X=\mathcal{R}_G+\mathcal{K}.
		\end{align}
		
		Next, we estimate $\mathcal{K}\in\mathbf{M}_{X}^{G}$ and $\mathcal{L}\in\mathbf{M}_X^G$. We have the following cases.\\
		(1) Let $\bm l\notin\bigcup_{\bm j\in \tilde{P}_1}\Omega_{\bm j}^1$. Then $U_{\bm l}$ is $0$-good. Similar to the proof of Lemma \ref{psq1}, we have for $\alpha\in[0,\alpha_1], $
		\begin{align*}
			&\|\mathcal{K}^{\{\bm l\}}\|_{\alpha_0}\lesssim N_1^{-\alpha_1+\alpha_0},\\
			&\|\mathcal{K}^{\{\bm l\}}\|_{\alpha+\alpha_0}\lesssim \delta_0^{28}<1,\\
			&\|\mathcal{L}^{\{\bm l\}}\|_{\alpha+\alpha_0}\lesssim\delta_0^{-2}.
		\end{align*}
		(2) Let $\bm l\in\left(\bigcup_{\bm j\in \tilde{P}_1}\Omega_{\bm k}^1\right)\bigcap G$. Then  there exist some $t\in[1,s]$ and $\bm j'\in\tilde{P}_t\setminus Q_t$ such that $\bm l\in \Omega_{\bm j'}^t$. Then by \eqref{tame} and \eqref{tba}, we obtain 
		\begin{align}\label{qla1s}
			\|\mathcal{Q}_{\bm l}\|_{\alpha_1}&\lesssim\|\mathcal{T}_{U_{\bm l}}^{-1}\|_{\alpha_1}\|\mathcal{W}_X\|_{\alpha_1}\lesssim \zeta_t^{\alpha_1}\delta_{t}^{-\frac{7}{3}}.
		\end{align}
		By \eqref{tame}, \eqref{smo2} and \eqref{tba}, we have for $\alpha\in(0,\alpha_1]$, 
		\begin{align*}
			\|\mathcal{Q}_{\bm l}\|_{\alpha+\alpha_0}&\lesssim\|\mathcal{T}_{U_{\bm l}}^{-1}\|_{\alpha+\alpha_0}\|\mathcal{W}_X\|_{0}+\|\mathcal{T}_{U_{\bm l}}^{-1}\|_{0}\|\mathcal{W}_X\|_{\alpha+\alpha_0}\\
			&\lesssim\tilde{\zeta}_t^{\alpha_0}\|\mathcal{T}_{U_{\bm l}}^{-1}\|_{\alpha}\|\mathcal{W}_X\|_{\alpha+\alpha_0}\lesssim \zeta_t^{\alpha}\tilde{\zeta}_t^{\alpha_0}\delta_t^{-\frac{7}{3}}.
		\end{align*}
		Note that if $\bm l'\in X\setminus U_{\bm l}$, then $\|\bm l-\bm l'\|\ge\frac{\tilde{\zeta}_t}{2}$. This implies $\mathcal{K}^{\{\bm l\}}(\bm l,\bm l')=0$ for $\|\bm l-\bm l'\|<\frac{\tilde{\zeta}_t}{2}$. By \eqref{smo1} and \eqref{qla1s}, we obtain
		\begin{align*}
			\|\mathcal{K}^{\{\bm l\}}\|_{\alpha_0}&\lesssim\tilde{\zeta}_t^{-\alpha_1+\alpha_0}\|\mathcal{K}^{\{\bm l\}}\|_{\alpha_1}\\
			&\le \tilde{\zeta}_t^{-\alpha_1+\alpha_0}\|\mathcal{Q}_{\bm l}\|_{\alpha_1}\\
			&\lesssim \zeta_t^{\alpha_1}\tilde{\zeta}_t^{-\alpha_1+\alpha_0}\delta_{t}^{-\frac{7}{3}}.
		\end{align*}
		Similarly, for $\alpha\in(0,\alpha_1],$  we obtain
		\begin{align*}
			\|\mathcal{K}^{\{\bm l\}}\|_{\alpha+\alpha_0}\le\|\mathcal{Q}_{\bm l}\|_{\alpha+\alpha_0}\lesssim\zeta_t^{\alpha}\tilde{\zeta}_t^{\alpha_0}\delta_t^{-\frac{7}{3}}.
		\end{align*}
		By the definition of $U_{\bm l}$, if $\|\bm l-\bm l'\|>2\tilde{\zeta}_t$, then $\bm l'\notin U_{\bm l}$. This implies $\mathcal{L}^{\{\bm l\}}(\bm l,\bm l')=0$ for $\|\bm l-\bm l'\|>2\tilde{\zeta}_t$. By \eqref{smo2}, \eqref{tb0} and \eqref{tba}, we have 
		\begin{align*}
			\|\mathcal{L}^{\{\bm l\}}\|_{\alpha_0}\lesssim\tilde{\zeta}_t^{\alpha_0}\|\mathcal{L}^{\{\bm l\}}\|_0\le \tilde{\zeta}_t^{\alpha_0}\|T_{U_{\bm l}}^{-1}\|_0<\tilde{\zeta}_t^{\alpha_0}\delta_t^{-\frac{31}{15}},
		\end{align*}
		and for $\alpha\in(0,\alpha_1]$,
		\begin{align*}
			\|\mathcal{L}^{\{\bm l\}}\|_{\alpha+\alpha_0}\lesssim \tilde{\zeta}_t^{\alpha_0}\|\mathcal{L}^{\{\bm l\}}\|_{\alpha}\le\tilde{\zeta}_t^{\alpha_0}\|T_{U_{\bm l}}^{-1}\|_{\alpha}<\zeta_t^{\alpha}\tilde{\zeta}_t^{\alpha_0}\delta_t^{-\frac{7}{3}}.
		\end{align*}
		To sum up, by \eqref{re}, we can get for $\alpha\in(0,\alpha_1],$
		\begin{align}
			\label{sk0}\|\mathcal{K}\|_0&\lesssim\sup_{\bm l\in G}\|\mathcal{K}^{\{\bm l\}}\|_{\alpha_0}\lesssim \max\{\zeta_1^{\alpha_1}\tilde{\zeta}_1^{-\alpha_1+\alpha_0}\delta_{1}^{-\frac{7}{3}},N_1^{-\alpha_1+\alpha_0}\}\le\frac{1}{2},\\
			\label{ska}\|\mathcal{K}\|_{\alpha}&\lesssim\sup_{\bm l\in G}\|\mathcal{K}^{\{\bm l\}}\|_{\alpha+\alpha_0}\lesssim\zeta_s^{\alpha}\tilde{\zeta}_s^{\alpha_0}\delta_s^{-\frac{7}{3}},
		\end{align}
		and
		\begin{align}
			\label{sl0}\|\mathcal{L}\|_{0}&\lesssim\sup_{\bm l\in G}\|\mathcal{L}^{\{\bm l\}}\|_{\alpha_0}\lesssim\tilde{\zeta}_s^{\alpha_0}\delta_s^{-\frac{31}{15}},\\
			\label{sla}\|\mathcal{L}\|_{\alpha}&\le\sup_{\bm l\in G}\|\mathcal{L}^{\{\bm l\}}\|_{\alpha+\alpha_0}\lesssim\zeta_s^{\alpha}\tilde{\zeta}_s^{\alpha_0}\delta_s^{-\frac{7}{3}}.
		\end{align}

		Finally, by recalling Lemma \ref{pa}, we have that $\mathcal{I}_G+\mathcal{K}_G$ is invertible and  
		\begin{align}\label{I+G3}
			\|(\mathcal{I}_G+\mathcal{K}_G)^{-1}\|_{\alpha}&\lesssim \min\{1,\|\mathcal{K}\|_\alpha\}\ \text{for}\ \alpha\in[0,\alpha_1].
		\end{align}
		From \eqref{eql1}, we have
		\begin{align*}
			(\mathcal{I}_G+\mathcal{K}_G)^{-1}\mathcal{L}\mathcal{T}_X=\mathcal{R}_G+(\mathcal{I}_G+\mathcal{K}_G)^{-1}\mathcal{K} \mathcal{R}_B.
		\end{align*}
		Let
		\begin{align*}
			\mathcal{M}=(\mathcal{I}_G+\mathcal{K}_G)^{-1}\mathcal{L}\in\mathbf{M}_X^G,\ \mathcal{N}=(\mathcal{I}_G+\mathcal{K}_G)^{-1}\mathcal{K} \mathcal{R}_B\in\mathbf{M}_B^G,
		\end{align*}
		then
		\begin{align*}
			\mathcal{M}\mathcal{T}_X=\mathcal{R}_G+\mathcal{N}.
		\end{align*}
		By recalling \eqref{tame} and \eqref{sk0}--\eqref{I+G3},  we have
		\begin{align}
			\label{sn01}\|\mathcal{N}\|_0&\le\|(\mathcal{I}_G+\mathcal{K}_G)^{-1}\|_0\|\mathcal{K}\|_0\lesssim 1,\\
			\label{sm01}\|\mathcal{M}\|_0&\le\|(\mathcal{I}_G+\mathcal{K}_G)^{-1}\|_0\|\mathcal{L}\|_0\lesssim\tilde{\zeta}_s^{\alpha_0}\delta_s^{-\frac{31}{15}},
		\end{align}
		and for $\alpha\in(0,\alpha_1]$,
		\begin{align}
			\nonumber\|\mathcal{N}\|_\alpha&\lesssim \left(\|(\mathcal{I}_G+\mathcal{K}_G)^{-1}\|_\alpha\|\mathcal{K}\|_0+\|(\mathcal{I}_G+\mathcal{K}_G)^{-1}\|_0\|\mathcal{K}\|_\alpha\right)\\
			\label{sna1}&\lesssim\zeta_s^{\alpha}\tilde{\zeta}_s^{\alpha_0}\delta_s^{-\frac{7}{3}},\\
			\nonumber\|\mathcal{M}\|_\alpha&\lesssim\left(\|(\mathcal{I}_G+\mathcal{K}_G)^{-1}\|_\alpha\|\mathcal{L}\|_0+\|(\mathcal{I}_G+\mathcal{K}_G)^{-1}\|_0\|\mathcal{L}\|_\alpha\right)\\
			\label{sma1}&\lesssim\zeta_s^{\alpha}\tilde{\zeta}_s^{2\alpha_0}\delta_s^{-\frac{22}{5}}.
		\end{align}
	\end{proof}
	
	We further have 
	\begin{lem}\label{tz2}
		Let \begin{align}\label{Z2}
			\mathcal{T}'=\mathcal{T}_X\mathcal{R}_B-\mathcal{T}_X\mathcal{R}_G \mathcal{N}\in\mathbf{M}_B^X,\ \mathcal{Z}=\mathcal{I}_X-\mathcal{T}_X\mathcal{R}_G \mathcal{M}\in\mathbf{M}_X^X. 
		\end{align}
		Then 
		\begin{align}\label{ztt2}
			\mathcal{Z}\mathcal{T}_X=\mathcal{T}', 
		\end{align}
		and  for $\alpha\in[0,\alpha_1],$
		\begin{align}\label{ez2}
			\|\mathcal{T}'\|_\alpha\lesssim\zeta_s^{\alpha}\tilde{\zeta}_s^{\alpha_0}\delta_t^{-\frac{7}{3}},\ \|\mathcal{Z}\|_\alpha\lesssim\zeta_s^{\alpha}\tilde{\zeta}_s^{2\alpha_0}\delta_s^{-\frac{22}{5}}.
		\end{align}
		Moreover, $\mathcal{R}_B\mathcal{T}_X^{-1}$ is a left inverse of $\mathcal{T}'$.
	\end{lem}
	\begin{proof}
		We first prove  \eqref{ez2}.  We have for $\alpha\in[0,\alpha_1]$,  
		\begin{align*}
			\|\mathcal{T}_X\|_\alpha\le\|\ep \mathcal{W}_X\|_\alpha+\|\mathcal{D}_X\|\lesssim1.		
		\end{align*}
		From \eqref{tame} and \eqref{sn01}--\eqref{sma1}, we get  for $\alpha\in[0,\alpha_1], $
		\begin{align*}
			\|\mathcal{T}'\|_\alpha&\lesssim \|\mathcal{T}_X\|_\alpha+\|\mathcal{T}_X\|_\alpha\|\mathcal{N}\|_0+\|\mathcal{T}_X\|_0\|\mathcal{N}\|_\alpha\lesssim\zeta_s^{\alpha}\tilde{\zeta}_s^{\alpha_0}\delta_s^{-\frac{7}{3}},\\
			\|\mathcal{Z}\|_\alpha&\lesssim \|\mathcal{T}_X\|_\alpha\|\mathcal{M}\|_0+\|\mathcal{T}_X\|_0\|\mathcal{M}\|_\alpha\lesssim\zeta_s^{\alpha}\tilde{\zeta}_s^{2\alpha_0}\delta_s^{-\frac{22}{5}}.
		\end{align*}
		The deduce  of both \eqref{ztt2} and the left inverse argument   is similar to that of Lemma \ref{tz}.
	\end{proof}
	
	\begin{lem}\label{t'v2}
		The operator $\mathcal{T}'$ defined in \eqref{Z2} has a left inverse $\mathcal{V}$ satisfying for $\alpha\in[0,\alpha_1],$
		\begin{align}\label{ev2}
			\|\mathcal{V}\|_\alpha\lesssim\zeta_{s+1}^{\alpha}\delta_{s+1}^{-\frac{31}{15}}.
		\end{align}
	\end{lem}		
	\begin{proof}		
		We introduce $Y=\lg_{\zeta_{s+1}}(\bm k)$ and let $\mathcal{E}=\mathcal{R}_{Y}\mathcal{T}'\in\mathbf{M}_B^Y$.
		We claim that $\mathcal{E}$ has a left inverse $\mathcal{V}$ satisfying 
		\begin{align*}
			\|\mathcal{V}\|_0<2\delta_{s+1}^{-\frac{31}{15}}.
		\end{align*}
		Let $\|\bm l-\bm l'\|\le \frac{\diam(Y)}{2}$ and $\mathcal{P}=\mathcal{T}'-\mathcal{E}$, which implies \begin{align*}
			\mathcal{P}(\bm l,\bm l')=0.
		\end{align*}
		Then by recalling \eqref{smo1} and \eqref{ez2}, we obtain for $\alpha\in[0,\alpha_1],$
		\begin{align*}
			\|\mathcal{P}\|_\alpha&\lesssim (\diam (Y))^{-\alpha_1+\alpha}\|\mathcal{P}\|_{\alpha_1}\le (\diam (Y))^{-\alpha_1+\alpha}\|\mathcal{T}'\|_{\alpha_1}\lesssim \zeta_{s+1}^{-\alpha_1+\alpha}\zeta_s^{\alpha_1}\tilde{\zeta}_s^{\alpha_0}\delta_s^{-\frac{7}{3}}.
		\end{align*}
		Thus by \eqref{indpa}, \eqref{s+1T-1} and $\alpha_1>2200\tau$, we have
		\begin{align*}
			\|\mathcal{P}\|_0\|\mathcal{R}_B\mathcal{T}_X^{-1}\|_0&\le\|\mathcal{P}\|_0\|\mathcal{T}_X^{-1}\|_0\lesssim \zeta_{s+1}^{-\alpha_1}\zeta_s^{\alpha_1}\tilde{\zeta}_s^{\alpha_0}\delta_{s}^{-\frac{7}{3}}\delta_{s+1}^{-\frac{31}{15}}<\frac{1}{2}.
		\end{align*}
		It then follows from Lemma \ref{pa} that $\mathcal{E}$ has a left inverse $\mathcal{V}\in\mathbf{M}_Y^B$ satisfying 
		\begin{align}\label{v02} 
			\|\mathcal{V}\|_0\le 2\|\mathcal{R}_B\mathcal{T}_X^{-1}\|_0<2\delta_{s+1}^{-\frac{31}{15}}.
		\end{align}
		If $\|\bm l-\bm l'\|\ge 2\diam(Y)$, we have
		\begin{align*}
			\mathcal{V}(\bm l,\bm l')=0.
		\end{align*}
		Using \eqref{smo2} and \eqref{v02}  yields for $\alpha\in[0,\alpha_1]$,
		\begin{align*}
			\|\mathcal{V}\|_\alpha\lesssim(\diam(Y))^{\alpha}\|\mathcal{V}\|_0\lesssim\zeta_{s+1}^{\alpha}\delta_{s+1}^{-\frac{31}{15}}.
		\end{align*}
		Finally, we have  
		\begin{align*}
			\mathcal{V}\mathcal{P}=\mathcal{R}_B\mathcal{V}\mathcal{R}_{Y}(\mathcal{T}'-R_Y \mathcal{T}')=0.
		\end{align*}
		Since  $\mathcal{T}'=\mathcal{D}+\mathcal{R}$ and $\mathcal{V}$ is a left inverse of $\mathcal{E}$, $\mathcal{V}$ is a left inverse of $\mathcal{T}'$. 
	\end{proof}
	
	We are ready to prove Theorem \ref{psqs}. 
	
	\begin{proof}[Proof of Theorem \ref{psqs}]
		Since \eqref{eql1}, \eqref{ztt2} and Lemma \ref{t'v2}, we  obtain
		\begin{align*}
			\left(\mathcal{M}-\mathcal{N}\mathcal{V}\mathcal{Z}+\mathcal{V}\mathcal{Z}\right)\mathcal{T}_X&=\mathcal{R}_G+\mathcal{N}+(\mathcal{R}_B-\mathcal{N})\mathcal{V}\mathcal{T}'\\
			&=\mathcal{R}_G\oplus\mathcal{R}_B=\mathcal{I}_X,
		\end{align*}
		which implies
		\begin{align*}
			\mathcal{T}_X^{-1}=\mathcal{M}-\mathcal{N}\mathcal{V}\mathcal{Z}+\mathcal{V}\mathcal{Z}.
		\end{align*}
		Hence 
		\begin{align*}
			\mathcal{R}_B\mathcal{T}_X^{-1}=\mathcal{V}\mathcal{Z},\ \mathcal{R}_G \mathcal{T}_X^{-1}=\mathcal{M}-\mathcal{N}\mathcal{R}_B\mathcal{T}_X^{-1}.
		\end{align*}
		Then for $\alpha\in(0,\alpha_1]$, we can obtain by using \eqref{tame}, \eqref{emn1s}, \eqref{ez2} and \eqref{ev2}  that 
		\begin{align*} 
			\|\mathcal{R}_B\mathcal{T}_X^{-1}\|_\alpha&\lesssim\left(\|\mathcal{V}\|_\alpha\|\mathcal{Z}\|_0+\|\mathcal{V}\|_0\|\mathcal{Z}\|_\alpha\right)\\
			&\lesssim\zeta_{s+1}^{\alpha}\tilde{\zeta}_{s}^{2\alpha_0}\delta_s^{-\frac{22}{5}}\delta_{s+1}^{-\frac{31}{15}},\\
			\|\mathcal{R}_G \mathcal{T}_X^{-1}\|_\alpha&\lesssim \|\mathcal{M}\|_\alpha+\|\mathcal{N}\|_\alpha\|\mathcal{R}_B \mathcal{T}_X^{-1}\|_0+\|\mathcal{N}\|_0\|\mathcal{R}_B \mathcal{T}_X^{-1}\|_\alpha\\
			&\lesssim\zeta_{s+1}^{\alpha}\tilde{\zeta}_{s}^{3\alpha_0}\delta_s^{-\frac{101}{15}}\delta_{s+1}^{-\frac{31}{15}}.
		\end{align*}
		Thus for  $\alpha\in(0,\alpha_1]$, we obtain
		\begin{align*}
			\|\mathcal{T}_X^{-1}\|_\alpha&\le\|\mathcal{R}_B \mathcal{T}_X^{-1}\|_\alpha+\|\mathcal{R}_G \mathcal{T}_X^{-1}\|_\alpha\\
			&\lesssim \zeta_{s+1}^{\alpha}\tilde{\zeta}_{s}^{3\alpha_0}\delta_s^{-\frac{101}{15}}\delta_{s+1}^{-\frac{31}{15}}\\
			&<\zeta_{s+1}^{\alpha}\delta_{s+1}^{-\frac{7}{3}},
		\end{align*} 
		which concludes the proof of Theorem \ref{psqs}. 
	\end{proof}
	\ \\
	
	\begin{itemize}
		\item[\textbf{Step 3}]: \textbf{Estimates of $\|\mathcal{T}_{\lg}^{-1}\|_{\alpha}$ for general $(s+1)$-good $\Lambda$}.
	\end{itemize}
	In this step, we will complete the verification of Theorem \ref{ind}.
	Assume that the finite set $\lg\subset\Z^d$ is $(s+1)$-$\good$, namely, 
	\begin{align}\label{s+1g}
		\left\{\begin{array}{l}
			\bm k'\in Q_{s'},\ \tilde{\Omega}_{\bm k'}^{s'}\subset\lg,\ \tilde{\Omega}_{\bm k'}^{s'}\subset\Omega_{\bm k}^{s'+1}\Rightarrow\tilde{\Omega}_{\bm k}^{s'+1}\subset\lg\ \text{for}\ s'<s+1,\\
			\{\bm k\in P_{s+1}:\ \tilde{\Omega}_{\bm k}^{s+1}\subset\lg\}\cap Q_{s+1}=\emptyset.
		\end{array}\right.
	\end{align}
	It remains to verify   \eqref{tsg01} and \eqref{tsg1} with $s$ being replaced with $s+1$.
	
	We will combine the estimates of $\mathcal{T}_{\tilde{\Omega}_{\bm k}^{s+1}}^{-1}$, smoothing property and rows estimate  to finish this verification.
	
	\begin{thm}\label{s+1gt}
		If $\lg$ is $(s+1)$-$\good$, then
		\begin{align*}
			\|\mathcal T_\lg^{-1}\|_0&<\delta_{s+1}^{-\frac{2}{15}}\times\sup\limits_{\{\bm k\in P_{s+1}:\ \tilde{\Omega}_{\bm k}^{s+1}\subset\lg\}}\left(\|\theta+\bm k\cdot\bm\omega-\theta_{s+1}\|_{\T}^{-1}\cdot\|\theta+\bm k\cdot\bm\omega+\theta_{s+1}\|_{\T}^{-1}\right),
		\end{align*}
		and for $\alpha\in(0,\alpha_1],$
		\begin{align*}
			\|\mathcal T_\lg^{-1}\|_\alpha&<\zeta_{s+1}^{\alpha}\delta_{s+1}^{-\frac{14}{3}}.
		\end{align*}
	\end{thm}
	\begin{proof}
		%We can prove Theorem \ref{s+1gt} now. 
		Define
		\begin{align}\label{tpt2}
			\tilde{P}_t=\{\bm k\in P_t:\ \exists \bm k'\in Q_{t-1}\ \text{s.t.}\ \tilde{\Omega}_{\bm k'}^{t-1}\subset \lg,\tilde{\Omega}_{\bm k'}^{t-1}\subset\Omega_{\bm k}^t\},\ 1\le t\le s+1.
		\end{align}
		Similar to the proof of Lemma \ref{psqs}, we have
		\begin{align*}
			\tilde{P}_{s+1}\cap Q_{s+1}=\emptyset, 
		\end{align*}
		and for any $\bm l\in \lg$, if 
		\begin{align*}
			\bm l\in\bigcup_{\bm k\in\tilde{P}_1}\Omega_{\bm k}^1,
		\end{align*}
		then there exists a $t\in[1,s+1]$ such that
		\begin{align*}
			\bm l\in\bigcup_{\bm k\in\tilde{P}_t\setminus Q_t}\Omega_{\bm k}^t.
		\end{align*}
		For every $\bm l\in \lg$, define 
		\begin{align*}
			U_{\bm l}=\left\{\begin{array}{ll}
				\lg_{\frac{1}{2}N_1}(\bm l)\cap \lg, &\text{if}\ \bm l\notin\bigcup_{\bm k\in \tilde{P}_1}\Omega_{\bm k}^1,\\
				\tilde{\Omega}_{\bm k}^t, &\text{if}\ \bm l\in\Omega_{\bm k}^t\ \text{for some}\ \bm k\in\tilde{P}_t\setminus Q_t.
			\end{array}\right.
		\end{align*}
		Define  $\mathcal{Q}_{\bm l}=\ep \mathcal{T}_{U_{\bm l}}^{-1}\mathcal{W}_{\lg\setminus U_{\bm l}}^{U_{\bm l}}\in\mathbf{M}_{\lg\setminus U_{\bm l}}^{U_{\bm l}}$. 
		We now vary $\bm l\in \lg$ and define
		\begin{align*}
			\mathcal{K}(\bm l,\bm l')=\left\{\begin{array}{ll}
				0, & \text{for}\ \bm l'\in U_{\bm l},\\
				\mathcal{Q}_{\bm l}(\bm l,\bm l'), &\text{for}\ \bm l'\in \lg\setminus U_{\bm l},
			\end{array}\right.
		\end{align*}
		and
		\begin{align*}
			\mathcal{L}(\bm l,\bm l')=\left\{\begin{array}{ll}
				\mathcal{T}_{U_{\bm l}}^{-1}(\bm l,\bm l'), & \text{for}\ \bm l'\in U_{\bm l}\\
				0, &\text{for}\ \bm l'\in \lg\setminus U_{\bm l}.
			\end{array}\right.
		\end{align*}
		
		Similar to those of Theorem \ref{1g}, we have for $\alpha\in(0,\alpha_1],$
		\begin{align}
			\label{s+1k0}	\|\mathcal{K}\|_0&\lesssim  \max\{\zeta_1^{\alpha_1}\tilde{\zeta}_1^{-\alpha_1+\alpha_0}\delta_{1}^{-\frac{7}{3}},N_1^{-\alpha_1+\alpha_0}\}\le\frac{1}{2},\\
			\label{s+1ka}	\|\mathcal{K}\|_{\alpha}&\lesssim\zeta_{s+1}^{\alpha}\tilde{\zeta}_{s+1}^{\alpha_0}\delta_{s+1}^{-\frac{7}{3}},
		\end{align}
		and
		\begin{align}
			\nonumber\|\mathcal{L}\|_{0}&\lesssim\tilde{\zeta}_{s+1}^{\alpha_0}\delta_{s+1}^{-\frac{1}{15}}\\
			\label{s+1l0}&\ \ \times\sup\limits_{\{\bm k\in P_{s+1}:\ \tilde{\Omega}_{\bm k}^{s+1}\subset\lg\}}\left(\|\theta+\bm k\cdot\bm\omega-\theta_{s+1}\|_{\T}^{-1}\cdot\|\theta+\bm k\cdot\bm\omega+\theta_{s+1}\|_{\T}^{-1}\right),\\
			\label{s+1la}\|\mathcal{L}\|_{\alpha}&\lesssim\zeta_{s+1}^{\alpha}\tilde{\zeta}_{s+1}^{\alpha_0}\delta_{s+1}^{-\frac{7}{3}}.
		\end{align}
		By recalling Lemma \ref{pa}, we have that $\mathcal{I}_{\lg}+\mathcal{K}$ is invertible and 
		\begin{align}
			\label{s+1I+K}\|(\mathcal{I}_\lg+\mathcal{K})^{-1}\|_{\alpha}&\lesssim \min\{1,\|\mathcal{K}\|_\alpha\}\ \text{for}\ \alpha\in[0,\alpha_1].
		\end{align}
		From \eqref{eq1}, we have
		\begin{align*}
			\mathcal{T}_\lg^{-1}=(\mathcal{I}_{\lg}+\mathcal{K})^{-1}\mathcal{L}.
		\end{align*}
		By recalling \eqref{tame} and \eqref{s+1k0}--\eqref{s+1I+K}, we have
		\begin{align*}
			\|\mathcal{T}_\lg^{-1}\|_0&\le\|(\mathcal{I}_{\lg}+\mathcal{K})^{-1}\|_0\|\mathcal{L}\|_0\\
			&<\delta_{s+1}^{-\frac{2}{15}}\times\sup\limits_{\{\bm k\in P_{s+1}:\ \tilde{\Omega}_{\bm k}^{s+1}\subset\lg\}}\left(\|\theta+\bm k\cdot\bm\omega-\theta_{s+1}\|_{\T}^{-1}\cdot\|\theta+\bm k\cdot\bm\omega+\theta_{s+1}\|_{\T}^{-1}\right),
		\end{align*}
		and for $\alpha\in(0,\alpha_1]$,
		\begin{align*}
			\|\mathcal{T}_\lg^{-1}\|_\alpha&\lesssim \left(\|(\mathcal{I}_\lg+\mathcal{K})^{-1}\|_\alpha\|\mathcal{L}\|_0+\|(\mathcal{I}_\lg+\mathcal{K})^{-1}\|_0\|\mathcal{L}\|_\alpha\right)\\
			&\lesssim  \zeta_{s+1}^{\alpha}\tilde{\zeta}_{s+1}^{2\alpha_0}\delta_{s+1}^{-\frac{22}{5}}<\zeta_{s+1}^{\alpha}\delta_{s+1}^{-\frac{14}{3}},
		\end{align*}
		which completes the proof. 
	\end{proof}
	
	\section{Power-law localization}\label{Loc}
	In this section, we will prove  Theorem \ref{apl} by combining  Green's function estimates and the Shnol's theorem.
	\begin{proof}[Proof of Theorem \ref{apl}]
		%We prove that for $0<|\ep|\le\ep_0$, $\theta\in\Theta_{\tau_1,\g_1}$, $\mathcal{H}(\theta)$ has the only pure point spectrum with polynomially decaying eigenfunctions. 
		Let $\ep_0$ be given by Theorem \ref{ge}. Fix $\theta\in\T\setminus\Theta_{\tau_1}$. Let $E\in\sigma(\mathcal{H}(\theta))$ be a generalized eigenvalue of $\mathcal{H}(\theta)$ and $\psi=\{\psi(\bm n)\}_{\bm n\in\Z^d}\ne0$ be the corresponding generalized eigenfunction satisfying 
		\begin{align*}
			|\psi(\bm n)|\le(1+\|\bm n\|)^d.
		\end{align*}
		From Shnol's theorem  (cf. \cite{Han19}), it suffices to show that $\psi$ decays polynomially and belongs to $\ell^2(\Z^d)$. For this purpose, note first that there exists some $\tilde{s}\in\N$ such that
		\begin{align}\label{4.1}
			\|2\theta+\bm n\cdot\bm\omega\|_{\T}>\frac{1}{\|n\|^{\tau_1}}\ \text{for all $n$ satisfying $\|n\|\ge N_{\tilde{s}}$}.
		\end{align}
		We claim that there exists some  $s_0>0$ such that for $s\ge s_0$,
		\begin{align}\label{4.2}
			\lg_{2N_s^{29}}\cap\left(\bigcup_{\bm k\in Q_s}\tilde{\Omega}_{\bm k}^{s}\right)\ne\emptyset.
		\end{align}
		Otherwise, there exists a  sequence $s_i\rightarrow+\infty$ (as $i\rightarrow\infty$) such that
		\begin{align}\label{4.3}
			\lg_{2N_{s_i}^{29}}\cap\left(\bigcup_{\bm k\in Q_{s_i}}\tilde{\Omega}_{\bm k}^{s_i}\right)=\emptyset.
		\end{align}
		Then we can enlarge $\tilde{\lg}_{N_{s_i}^{29}}$ to $\tilde{\lg}_i$ so that 
		\begin{align*}
			\lg_{N_{s_i}^{29}}\subset\tilde{\lg}_i\subset\lg_{N_{s_i}^{29}+50N_{s_i}^5}
		\end{align*}
		and
		\begin{align*}
			\tilde{\lg}_i\cap\tilde{\Omega}_{\bm k}^{s'}\ne\emptyset\Rightarrow\tilde{\Omega}_{\bm k}^{s'}\subset\tilde{\lg}_i\ \text{for $s'\le s_i$ and $\bm k\in P_{s'}$}.
		\end{align*}
		From \eqref{4.3}, we have
		\begin{align*}
			\tilde{\lg}_i\cap\left(\bigcup_{\bm k\in Q_{s_i}}\tilde{\Omega}_{\bm k}^{s_i}\right)=\emptyset,
		\end{align*}
		which shows that $\tilde{\lg}_i$ is $s_i$-good. Let $\tilde{\lg}_{i,0}=\lg_{\frac{1}{2}N_{s_i}^{29}}\cap\tilde{\lg}_i$. Using Poisson's identity  yields for $\bm n\in\lg_{N_{s_i}}$, 
		\begin{align*}
			|\psi(\bm n)|\le&\sum_{\bm n'\in\tilde{\lg}_i,\bm n''\notin\tilde{\lg}_i}|\mathcal{T}_{\tilde{\lg}_i}^{-1}(\bm n,\bm n')|\cdot|\mathcal{W}(\bm n',\bm n'')|\cdot|\psi(\bm n'')|\\
			\le &(I)+(II),
		\end{align*}
		where
		\begin{align*}
			(I)&=\sum_{\bm n'\in\tilde{\lg}_{i,0},\bm n''\notin\tilde{\lg}_i}|\mathcal{T}_{\tilde{\lg}_i}^{-1}(\bm n,\bm n')|\cdot|\mathcal{W}(\bm n',\bm n'')|\cdot|\psi(\bm n'')|,\\
			(II)&=\sum_{\bm n'\in\tilde{\lg}_i\setminus\tilde{\lg}_{i,0},\bm n''\notin\tilde{\lg}_i}|\mathcal{T}_{\tilde{\lg}_i}^{-1}(\bm n,\bm n')|\cdot|\mathcal{W}(\bm n',\bm n'')|\cdot|\psi(\bm n'')|.
		\end{align*}
		For $(I)$, we have by Theorem \ref{ge}, \eqref{oddgf}, and $|\psi(\bm n)|\le(1+\|\bm n\|)^d$ that 
		\begin{align*}
			(I)&\lesssim(N_{s_i}^{29})^{-\alpha_1+2d}\|\mathcal{T}_{\tilde{\lg}_i}\|_{d}\|\mathcal{W}\|_{\alpha_1}\\
			&<\delta_{s_i}^{\frac{\alpha_1}{36}}\rightarrow0\ \text{as $i\rightarrow\infty$}.
		\end{align*}
		For $(II)$, we also have by Theorem \ref{ge}, \eqref{oddgf} and $|\psi(\bm n)|\le(1+\|\bm n\|)^d$ that 
		\begin{align*}
			(II)&\lesssim(N_{s_i}^{29})^{-\alpha_1+2d}\|\mathcal{T}_{\tilde{\lg}_i}^{-1}\|_{\alpha_1}\|\mathcal{W}\|_{d}\\
			&<\delta_{s_i}^{\frac{\alpha_1}{36}}\rightarrow0\ \text{as $i\rightarrow\infty$}.
		\end{align*}
		It follows that $\psi(\bm n)=0\ {\rm for}\ \forall \bm n\in\Z^d$, which  contradicts $\psi\ne0$.  And the Claim is proved.
		
		Next define
		\begin{align*}
			U_s=\lg_{8N_{s+1}^{29}}\setminus\lg_{4N_s^{29}},\ U_s^{*}=\lg_{10N_{s+1}^{29}}\setminus\lg_{3N_s^{29}}.
		\end{align*}
		We can also enlarge $U_s^{*}$ to $\tilde{U}_s^*$ so that
		\begin{align*}
			U_s^*\subset \tilde{U}_s^*\subset\lg_{50N_s^5}(U_s^*)
		\end{align*}
		and
		\begin{align*}
			\tilde{U}_s^*\cap\tilde{\Omega}_{\bm k}^{s'}\ne\emptyset\Rightarrow\tilde{\Omega}_{\bm k}^{s'}\subset\tilde{U}_s^*\ \text{for $s'\le s$ and $\bm k\in P_{s'}$}.
		\end{align*}
		Let $\bm n$ satisfy $\|\bm n\|>\max(4N_{\tilde{s}}^{29},4N_{s_0}^{29})$. Then there exists some $s\ge\max(\tilde{s},s_0)$ such that
		\begin{align}\label{4.4}
			\bm n\in U_s\subset \tilde{U}_s^*.
		\end{align}
		Without loss of generality and since \eqref{4.2}, we can assume
		\begin{align*}
			\lg_{2N_s^{29}}\cap\tilde{\Omega}_{\bm k}^s\ne\emptyset
		\end{align*}
		for some $\bm k\in Q_s^+$. Then for $\bm k\ne\bm k'\in Q_s^+$, we have
		\begin{align*}
			\|\bm k-\bm k'\|>\left(\frac{\g}{2\delta_s}\right)^{\frac{1}{\tau}}\gtrsim N_{s+1}^{30}\gg\diam (\tilde{U}_s^*).
		\end{align*}
		Therefore,
		\begin{align*}
			\tilde{U}_s^*\cap\left(\bigcup_{\bm l\in Q_s^+}\tilde{\Omega}_{\bm l}^s\right)=\emptyset.
		\end{align*}
		Now, if there exists some $\bm l\in Q_s^-$such that
		\begin{align*}
			\tilde{U}_s^*\cap\tilde{\Omega}_{\bm l}^s\ne\emptyset,
		\end{align*}
		then
		\begin{align*}
			N_{s}\le N_s^{29}-100N_s^5\le\|\bm l\|-\|\bm k\|\le\|\bm l+\bm k\|\le\|\bm l\|+\|\bm k\|<11N_{s+1}^{29}.
		\end{align*}
		Recalling
		\begin{align*}
			Q_s\subset P_s\subset \Z^d+\frac{1}{2}\sum_{i=0}^{s-1}\bm l_i,
		\end{align*}
		we have $\bm l+\bm k\in\Z^d$. According to \eqref{4.1}, we obtain 
		\begin{align*}
			\frac{1}{(11N_{s+1}^{29})^{\tau_1}}<\|2\theta+(\bm l+\bm k)\cdot\bm\omega\|_{\T}\le\|\theta+\bm l\cdot\bm\omega-\theta_s\|_{\T}+\|\theta+\bm k\cdot\bm\omega+\theta_s\|_{\T}<2\delta_s,
		\end{align*}
		which contradicts
		\begin{align*}
			\delta_s^{-1}\gtrsim N_{s+1}^{30\tau}\gg N_{s+1}^{29\tau_1}.
		\end{align*}
		We thus have shown
		\begin{align*}
			\tilde{U}_s^*\cap\left(\bigcup_{\bm l\in Q_s}\tilde{\Omega}_{\bm l}^s\right)=\emptyset.
		\end{align*}
		This implies that $\tilde{U}_s^*$ is $s$-$\good$. 
		
		Finally, by recalling \eqref{4.4}, we  can set
		\begin{align*}
			\hat{U}_s=\lg_{\frac{1}{2}N_s^{29}}(U_s).
		\end{align*}
		Then
		\begin{align*}
			|\psi(\bm n)|\le&\sum_{\bm n'\in\tilde{U}_s^*,\bm n''\notin\tilde{U}_s^*}|\mathcal{T}_{\tilde{U}_s^*}^{-1}(\bm n,\bm n')|\cdot|\mathcal{W}(\bm n',\bm n'')|\cdot|\psi(\bm n'')|\\
			\le &(III)+(IV),
		\end{align*}
		where
		\begin{align*}
			(III)&=\sum_{\bm n'\in\hat{U}_s,\bm n''\notin\tilde{U}_s^*}|\mathcal{T}_{\tilde{U}_s^*}^{-1}(\bm n,\bm n')|\cdot|\mathcal{W}(\bm n',\bm n'')|\cdot|\psi(\bm n'')|\\
			(IV)&=\sum_{\bm n'\in\tilde{U}_s^*\setminus\hat{U}_s,\bm n''\notin\tilde{U}_s^*}|\mathcal{T}_{\tilde{U}_s^*}^{-1}(\bm n,\bm n')|\cdot|\mathcal{W}(\bm n',\bm n'')|\cdot|\psi(\bm n'')|.
		\end{align*}
		For $(III)$, we have by Theorem \ref{ge}, \eqref{oddgf} and $|\psi(\bm n)|\le(1+\|\bm n\|)^d$ that 
		\begin{align*}
			(III)&\lesssim(N_{s}^{29})^{-\alpha_1+d}\|\mathcal T_{\tilde{\lg}_i}\|_{d}\|\mathcal W\|_{\alpha_1}(1+\|\bm n\|)^d\\
			&\lesssim (N_{s+1})^{\frac{29}{30}(-\alpha_1+2d)+140\tau}(1+\|\bm n\|)^d.
		\end{align*}
		For $(IV)$, we also have by Theorem \ref{ge}, \eqref{oddgf} and $|\psi(\bm n)|\le(1+\|\bm n\|)^d$ that 
		\begin{align*}
			(IV)&\lesssim(N_{s}^{29})^{-\alpha_1+d}\|\mathcal T_{\tilde{\lg}_i}^{-1}\|_{\alpha_1}\|\mathcal W\|_{d}(1+\|\bm n\|)^d\\
			&\lesssim (N_{s+1})^{-\frac{13}{15}\alpha_1+\frac{19}{15}d+140\tau}(1+\|\bm n\|)^d.
		\end{align*}
		Combining the above estimates and since $\alpha_1>2200\tau$, $\|\bm n\|\le 8N_{s+1}^{29}$, we have
		\begin{align*}
			|\psi(\bm n)|<(1+\|\bm n\|)^{-\frac{\alpha_1}{60}}.
		\end{align*}
		
		We complete the proof of arithmetic power-law localization. 
	\end{proof}
	
	\section{Dynamical localization}
	In this section, we  prove Theorem \ref{adl}  concerning dynamical localization.
	\begin{proof}[Proof of Theorem \ref{adl}]
		%Let $\ep_0$ be  given by Theorem \ref{ge} holds true. 
		Since power-law localization holds true for $\theta\in\Theta_{\tau_1,A}^*$ by Theorem \ref{apl}, let $\{\varphi_{\bm q}(\theta),E_{\bm q}(\theta)\}_{\bm q\in\Z^d}$ denote a complete set of eigenfunctions and corresponding eigenvalues of $\mathcal{H}(\theta)$. For simplicity, we omit all dependences   on $\theta$. Then
		\begin{align*}
			\delta_{\bm 0}=\sum_{\bm q}\varphi_{\bm q}(\bm 0)\varphi_{\bm q}
		\end{align*}
		and hence
		\begin{align*}
			e^{\sqrt{-1}t\mathcal{H}}\delta_{\bm 0}=\sum_{\bm q}e^{\sqrt{-1}t E_{\bm q}}\varphi_{\bm q}(\bm 0)\varphi_{\bm q}.
		\end{align*}
		Thus, it suffices  to estimate 
		\begin{align}\label{dyn}
			\sum_{\bm q}\left(\sum_{\bm x}(1+\|\bm x\|)^{p}|\varphi_{\bm q}(\bm x)|\right)|\varphi_{\bm q}(\bm 0)|.
		\end{align}
		Let $I_0=\emptyset$ and $I_j=\{\bm q:\ |\varphi_{\bm q}(\bm 0)|>N_j^{-15\alpha_1}\}\ (j\ge1)$. Then
		\begin{align}\label{dyn2}
			\eqref{dyn}=\sum_{j=1}^{\infty}\sum_{\bm q\in I_j\setminus I_{j-1}}\left(\sum_{\bm x}(1+\|\bm x\|)^{p}|\varphi_{\bm q}(\bm x)|\right)|\varphi_{\bm q}(\bm 0)|.
		\end{align}
		We claim that for $\bm q\in I_j$ and $s\ge j$,
		\begin{align}\label{dynnon}
			\lg_{2N_s^{29}}\cap\left(\bigcup_{\bm k\in Q_s}\tilde{\Omega}_{\bm k}^{s}\right)\ne\emptyset.
		\end{align}
		Otherwise, there exists some $s$-$\good$ set $\lg$ such that
		\begin{align*}
			\lg_{N_s^{29}}\subset\lg\subset\lg_{N_s^{29}+50N_s^{5}}.
		\end{align*}
		Then since Theorem \ref{ge} and \eqref{oddgf},  we get a contradiction with 
		\begin{align*}
			|\varphi_{\bm q}(\bm 0)|&\le\sum_{\bm n'\in\lg,\bm n''\notin\lg}|\mathcal{T}_{\lg}^{-1}(\bm 0,\bm n')|\cdot|\mathcal{W}(\bm n',\bm n'')|\cdot|\varphi_{\bm q}(\bm n'')|\\
			&<N_s^{-15\alpha_1}\le N_j^{-15\alpha_1}.
		\end{align*}
		
		Now assume
		\begin{align}\label{A}
			\delta_{m}^{\frac{1}{30}}<A\le \delta_{m-1}^{\frac{1}{30}}\ (\delta_{-1}:=+\infty).
		\end{align}
		If $\bm q\in I_j$, then by \eqref{TA} and similar to the proof of Theorem \ref{apl},  we can prove that for $s\ge\max(m,j)$, there is no bad  (enlarged resonant) block of the $s$-th induction step  inside $\lg_{10N_{s+1}^{29}}\setminus\lg_{3N_s^{29}}$, which proves $|\varphi_{\bm q}(\bm x)|<(1+\|\bm x\|)^{-\frac{\alpha_1}{60}}$ for $\|\bm x\|\ge\max(4N_{m}^{29},4N_{j}^{29})$. From the Hilbert-Schmidt argument, we have
		\begin{align*}
			C(d)N_{\max(m,j)}^{29d}&\ge\sum_{\|x\|\le 4N_{\max(m,j)}^{29}}\sum_{\bm q}|\varphi_{\bm q}(\bm x)|^2\\
			&\ge\sum_{\bm q\in I_j}\sum_{\|\bm x\|\le 4N_{\max(m,j)}^{29}}|\varphi_{\bm q}(\bm x)|^2\\
			&=\# I_j-\sum_{\bm q\in I_j}\sum_{\|\bm x\|> 4N_{\max(m,j)}^{29}}|\varphi_{\bm q}(\bm x)|^2\\
			&\ge\frac{1}{2}\# I_j.
		\end{align*}
		Thus $\# I_j\le C(d)N_{\max(m,j)}^{29d}$.
		
		To estimate \eqref{dyn2}, using $|\varphi_{\bm q}(\bm x)|<(1+\|\bm x\|)^{-\frac{\alpha_1}{60}}$ for $\bm q\in I_m$ and $\|\bm x\|\ge 4N_m^{29}$ implies 
		\begin{align}
			\nonumber&\sum_{j=1}^{m}\sum_{\bm q\in I_j\setminus I_{j-1}}\left(\sum_{\bm x}(1+\|\bm x\|)^{p}|\varphi_{\bm q}(\bm x)|\right)|\varphi_{\bm q}(\bm 0)|\\
			\nonumber\le&\sum_{\bm q\in I_m}\left(\sum_{\bm x}(1+\|\bm x\|)^{p}|\varphi_{\bm q}(\bm x)|\right)\\
			\nonumber\le&\# I_m \cdot\sup_{\bm q\in I_m}\left(\sum_{\|\bm x\|\le 4N_m^{29}}+\sum_{\|\bm x\|>4N_m^{29}}\right)(1+\|\bm x\|)^{p}|\varphi_{\bm q}(\bm x)|\\
			\label{dyns}\le& C(\alpha_1,p,d)N_m^{29(p+2d)}.
		\end{align}
		Using $|\varphi_{\bm q}(\bm x)|<N_j^{-15\alpha_1}\lesssim(1+\|\bm x\|)^{-\frac{\alpha_1}{60}}$ for $j\ge m$, $\bm q\in I_j$ and $\|\bm x\|\ge4 N_j^{29}$ yields 
		\begin{align*}
			&\sum_{\bm q\in I_j\setminus I_{j-1}}\left(\sum_{\bm x}(1+\|\bm x\|)^{p}|\varphi_{\bm q}(\bm x)|\right)|\varphi_{\bm q}(\bm 0)|\\
			\le&\# I_j\cdot\sup_{q\in I_j}\left(\sum_{\|\bm x\|\le 4N_j^{29}}+\sum_{\|\bm x\|>4N_j^{29}}\right)(1+\|\bm x\|)^{p}\cdot|\varphi_{\bm q}(\bm x)|\cdot N_{j-1}^{-15\alpha_1}\\
			\le & C(\alpha_1,p,d)(N_j^{29})^{p+2d}N_{j-1}^{-15\alpha_1}\le C(\alpha_1,p,d)N_j^{-\frac{\alpha_1}{2}+30p+60d},
		\end{align*}
		where $N_0=1$. Summing up $j$ for $j\ge m$ gives
		\begin{align}
			\nonumber&\sum_{j=m}^{\infty}\sum_{\bm q\in I_j\setminus I_{j-1}}\left(\sum_{\bm x}(1+\|\bm x\|)^{p}|\varphi_{\bm q}(\bm x)|\right)|\varphi_{\bm q}(\bm 0)|\\
			\label{dynb}\le&\left\{\begin{array}{ll}
				C(\alpha_1,p,d)N_m^{-\frac{\alpha_1}{2}+30p+60d} & \text{if $m\ge1$},\\
				C(\alpha_1,p,d)N_1^{29(p+2d)} & \text{if $m=0$}.
			\end{array}\right.
		\end{align}
		From \eqref{dyns} and \eqref{dynb}, we obtain
		\begin{align*}
			\eqref{dyn2}&\le C(\alpha_1,p,d)\max(N_m^{29(p+2d)},N_1^{29(p+2d)})\\
			&\le C(\alpha_1,p,d)\max(A^{-\frac{29(p+2d)}{\tau}},\ep_0^{-\frac{29(p+2d)}{\tau}}),
		\end{align*}
		where we have used  \eqref{A}. 
		
		Hence we finish the proof of dynamical localization.
	\end{proof}

	\section{$\left(\frac{1}{2}-\right)$ H\"older contuinity of the IDS} 
	
	In this section, we prove the finite volume version of the  $\left(\frac{1}{2}-\right)$-H\"older continuity  of the  IDS.
	\begin{proof}[Proof of Thoerem \ref{tids}]
		Let $\mathcal{T}$ be given by \eqref{T}. Fix $\theta\in\T$, $E\in\R $  and $\mu>0$ as in Theorem \ref{tids}. %\in\left[\frac{1}{\alpha_1},\frac{1}{16}\right)$, . 
		Let $\ep_0$ be such  that  Theorem \ref{ge} holds true for  $0<|\ep|< \ep_0$. Let 
		\begin{align}\label{et}
			0<\et<\et_0=\delta_1^{\frac{17}{135\mu}}.
		\end{align}
		
		Denote by $\{\xi_r:\ r=1,\cdots,R\}\subset {\rm span}_{\R}\{\delta_{\bm n}:\ \bm n\in\lg_N\}$ the $\ell^2$-orthonormal eigenvectors of $\mathcal T_{\lg_N}$ with  eigenvalues belonging to $[-\et,\et]$. We aim to prove that for sufficiently large $N$ (depending on $\et$),
		\begin{align*}
			R\le (\# \lg_N)\et^{\frac{1}{2}-\mu}.
		\end{align*}
		From \eqref{et}, we can choose $s\ge1$ such that
		\begin{align}\label{ets}
			\delta_{s+1}^{\frac{17}{135\mu}}\le \et<\delta_{s}^{\frac{17}{135\mu}}.
		\end{align}
		Enlarge $\lg_N$ to $\tilde{\lg}_N$ so that
		\begin{align*}
			\lg_N\subset \tilde{\lg}_N\subset\lg_{N+50N_s^{5}}
		\end{align*}
		and
		\begin{align*}
			\tilde{\lg}_N\cap\tilde{\Omega}_{\bm k}^{s'}\ne\emptyset\Rightarrow\tilde{\Omega}_{\bm k}^{s'}\subset\tilde{\lg}_N\ \text{for}\ s'\le s\ \text{and}\ \bm k\in P_{s'}.
		\end{align*}
		Define further
		\begin{align*}
			\mathscr{K}=\left\{\bm k\in P_s\cap\tilde{\lg}_N:\  \min_{\sigma=\pm1}(\|\theta+\bm k\cdot\bm\omega+\sigma\theta_s\|_{\T})<\et^{\frac{1}{2}-\frac{\mu}{2}}\right\}
		\end{align*}
		and
		\begin{align*}
			\tilde{\lg}_N^{'}=\tilde{\lg}_N\setminus\bigcup_{\bm k\in P_s\cap\tilde{\lg}_N}\Omega_{\bm k}^s,\ \tilde{\lg}_N^{''}=\tilde{\lg}_N\setminus\bigcup_{\bm k\in\mathscr{K}}\Omega_{\bm k}^s.
		\end{align*}
		Thus $\tilde{\lg}_N^{'}\subset\tilde{\lg}_N^{''}$ and
		\begin{align*}
			\tilde{\lg}_N^{''}\setminus\tilde{\lg}_N^{'}\subset\bigcup_{\bm k\in(P_s\cap\tilde{\lg}_N)\setminus\mathscr{K}}\Omega_{\bm k}^s\subset\bigcup_{\bm k\in(P_s\cap\tilde{\lg}_N)\setminus\mathscr{K}}\tilde{\Omega}_{\bm k}^s\subset\tilde{\lg}_N^{''}.
		\end{align*}
		Define
		\begin{align*}
			\tilde{P}_t=\{\bm k\in P_t\cap\tilde{\lg}_N:\ \exists \bm k'\in Q_{t-1}\ \text{s.t.}\ \tilde{\Omega}_{\bm k'}^{t-1}\subset \tilde{\lg}_N^{'},\tilde{\Omega}_{\bm k'}^{t-1}\subset\Omega_{\bm k}^t\},\ t\in[1,s].
		\end{align*}
		Similar to the proof of Lemma \ref{psqs}, we have
		\begin{align*}
			\tilde{P}_{s}=\emptyset, 
		\end{align*}
		and for any $\bm l\in \tilde{\lg}_N^{'}$, if 
		\begin{align*}
			\bm l\in\bigcup_{\bm j\in\tilde{P}_1}\Omega_{\bm k}^1,
		\end{align*}
		there exists some  $t\in[1,s-1]$ such that
		\begin{align*}
			\bm l\in\bigcup_{\bm k\in\tilde{P}_t\setminus Q_t}\Omega_{\bm k}^t.
		\end{align*}
		For every $\bm l\in \tilde{\lg}_{N}^{''}$, define %its block in $\tilde{\lg}_{N}^{''}$:
		\begin{align*}
			U_{\bm l}=\left\{\begin{array}{ll}
				\lg_{\frac{1}{2}N_1}(\bm l)\cap \tilde{\lg}_{N}^{''}, &\text{if}\ \bm l\notin\bigcup_{\bm k\in \left(\tilde{P}_1\cup(P_s\cap\tilde{\lg}_N)\right)\setminus\mathscr{K}}\Omega_{\bm k}^1,\\
				\tilde{\Omega}_{\bm k}^t, &\text{if}\ \bm l\in\Omega_{\bm k}^t\ \text{for some}\ \bm k\in\tilde{P}_t\setminus Q_t,\\
				\tilde{\Omega}_{\bm k}^s, &\text{if}\ \bm l\in\Omega_{\bm k}^s\ \text{for some}\ \bm k\in(P_s\cap\tilde{\lg}_N)\setminus\mathscr{K}.
			\end{array}\right.
		\end{align*}
		Define  $\mathcal{Q}_{\bm l}=\ep \mathcal{T}_{U_{\bm l}}^{-1}\mathcal{W}_{\tilde{\lg}_{N}^{''}\setminus U_{\bm l}}^{U_{\bm l}}\in\mathbf{M}_{\tilde{\lg}_{N}^{''}\setminus U_{\bm l}}^{U_{\bm l}}$. 
		Varying  $\bm l\in \tilde{\lg}_{N}^{''}$ leads to 
		\begin{align*}
			\mathcal{K}(\bm l,\bm l')=\left\{\begin{array}{ll}
				0, & \text{for}\ \bm l'\in U_{\bm l}\\
				\mathcal{Q}_{\bm l}(\bm l,\bm l'), &\text{for}\ \bm l'\in \tilde{\lg}_{N}^{''}\setminus U_{\bm l},
			\end{array}\right.
		\end{align*}
		and
		\begin{align*}
			\mathcal{L}(\bm l,\bm l')=\left\{\begin{array}{ll}
				\mathcal T_{U_{\bm l}}^{-1}(\bm l,\bm l'), & \text{for}\ \bm l'\in U_{\bm l}\\
				0, &\text{for}\ \bm l'\in \tilde{\lg}_{N}^{''}\setminus U_{\bm l}.
			\end{array}\right.
		\end{align*}
		We estimate $\mathcal{K}\in\mathbf{M}_{\tilde{\lg}_{N}^{''}}^{\tilde{\lg}_{N}^{''}}$ and $\mathcal{L}\in\mathbf{M}_{\tilde{\lg}_{N}^{''}}^{\tilde{\lg}_{N}^{''}}$. Similar to the proof of Lemma \ref{psqs}, we can obtain\\
		(1) If $\bm l\notin\bigcup_{\bm k\in\left(\tilde{P}_1\cup(P_s\cap\tilde{\lg}_N)\right)\setminus\mathscr{K}}\Omega_{\bm k}^1$, then
		\begin{align*}
			&\|\mathcal{K}^{\{\bm l\}}\|_{\alpha_0}\lesssim N_1^{-\alpha_1+\alpha_0},\ \|\mathcal{L}^{\{\bm l\}}\|_{\alpha_0}\lesssim\delta_0^{-2}.
		\end{align*}
		\ \\
		(2) If $\bm l\in\Omega_{\bm k}^t$ for some $\bm k\in\tilde{P}_t\setminus Q_t$ and $t\in[1,s-1]$, then 
		\begin{align*}
			&\|\mathcal{K}^{\{\bm l\}}\|_{\alpha_0}\lesssim \zeta_t^{\alpha_1}\tilde{\zeta}_t^{-\alpha_1+\alpha_0}\delta_{t}^{-\frac{7}{3}},\ \|\mathcal{L}^{\{\bm l\}}\|_{\alpha_0}\lesssim\tilde{\zeta}_t^{\alpha_0}\delta_t^{-\frac{31}{15}}.
		\end{align*}
		\ \\
		(3) If $\bm l\in\Omega_{\bm k}^s$ for some $\bm k\in(P_s\cap\tilde{\lg}_N)\setminus\mathscr{K}$, then by \eqref{tame} and Remark \ref{remt}, we get  
		\begin{align}\label{idsql}
			\|\mathcal{Q}_{\bm l}\|_{\alpha_1}&\lesssim\|\mathcal T_{U_{\bm l}}^{-1}\|_{\alpha_1}\|\mathcal W_\lg\|_{\alpha_1}\lesssim \zeta_s^{\alpha_1}\delta_{s}^{-\frac{1}{3}}\et^{\mu-1}.
		\end{align}
		Note that if $\bm l'\in \lg\setminus U_{\bm l}$, then $\|\bm l-\bm l'\|\ge\frac{\tilde{\zeta}_s}{2}$. This implies $\mathcal{K}^{\{\bm l\}}(\bm l,\bm l')=0$ for $\|\bm l-\bm l'\|<\frac{\tilde{\zeta}_s}{2}$. By \eqref{smo1} and \eqref{idsql}, we obtain
		\begin{align*}
			\|\mathcal{K}^{\{\bm l\}}\|_{\alpha_0}&\lesssim \tilde{\zeta}_s^{-\alpha_1+\alpha_0}\|\mathcal{K}^{\{\bm l\}}\|_{\alpha_1}\le \tilde{\zeta}_s^{-\alpha_1+\alpha_0}\|\mathcal{Q}_{\bm l}\|_{\alpha_1}\lesssim \zeta_s^{\alpha_1}\tilde{\zeta}_s^{-\alpha_1+\alpha_0}\delta_{s}^{-\frac{1}{3}}\et^{\mu-1}.
		\end{align*}
		By the definition of $U_{\bm l}$, if $\|\bm l-\bm l'\|>2\tilde{\zeta}_s$, then $l'\notin U_{\bm l}$. This implies $\mathcal{L}^{\{\bm l\}}(\bm l,\bm l')=0$ for $\|\bm l-\bm l'\|>2\tilde{\zeta}_s$. By \eqref{smo2} and \eqref{tb0}, we have
		\begin{align*}
			\|\mathcal{L}^{\{\bm l\}}\|_{\alpha_0}&\lesssim\tilde{\zeta}_s^{\alpha_0}\|\mathcal{L}^{\{\bm l\}}\|_0\le \tilde{\zeta}_s^{\alpha_0}\|\mathcal{T}_{U_{\bm l}}^{-1}\|_0\\
			&\lesssim\tilde{\zeta}_s^{\alpha_0}\delta_s^{-\frac{1}{15}}\|\theta+\bm k\cdot\bm\omega-\theta_s\|_{\T}^{-1}\cdot\|\theta+\bm k\cdot\bm\omega+\theta_s\|_{\T}^{-1}\\
			&\lesssim \tilde{\zeta}_s^{\alpha_0}\delta_s^{-\frac{1}{15}}\et^{\mu-1}.
		\end{align*}
		By \eqref{re}, \eqref{ets} and the definition of $\mu$, we can get
		\begin{align*}
			\|\mathcal{K}\|_0\lesssim &\sup_{\bm l\in \tilde{\lg}_{N}^{''}}\|\mathcal{K}^{\{\bm l\}}\|_{\alpha_0}\lesssim \max\{\zeta_s^{\alpha_1}\tilde{\zeta}_s^{-\alpha_1+\alpha_0}\delta_{s}^{-\frac{1}{3}}\et^{\mu-1},\zeta_1^{\alpha_1}\tilde{\zeta}_1^{-\alpha_1+\alpha_0}\delta_{1}^{-\frac{7}{3}},N_1^{-\alpha_1+\alpha_0}\}\le\frac{1}{2},
		\end{align*}
		and 
		\begin{align*}
			\|\mathcal{L}\|_{0}&\lesssim \sup_{\bm l\in \tilde{\lg}_{N}^{''}}\|\mathcal{L}^{\{\bm l\}}\|_{\alpha_0}\lesssim \tilde{\zeta}_s^{\alpha_0}\delta_{s}^{-\frac{1}{15}}\et^{\mu-1}.
		\end{align*}
		Recalling Lemma \ref{pa}, we have that $\mathcal{I}_{\tilde{\lg}_{N}^{''}}+\mathcal{K}$ is invertible and  
		\begin{align*}
			\|(\mathcal{I}_{\tilde{\lg}_{N}^{''}}+\mathcal{K})^{-1}\|_{0}&\le 2.
		\end{align*}
		From \eqref{eq1}, we have
		\begin{align*}
			\mathcal{T}_{\tilde{\lg}_{N}^{''}}^{-1}=(\mathcal{I}_{\tilde{\lg}_{N}^{''}}+\mathcal{K})^{-1}\mathcal{L}.
		\end{align*}
		Recalling \eqref{tame}, we have
		\begin{align}
			\nonumber\|\mathcal{T}_{\tilde{\lg}_{N}^{''}}^{-1}\|_0&\le\|(\mathcal{I}_{\tilde{\lg}_{N}^{''}}+\mathcal{K})^{-1}\|_0\|\mathcal{L}\|_0\\
			\nonumber&\lesssim\tilde{\zeta}_s^{\alpha_0}\delta_{s}^{-\frac{1}{15}}\et^{\mu-1}\\
			\label{et-1}&<\frac{1}{2}\delta_{s}^{-\frac{13}{180}}\et^{\mu-1}<\frac{1}{2}\et^{-1}, 
		\end{align}
		where the last inequality follows from \eqref{et} and \eqref{ets}.
		
		Finally,  similar to  the proof  in \cite{CSZ24a}, we obtain 
		\begin{align*}
			R\lesssim N_s^{3d}\cdot\et^{\frac{1}{2}-\frac{\mu}{2}}\#\lg_N\le\et^{\frac{1}{2}-\mu}\#\lg_N.
		\end{align*}
		We finish the proof of Theorem \ref{tids}.
	\end{proof}
	
	\section{Absence of eigenvaalues} \label{Abs}
	To accomplish the proof of Theorem \ref{aps},  we first need 
	\begin{lem}\label{0e}
		Under the assumptions of Theorem \ref{ge},   for a.e. $\theta\in\T$, there is an integer $s_0(\theta)>0$ such that $Q_s\cap\lg_{2N_s^{29}}=\emptyset$ for $s\ge s_0(\theta)$.
	\end{lem}
	\begin{proof}
		Define
		\begin{align*}
			B^s=\{\theta\in\T:\ Q_s\cap\lg_{2N_s^{29}}\ne\emptyset\}. 
		\end{align*}
		and  $\mathscr{K}=\cap_{i=0}^{\infty}\cup_{j\ge i}B^i$. It suffices to show that $\mathscr K$ has zero Lebesgue measure.  %will be the set of measure zero we need to prove the lemma. To prove that this set has measure zero we define
		For this purpose, write 
		\begin{align*}
			B_{\bm k}^s=\{\theta\in\T: \ \bm k\in Q_s\cap\lg_{2N_s^{29}}\}\subset\{\theta\in\T:\ \bm k\in\lg_{2N_s^{29}},\min_{\sigma=\pm 1}\|\theta+\bm k\cdot\bm\omega+\sigma\theta_s\|_{\T}<\delta_s\}. 
		\end{align*}
		Then $B^s=\cup_{\bm k} B_{\bm k}^s$. By the Borel-Cantelli theorem,  it remains to prove  $\sum\limits_{s\geq 0} m(B^s)<\infty$,   where $m(\cdot)$ is the Lebesgue measure.  Indeed, since
		\begin{align*}
			m(B_{\bm k}^s)\le m(\{\theta\in\T:\ \bm k\in\lg_{2N_s^{29}},\min_{\sigma=\pm 1}\|\theta+\bm k\cdot\bm\omega+\sigma\theta_s\|_{\T}<\delta_s\})<4\delta_s,
		\end{align*}
		we  obtain
		\begin{align*}
			\sum_{s} m(B^s)\le \sum_{s}(\sum_{\bm k}m(B_{\bm k}^s))\le C(d)\sum_{s} (N_s^{29d}\delta_s)<+\infty. 
		\end{align*}
		
		This proves  Lemma \ref{0e}.
	\end{proof}
	
	Now we can prove Theorem \ref{aps}. 
	\begin{proof}[Proof of Theorem \ref{aps}]
		%Let $0<|\ep|<\ep_0$ be as in Theorem \ref{ge}. 
		Suppose $\tilde{\mathcal{H}}$ has  some eigenvalue $E$. Then there must be $\psi=\{\psi_l\}_{l\in\Z}\in\ell^2(\Z),\ \psi\neq 0$ so that
		\begin{align*}
			\sum_{l'\in\Z}\hat{v}( l- l')\psi_{l'}+(\ep u(\bm x+l\bm \omega)-E)\psi_l=0.
		\end{align*}
		Define
		\begin{align*}
			F(\theta)=\sum_{l\in\Z}\psi_l e^{2\pi\sqrt{-1}l\theta}
		\end{align*}
		and
		\begin{align*}
			\xi_{\bm n}(\theta)=e^{2\pi\sqrt{-1}\bm n\cdot \bm x}F(\theta+\bm n\cdot\bm\omega).
		\end{align*}
		We have
		\begin{align}\label{pai}
			\|F\|_{L^2(\T)}=\|\psi\|_{\ell^2(\Z)}>0
		\end{align}
		and %by direction  computation
		\begin{align}\label{eq6}
			(v(\theta)-E)F(\theta)+\ep\sum_{\bm k\in\Z^d}\phi(\bm k)\xi_{\bm k}(\theta)=0.
		\end{align}
		Then 
		\begin{align*}
			\int_{\T}\sum_{\bm n\in\Z^d}\frac{|\xi_{\bm n}(\theta)|^2}{(1+\|\bm n\|)^{2d}}d\theta=\sum_{\bm n\in\Z^d}\frac{\|F\|_{L^2(\T)}^2}{(1+\|\bm n\|)^{2d}}\le C(d)\|F\|_{L^2(\T)}^2<+\infty.
		\end{align*}
		This implies that there is a set $\mathscr{L}$ with $m(\mathscr{L})=0$ such that for $\theta\in\T\setminus\mathscr{L}$, we have $\sum_{\bm n\in\Z^d}\frac{|\xi_{\bm n}(\theta)|^2}{(1+\|\bm n\|)^{2d}}<+\infty$ and 
		\begin{align}\label{xizz}
			|\xi_{\bm n}(\theta)|\le C(\theta,d)(1+\|\bm n\|)^d,\ C(\theta,d)>0.
		\end{align}
		
		Now we let $\theta=\theta+\bm n\cdot\bm\omega$ in \eqref{eq6}. Then
		\begin{align*}
			(v(\theta+\bm n\cdot\bm\omega)-E)F(\theta+\bm n\cdot\bm\omega)+\ep\sum_{\bm k\in\Z^d}\phi(\bm k)e^{2\pi\sqrt{-1}\bm k\cdot \bm x}F(\theta+(\bm n+\bm k)\cdot\bm\omega)=0.
		\end{align*}
		Multiplying by $e^{2\pi\sqrt{-1}\bm n\cdot\bm x}$ in the above equality implies
		\begin{align*}
			(v(\theta+\bm n\cdot\bm\omega)-E)\xi_{\bm n}(\theta)+\ep\sum_{\bm k\in\Z^d}\phi(\bm n-\bm k)\xi_{\bm k}(\theta)=0.
		\end{align*}
		Moreover, $\mathcal H(\theta)\xi(\theta)=E\xi(\theta)$ with $\xi(\theta)=\{\xi_{\bm n}(\theta)\}_{\bm n\in\Z^d}$. Hence, we have by  using the  Poisson's identity for $\bm n\in\lg\subset\Z^d$,
		\begin{align}\label{poi}
			\xi_{\bm n}(\theta)=-\ep\sum_{\bm n'\in\lg,\bm n''\notin\lg}\mathcal{T}_{\lg}^{-1}(E;\theta)(\bm n,\bm n')\cdot\phi(\bm n'-\bm n'')\cdot\xi_{\bm n''}(\theta).	
		\end{align}
		By \eqref{tl}, we can enlarge ${\lg}_{N_{s}^{29}}$ to $\tilde{\lg}_s$ satisfying
		\begin{align*}
			\lg_{N_{s}^{29}}\subset\tilde{\lg}_s\subset\lg_{N_{s}^{29}+50N_{s}^{5}}
		\end{align*}
		and
		\begin{align*}
			\tilde{\lg}_s\cap\tilde{\Omega}_{\bm k}^{s'}\ne\emptyset\Rightarrow\tilde{\Omega}_{\bm k}^{s'}\subset\tilde{\lg}_s\ \text{for $s'\le s$ and $\bm k\in P_{s'}$}.
		\end{align*}
		From Lemma \ref{0e}, we can get a set $\mathscr{K}$ with $m(\mathscr{K})=0$ such that for $\theta\in\T\setminus\mathscr{K}$, there exists an integer $s(\theta)$ such that $Q_s\cap\lg_{2N_s^{29}}=\emptyset$ for $s\ge s_0(\theta)$. For $s\ge s_0(\theta)$, we have
		\begin{align*}
			\tilde{\lg}_s \cap Q_s\subset \lg_{2N_s^{29}}\cap Q_s=\emptyset,
		\end{align*}
		which implies $\tilde{\lg}_s$ is $s$-$\good$. Let $\tilde{\lg}_{s,0}=\lg_{\frac{1}{2}N_s^{29}}\cap\tilde{\lg}_s$. Recalling \eqref{poi}, one has for $\theta\in\T\setminus(\mathscr{K}\cup\mathscr{L})$, $s\ge s_0(\theta)$,
		\begin{align*}
			|F(\theta)|=|\xi_{\bm 0}(\theta)|&\le\sum_{\bm n'\in\tilde{\lg}_s,\bm n''\notin\tilde{\lg}_s}|\mathcal{T}_{\tilde{\lg}_s}^{-1}(E;\theta)(\bm 0,\bm n')|\cdot|\phi(\bm n'-\bm n'')|\cdot|\xi_{\bm n''}(\theta)|\\
			&\le(I)+(II), 
		\end{align*}
		where
		\begin{align*}
			(I)&=\sum_{\bm n'\in\tilde{\lg}_{s,0},\bm n''\notin\tilde{\lg}_s}|\mathcal{T}_{\tilde{\lg}_s}^{-1}(E;\theta)(\bm 0,\bm n')|\cdot|\phi(\bm n'-\bm n'')|\cdot|\xi_{\bm n''}(\theta)|,\\
			(II)&=\sum_{\bm n'\in\tilde{\lg}_s\setminus\tilde{\lg}_{s,0},\bm n''\notin\tilde{\lg}_s}|\mathcal{T}_{\tilde{\lg}_s}^{-1}(E;\theta)(\bm 0,\bm n')|\cdot|\phi(\bm n'-\bm n'')|\cdot|\xi_{\bm n''}(\theta)|.
		\end{align*}
		For $(I)$, we have by Theorem \ref{ge}, \eqref{oddgf} and \eqref{xizz} that 
		\begin{align*}
			(I)&\le C(\theta,\alpha_1,d)(N_s^{29})^{-\alpha_1+d}\|\mathcal T_{\tilde{\lg}_s}^{-1}\|_{d}\|\phi\|_{\alpha_1}\\
			&\le  C(\theta,\alpha_1,d)\delta_s^{\frac{\alpha_1}{36}}\rightarrow0\ \text{as $s\rightarrow\infty$}.
		\end{align*}
		For $(II)$, we also have by Theorem \ref{ge}, \eqref{oddgf} and \eqref{xizz} that 
		\begin{align*}
			(II)&\le C(\theta,\alpha_1,d)(N_s^{29})^{-\alpha_1+d}\|\mathcal T_{\tilde{\lg}_s}^{-1}\|_{\alpha_1}\|\phi\|_{d}\\
			&\le  C(\theta,\alpha_1,d)\delta_s^{\frac{\alpha_1}{36}}\rightarrow0\ \text{as $s\rightarrow\infty$}.
		\end{align*}
		This implies $F(\theta)=0$ for $\theta\in\T\setminus(\mathscr{K}\cup\mathscr{L})$. Thus $\|F\|_{L^2(\T)}=0$, which contradicts \eqref{pai}.
		
		This proves Theorem \ref{aps}.
	\end{proof}

	\section*{Acknowledgments}
This work  was  partially supported by NSFC  (No. 12522110). The authors would like to thank the reviewer for helpful suggestions.
\section*{Data Availability}
The manuscript has no associated data.
\section*{Declarations}
{\bf Conflicts of interest} \ The authors  state  that there is no conflict of interest.

	\newpage

	\appendix{}
	\section{}\label{app}
	\begin{proof}[Proofs  concerning Remark \ref{v1}]
		First, we verify functions in \textit{Example 1} belong to class $\mathscr V.$
		We have 
		\begin{align*}
			2\le\frac{|\sin\pi z|}{\|z\|_\T}\le\pi.
		\end{align*}
		By the continuity of $\frac{|\sin\pi z|}{\|z\|_\T}$ and the compactness of $\T$, there is some $R>0$ such that
		\begin{align*}
			1\le\frac{|\sin\pi z|}{\|z\|_\T}\le 4,\ \forall z\in\D_{2R}.
		\end{align*}
		Therefore, for $z,z'\in\D_R$,
		\begin{align*}
			|\cos2\pi z-\cos2\pi z'|&=2|\sin\pi(z+z')|\cdot|\sin\pi(z-z')|\ge2\|z+z'\|_\T\|z-z'\|_\T,\\
			|\cos2\pi z-\cos2\pi z'|&=2|\sin\pi(z+z')|\cdot |\sin\pi(z-z')|\le32\|z+z'\|_\T\|z-z'\|_\T.
		\end{align*}
		For 
		\begin{align*}
			v_1(z)=\cos2\pi z+\lambda_2\cos^2 2\pi z+\cdots+\lambda_n\cos^n 2\pi z,
		\end{align*}
		we have
		\begin{align*}
			v_1(z)-v_1(z')=(\cos2\pi z-\cos2\pi z')\left(1+\sum_{k=2}^{n}\lambda_{k}\left(\sum_{l=0}^{k-1}\cos^{l}2\pi z\cos^{k-1-l}2\pi z'\right)\right).
		\end{align*}
		Let
		\begin{align*}
			u_1(z,z')=\sum_{k=2}^{n}\lambda_{k}\left(\sum_{l=0}^{k-1}\cos^{l}2\pi z\cos^{k-1-l}2\pi z'\right).
		\end{align*}
		From $\sup\limits_{z\in\T}|\cos2\pi z|\le1$ and $2|\lambda_2|+\cdots+n|\lambda_n|<1$, we get
		\begin{align*}
			\sup_{(z,z')\in\T\times\T}|u_1(z,z')|\le\sum_{k=2}^n k|\lambda_k|<1.
		\end{align*}
		By the continuity of $u_1$ and the compactness of $\T$, there is some $0<R_1<R$ such that
		\begin{align*}
			\sup_{(z,z')\in\D_{R_1}\times\D_{R_1}}|u_1(z,z')|<1.
		\end{align*}
		Hence, for $z,z'\in \mathbb D_{R_1}$,
		\begin{align*}
			|v_1(z)-v_1(z')|&\ge2\left(1-\sup_{(z,z')\in\D_{R_1}\times\D_{R_1}}|u_1(z,z')|\right)\|z+z'\|_\T\|z-z'\|_\T,\\
			|v_1(z)-v_1(z')|&\le64\|z+z'\|_\T\|z-z'\|_\T,
		\end{align*}
		which shows $v_1\in\mathscr V.$

		Next, we verify functions in \textit{Example 2} belong to class $\mathscr V.$   For
		\begin{align*}
			v_2(z)=\cos2\pi z+\epsilon f(z),
		\end{align*}
		we  get
		\begin{align*}
			v_2(z)-v_2(z')=(\cos2\pi z-\cos2\pi z')\left(1+\epsilon\frac{f(z)-f(z')}{\cos2\pi z-\cos2\pi z'}\right).
		\end{align*}
		Let
		\begin{align*}
			u_2(z,z')=\frac{f(z)-f(z')}{\cos2\pi z-\cos2\pi z'}.
		\end{align*}
		Since $f$ is $1$-periodic and even, we have $f(z)=f(z+j)$ and $f(z)=f(-z+j)$ for $\forall j\in\Z$, which implies $u_2(z,z')$ only has removable singular points. Hence $u_2(z,z')$ can be  continuous on $\D_R\times \D_R$ and there exists some $C=C(f)>0$ such that
		\begin{align*}
			\sup_{(z,z')\in\D_{R}\times\D_{R}}\left|u_2(z,z')\right|\le C.
		\end{align*}
		Therefore, for $z,z'\in \mathbb D_{R}$, one has since  $|\epsilon| C\le\frac{1}{2},$
		\begin{align*}
			\|z+z'\|_\T\|z-z'\|_\T\le|v_2(z)-v_2(z')|\le48\|z+z'\|_\T\|z-z'\|_\T,
		\end{align*}
		which implies $v_2\in\mathscr V$ for $0<|\epsilon|\ll1.$
	\end{proof}

	The following elementary inequality plays an important role in the proof of our tame estimate.
	\begin{lem}\label{ee}
		Let $K(n,\alpha)$ be given by \eqref{kns}. For $\alpha\ge0$, $x_i\ge0$ and $1\le i\le n$, we have
		\begin{align}\label{kn}
			\left(\sum_{i=1}^n x_i\right)^\alpha\le K(n,\alpha)\left(\sum_{i=1}^n x_i^\alpha\right). 
		\end{align}
	\end{lem}
	\begin{proof}[Proof of Lemma \ref{ee}]
		The proof of case $0\leq \alpha\leq 1$ is trivial and we omit the details.

		In the following, we only consider the case $\alpha>1.$  We proceed by   induction on $n.$
		The lemma holds trivially for $n=1$. Now, we assume \eqref{ee} holds true for $n=m$ and we will prove   it  for  $n=m+1$.  When $n=m+1$, we define for  $\alpha>1$ the function 
		\begin{align*}
			f_\alpha(x_1,\cdots,x_m,x_{m+1})=\left(\sum_{i=1}^{m+1} x_i\right)^\alpha-K(m+1,\alpha)\left(\sum_{i=1}^{m+1} x_i^\alpha\right),\ x_i\ge0,\ 1\le i\le m+1. 
		\end{align*}
		%where $K(n,\alpha)$ is defined on \eqref{kns}. 
		We have
		\begin{align*}
			\p_{x_{m+1}}f_\alpha(x_1,\cdots,x_m,x_{m+1})=\alpha\left(\sum_{i=1}^{m+1} x_i\right)^{\alpha-1}-\alpha K(m+1,\alpha)x_{m+1}^{\alpha-1}.
		\end{align*}
		Given $x_i^0\ge0$ ($1\le i\le m$), we have 
		\begin{align*}
			\p_{x_{m+1}}f_\alpha(x_1^0,\cdots,x_m^0,x_{m+1})=\alpha\left(\left(\sum_{i=1}^m x_i^0+x_{m+1}\right)^{\alpha-1}-((m+1)x_{m+1})^{\alpha-1}\right).
		\end{align*}
		Since $x^{\alpha-1}$ ($\alpha>1$) is non-decreasing  on $[0,+\infty)$, we obtain 
		\begin{align*}
			\p_{x_{m+1}}f_\alpha(x_1^0,\cdots,x_m^0,x_{m+1})&\ge0,\ 0\le x_{m+1}\le\frac{\sum_{i=1}^m x_i^0}{m},\\
			\p_{x_{m+1}}f_\alpha(x_1^0,\cdots,x_m^0,x_{m+1})&\le0,\ x_{m+1}>\frac{\sum_{i=1}^m x_i^0}{m}.
		\end{align*}
		%Then $f_\alpha(x_1^0,\cdots,x_m^0,x_{m+1})$ is increasing on $\left[0,\frac{\sum_{i=1}^m x_i^0}{m}\right]$ and decreasing on $\left(\frac{\sum_{i=1}^m x_i^0}{m},+\infty\right)$. 
		As a result,  and since \eqref{ee} holds for $n=m$,  we get for all $x_{m+1}\geq 0, $
		\begin{align*}
			f_\alpha(x_1^0,\cdots,x_m^0,x_{m+1})&\le f_\alpha\left(x_1^0,\cdots,x_m^0,\frac{\sum_{i=1}^m x_i^0}{m}\right)\\
			&=\left(\frac{m+1}{m}\right)^{\alpha-1}\left(\left(\sum_{i=1}^m x_i^0\right)^\alpha-K(m,\alpha)\sum_{i=1}^m (x_i^0)^\alpha\right)\\
			&\le0.
		\end{align*}
		Therefore, for  $\forall x_i\in[0,+\infty)$, $1\le i\le m+1$, we have 
		\begin{align*}
			f_\alpha(x_1,\cdots,x_m,x_{m+1})\le\sup_{x_i^0\ge0 \atop 1\le i\le m}f_\alpha(x_1^0,\cdots,x_m^0,x_{m+1})\le0, 
		\end{align*}
		which proves \eqref{ee} for $n=m+1.$
		
		We thus complete the proof. 
		
		%Similarly, we can obtain $f_\alpha(x_1,\cdots,x_m,x_{m+1})\le0$ for $0\le \alpha\le1$. By induction, $\forall n\in\N$,
		%\begin{align*}
		%\left(\sum_{i=1}^n x_i\right)^\alpha\le K(n,\alpha)\left(\sum_{i=1}^n x_i^\alpha\right).
		%\end{align*}
	\end{proof}
	
	\begin{proof}[Proof of Lemma \ref{tp}]
		Using  \eqref{ee} yields for any $\bm l, \bm k_1, \cdots, \bm k_{n-1}\in \Z^d, $
		\begin{align*}
			(1+\|\bm l\|)^\alpha
			& \leq K(n, \alpha)\left((1+\|\bm l-\bm k_1\|)^\alpha+(1+\|\bm k_1-\bm k_2\|)^\alpha+\cdots+(1+\|\bm k_{n-1}\|)^\alpha\right).
		\end{align*}
		Then from the definition \eqref{norm}, we get
		\begin{align*}
			\|\prod_{i=1}^n \mathcal{M}_i\|_\alpha&=\sum_{\bm l\in\Z^d}\left(\sup_{\bm k\in\Z^d}\left|\left(\prod_{i=1}^n \mathcal{M}_i\right)(\bm k+\bm l,\bm k)\right|\right)(1+\|\bm l\|)^\alpha\\
			&\le K(n,\alpha)\sum_{\bm k_{n-1}\in\Z^d}\cdots\sum_{\bm k_2\in\Z^d}\sum_{\bm k_{1}\in\Z^d}\sum_{\bm l\in\Z^d}\left(\sup_{\bm k\in\Z^d}|\mathcal{M}_1(\bm k+\bm l,\bm k+\bm k_1)|\right)\\
			&\ \ \cdot\left(\sup_{\bm k\in\Z^d}|\mathcal{M}_2(\bm k+\bm k_1,\bm k+\bm k_2)|\right)\cdots\left(\sup_{\bm k\in\Z^d}|\mathcal{M}_n(\bm k+\bm k_{n-1},\bm k)|\right)\\
			&\ \ \times \left((1+\|\bm l-\bm k_1\|)^\alpha+(1+\|\bm k_1-\bm k_2\|)^\alpha+\cdots+(1+\|\bm k_{n-1}\|)^\alpha\right)\\
			&\le K(n,\alpha)\sum_{i=1}^n\left(\prod_{j\ne i}\|\mathcal{M}_j\|_0\right)\|\mathcal{M}_i\|_\alpha.
		\end{align*}
	\end{proof}
	
	\begin{proof}[Proof of Lemma \ref{chi}]
		By Hadamard's inequality, we have for any $\bm i,\bm j\in\lg, $
		\begin{align*}
			|\ji \delta_{\bm i}, \mathcal{S}_{\lg}^* \delta_{\bm j}\jd|&\le \prod_{\bm l\ne\bm i}\left(\sum_{\bm k\ne\bm j}|\ji \delta_{\bm l}, \mathcal{S}_{\lg}\delta_{\bm k}\jd|^2\right)^{\frac{1}{2}}\\
			&\le \prod_{\bm l\ne\bm i}\left(\sum_{\bm k\ne\bm j}|\ji \delta_{\bm l}, \mathcal{S}_{\lg}\delta_{\bm k}\jd|\right)\ ({\rm since}\ \eqref{ee})\\
			&\le \|\mathcal{S}_{\lg}\|_0^{\#\lg-1}.
		\end{align*}
		Moreover, we have 
		\begin{align*}
			\|\mathcal{S}_{\lg}^*\|_0\le\sum_{\bm i, \bm j\in\lg}|\ji \delta_{\bm i}, \mathcal{S}_{\lg}^* \delta_{\bm j}\jd|\le (\#\lg)^2\|\mathcal{S}_{\lg}\|_0^{\#\lg-1}. 
		\end{align*}
	\end{proof}
	
	\begin{proof}[Proof concerning  Remark \ref{dsq1}]
		Let $\bm i\in Q_0^+$ and $\bm j\in\tilde{Q}_0^-$ satisfy
		\begin{align*}
			\|\theta+\bm i\cdot\bm \omega+\theta_0\|_{\T}<\delta_0,\ \|\theta+\bm j\cdot\bm \omega-\theta_0\|_{\T}<\delta_0^{\frac{2}{3}}.
		\end{align*}
		Then \eqref{dc} implies that $1,\omega_1,\cdots,\omega_d$ are rationally independent and $\{\bm k\cdot\bm\omega\}_{k\in\Z^d}$ is dense in $\T$. Thus, there exist a $\bm k\in\Z^d$ such that $\|2\theta+\bm k\cdot\bm\omega\|_{\T}$ is sufficiently small with
		\begin{align*}
			\|\theta+(\bm k-\bm j)\cdot\bm \omega+\theta_0\|_{\T}&\le\|2\theta+\bm k\cdot\bm\omega\|_{\T}+\|\theta+\bm j\cdot\bm \omega-\theta_0\|_{\T}<\delta_0^{\frac{2}{3}},\\
			\|\theta+(\bm k-\bm i)\cdot\bm \omega-\theta_0\|_{\T}&\le\|2\theta+\bm k\cdot\bm\omega\|_{\T}+\|\theta+\bm i\cdot\bm \omega+\theta_0\|_{\T}<\delta_0.
		\end{align*}
		We obtain $\bm k-\bm j\in\tilde{Q}_0^+$ and $\bm k-\bm i\in Q_0^-$, which implies
		\begin{align*}
			\text{dist}\left(\tilde{Q}_0^+,Q_0^-\right)\le\text{dist}\left(\tilde{Q}_0^-,Q_0^+\right).
		\end{align*}
		The similar argument shows
		\begin{align*}
			\text{dist}\left(\tilde{Q}_0^+,Q_0^-\right)\ge\text{dist}\left(\tilde{Q}_0^-,Q_0^+\right).
		\end{align*}
		We have shown
		\begin{align*}
			\text{dist}\left(\tilde{Q}_0^+,Q_0^-\right)=\text{dist}\left(\tilde{Q}_0^-,Q_0^+\right).
		\end{align*}
	\end{proof}
	
	\begin{lem}[Schur complement lemma]\label{scl}
		Let $\lg_1$ and $\lg_2$ be finite subsets of $\Z^d$ with $\lg_1\cap\lg_2=\emptyset$. Suppose $\mathcal{A}\in\mathbf{M}_{\lg_1}^{\lg_1}$, $\mathcal{B}\in\mathbf{M}_{\lg_2}^{\lg_1}$, $\mathcal{C}\in\mathbf{M}_{\lg_1}^{\lg_2}$, $\mathcal{D}\in\mathbf{M}_{\lg_2}^{\lg_2}$ and
		\begin{align*}
			\mathcal{M}=\left(\begin{array}{cc}
				\mathcal{A} & \mathcal{B}\\
				\mathcal{C} & \mathcal{D}
			\end{array}\right)\in\mathbf{M}_{\lg_1\cup\lg_2}^{\lg_1\cup\lg_2}.
		\end{align*}
		Assume further that $\mathcal{A}$ is invertible and $\|\mathcal{B}\|_0,\|\mathcal{C}\|_0\le1$. Then we have
		\begin{itemize}
			\item[(1)]
			\begin{align*}
				\det \mathcal{M}=\det \mathcal{A}\cdot\det \mathcal{S},
			\end{align*}
			where
			\begin{align*}
				\mathcal{S}=\mathcal{D}-\mathcal{C}\mathcal{A}^{-1}\mathcal{B}\in\mathbf{M}_{\lg_2}^{\lg_2}
			\end{align*}
			is called the Schur complement of $\mathcal{A}$.
			\item[(2)] $\mathcal{M}$ is invertible iff $\mathcal{S}$ is invertible and
			\begin{align}\label{sc}
				\|\mathcal{S}^{-1}\|_0\le\|\mathcal{M}^{-1}\|_0<4(1+\|\mathcal{A}^{-1}\|_0)^2(1+\|\mathcal{S}^{-1}\|_0).
			\end{align}
		\end{itemize}
	\end{lem}
	\begin{proof}[Proof of Lemma \ref{scl}]
		(1) Since $\mathcal{A}$ is invertible, we have
		\begin{align*}
			\left(\begin{array}{cc}
				\mathcal{I}_{\lg_1} & \bm 0\\
				-\mathcal{C}\mathcal{A}^{-1}& \mathcal{I}_{\lg_2}
			\end{array} \right)\mathcal{M} \left(\begin{array}{cc}
				\mathcal{I}_{\lg_1} & -\mathcal{A}^{-1}\mathcal{B}\\
				\bm 0 & \mathcal{I}_{\lg_2}
			\end{array}\right)=\left(\begin{array}{cc}
				\mathcal{A} & \bm 0\\
				\bm 0 & \mathcal{S}
			\end{array}\right),
		\end{align*}
		which implies
		\begin{align*}
			\det\mathcal{M}=\det\mathcal{A}\cdot\det\mathcal{S}.
		\end{align*}
		
		(2) Direct computation shows
		\begin{align*}
			\mathcal{M}^{-1}=\left(\begin{array}{cc}
				\mathcal{A}^{-1}+\mathcal{A}^{-1}\mathcal{B}\mathcal{S}^{-1}\mathcal{C}\mathcal{A}^{-1} & -\mathcal{A}^{-1}\mathcal{B}\mathcal{S}^{-1}\\
				-\mathcal{S}^{-1}\mathcal{C}\mathcal{A}^{-1} & \mathcal{S}^{-1}
			\end{array}\right),
		\end{align*}
		
		which combines \eqref{tame} implying \eqref{sc}.
	\end{proof}
	
	\begin{lem}\label{ef}
		Let $\bm l\in\frac{1}{2}\Z^d$ and let $\lg\subset\Z^d+\bm l$ be a finite set which is symmetrical about the origin (i.e., $\bm n\in\lg\Leftrightarrow-\bm n\in\lg$). Then
		\begin{align*}
			\det \mathcal{T}_{\lg}(z)=\det((v(z+\bm n\cdot\bm\omega)-E)\delta_{\bm n,\bm n'}+\ep \mathcal{W})_{\bm n\in\lg}
		\end{align*}
		is an even function of $z$.
	\end{lem}
	\begin{proof}[Proof of Lemma \ref{ef}]
		Define the unitary map
		\begin{align*}
			\mathcal{U}_\lg:\ell^2(\lg)\rightarrow\ell^2(\lg)\ \text{with}\ (\mathcal{U}_{\lg}\psi)(\bm n)=\psi(-\bm n).
		\end{align*}
		Then
		\begin{align*}
			\mathcal{U}_{\lg}^{-1}\mathcal{T}_{\lg}(z)\mathcal{U}_{\lg}=((v(z-\bm n\cdot\bm\omega)-E)\delta_{\bm n,\bm n'}+\ep \mathcal{W})_{n\in\lg}=\mathcal{T}_{\lg}(-z),
		\end{align*}
		which implies
		\begin{align*}
			\det \mathcal{T}_{\lg}(z)=\det \mathcal{T}_{\lg}(-z).
		\end{align*}
	\end{proof}
	
	\begin{lem}\label{det1}
		Let $\mathcal{A},\mathcal{B}:\ell^2(\Z^d)\rightarrow\ell^2(\Z^d)$ be linear operators and let $\lg$ be a finite subset of $\Z^d$. If $\|\mathcal{A}_\lg\|_0\le M$ and $\|\mathcal{B}_{\lg}\|_0\le \ep$, then
		\begin{align}\label{detd}
			\left|\det(\mathcal{A}_{\lg}+\mathcal{B}_{\lg})-\det \mathcal{A}_{\lg}\right|\le \ep (\#\lg)^2(M+\ep)^{\#\lg-1}.
		\end{align}
	\end{lem}
	\begin{proof}
		Let $f(t)=\det(\mathcal{A}_{\lg}+t\mathcal{B}_{\lg})$. Then
		\begin{align*}
			f'(t)={\rm tr}(\mathcal{B}_{\lg}(\mathcal{A}_{\lg}+t\mathcal{B}_{\lg})^*).
		\end{align*}
		Since $\|\mathcal{A}_{\lg}\|_0\le M$ and $|\mathcal{B}_{\lg}\|_0\le \ep$, we have
		\begin{align*}
			\|\mathcal{A}_{\lg}+t\mathcal{B}_{\lg}\|_0\le \|\mathcal{A}_{\lg}\|_0+|t|\cdot\|\mathcal{B}_{\lg}\|_0\le M+\ep |t|.
		\end{align*}
		By Lemma \ref{chi}, we get for any $\bm i,\bm j\in\lg$,
		\begin{align*}
			|\ji \delta_{\bm i}, (\mathcal{A}_{\lg}+t\mathcal{B}_{\lg})^* \delta_{\bm j}\jd|\le \|\mathcal{A}_{\lg}+t\mathcal{B}_{\lg}\|_0^{\#\lg-1}\le (M+\ep |t|)^{\#\lg-1}.
		\end{align*}
		Therefore,
		\begin{align*}
			|f'(t)|&\le \sum_{\bm i\in\lg}|\ji \delta_{\bm i},\mathcal{B}_{\lg}(\mathcal{A}_{\lg}+t\mathcal{B}_{\lg})^* \delta_{\bm i}\jd|\\
			&\le\ep (\#\lg)^2\max_{\bm i,\bm j\in\lg}|\ji \delta_{\bm i}, (\mathcal{A}_{\lg}+t\mathcal{B}_{\lg})^* \delta_{\bm j}\jd|\\
			&\le \ep(\#\lg)^2(M+\ep |t|)^{\#\lg-1}.
		\end{align*}
		According to the mean-value theorem, we obtain for some $\xi\in(0,1),$
		\begin{align*}
			|\det(\mathcal{A}_{\lg}+\mathcal{B}_{\lg})-\det \mathcal{A}_{\lg}|=|f(1)-f(0)|=|f'(\xi)|\le \ep (\#\lg)^2(M+\ep)^{\#\lg-1}.
		\end{align*}
	\end{proof}
	
	\bibliographystyle{alpha}
	%\bibliography{Powerlaw}

\begin{thebibliography}{CSZ24b}

\bibitem[AJ10]{AJ10}
A.~Avila and S.~Jitomirskaya.
\newblock Almost localization and almost reducibility.
\newblock {\em J. Eur. Math. Soc. (JEMS)}, 12(1):93--131, 2010.

\bibitem[AM93]{AM93}
M.~Aizenman and S.~Molchanov.
\newblock Localization at large disorder and at extreme energies: an elementary
  derivation.
\newblock {\em Comm. Math. Phys.}, 157(2):245--278, 1993.

\bibitem[Amo09]{Amo09}
S.~Amor.
\newblock H\"{o}lder continuity of the rotation number for quasi-periodic
  co-cycles in {${\rm SL}(2,\Bbb R)$}.
\newblock {\em Comm. Math. Phys.}, 287(2):565--588, 2009.

\bibitem[AYZ17]{AYZ17}
A.~Avila, J.~You, and Q.~Zhou.
\newblock Sharp phase transitions for the almost {M}athieu operator.
\newblock {\em Duke Math. J.}, 166(14):2697--2718, 2017.

\bibitem[BB12]{BB12}
M.~Berti and P.~Bolle.
\newblock Sobolev quasi-periodic solutions of multidimensional wave equations
  with a multiplicative potential.
\newblock {\em Nonlinearity}, 25(9):2579--2613, 2012.

\bibitem[BB13]{BB13}
M.~Berti and P.~Bolle.
\newblock Quasi-periodic solutions with {S}obolev regularity of {NLS} on {$\Bbb
  T^d$} with a multiplicative potential.
\newblock {\em J. Eur. Math. Soc. (JEMS)}, 15(1):229--286, 2013.

\bibitem[BB20]{BB20}
M.~Berti and P.~Bolle.
\newblock {\em Quasi-periodic solutions of nonlinear wave equations on the
  {$d$}-dimensional torus}.
\newblock EMS Monographs in Mathematics. EMS Publishing House, Berlin, [2020]
  \copyright 2020.

\bibitem[BCP15]{BCP15}
M.~Berti, L.~Corsi, and M.~Procesi.
\newblock An abstract {N}ash-{M}oser theorem and quasi-periodic solutions for
  {NLW} and {NLS} on compact {L}ie groups and homogeneous manifolds.
\newblock {\em Comm. Math. Phys.}, 334(3):1413--1454, 2015.

\bibitem[BG00]{BG00}
J.~Bourgain and M.~Goldstein.
\newblock On nonperturbative localization with quasi-periodic potential.
\newblock {\em Ann. of Math. (2)}, 152(3):835--879, 2000.

\bibitem[BGS02]{BGS02}
J.~Bourgain, M.~Goldstein, and W.~Schlag.
\newblock Anderson localization for {S}chr\"{o}dinger operators on {$\bold
  Z^2$} with quasi-periodic potential.
\newblock {\em Acta Math.}, 188(1):41--86, 2002.

\bibitem[BJ02]{BJ02}
J.~Bourgain and S.~Jitomirskaya.
\newblock Absolutely continuous spectrum for 1{D} quasiperiodic operators.
\newblock {\em Invent. Math.}, 148(3):453--463, 2002.

\bibitem[Bou97]{Bou97}
J.~Bourgain.
\newblock On {M}elnikov's persistency problem.
\newblock {\em Math. Res. Lett.}, 4(4):445--458, 1997.

\bibitem[Bou98]{Bou98}
J.~Bourgain.
\newblock Quasi-periodic solutions of {H}amiltonian perturbations of 2{D}
  linear {S}chr\"{o}dinger equations.
\newblock {\em Ann. of Math. (2)}, 148(2):363--439, 1998.

\bibitem[Bou00]{Bou00}
J.~Bourgain.
\newblock H\"{o}lder regularity of integrated density of states for the almost
  {M}athieu operator in a perturbative regime.
\newblock {\em Lett. Math. Phys.}, 51(2):83--118, 2000.

\bibitem[Bou05]{Bou05}
J.~Bourgain.
\newblock {\em Green's function estimates for lattice {S}chr\"{o}dinger
  operators and applications}, volume 158 of {\em Annals of Mathematics
  Studies}.
\newblock Princeton University Press, Princeton, NJ, 2005.

\bibitem[Bou07]{Bou07}
J.~Bourgain.
\newblock Anderson localization for quasi-periodic lattice {S}chr\"{o}dinger
  operators on {$\Bbb Z^d$}, {$d$} arbitrary.
\newblock {\em Geom. Funct. Anal.}, 17(3):682--706, 2007.

\bibitem[CD93]{CD93}
V.~A. Chulaevsky and E.~I. Dinaburg.
\newblock Methods of {KAM}-theory for long-range quasi-periodic operators on
  {${\bf Z}^\nu$}. {P}ure point spectrum.
\newblock {\em Comm. Math. Phys.}, 153(3):559--577, 1993.



\bibitem[CSZ23]{CSZ23}
H.~Cao, Y.~Shi, and Z.~Zhang.
\newblock Localization and regularity of the integrated density of states for
  {S}chr\"{o}dinger operators on {$\Bbb Z^d$} with {$C^2$}-cosine like
  quasi-periodic potential.
\newblock {\em Comm. Math. Phys.}, 404(1):495--561, 2023.

\bibitem[CSZ24a]{CSZ24b}
H.~Cao, Y.~Shi, and Z.~Zhang.
\newblock On the spectrum of quasi-periodic {S}chr\"{o}dinger operators on
  $\mathbb{Z}^d$ with {${C}^2$}-cosine type potentials.
\newblock {\em Comm. Math. Phys.}, 405(8):Paper No. 174, 84, 2024.

\bibitem[CSZ24b]{CSZ24a}
H.~Cao, Y.~Shi, and Z.~Zhang.
\newblock Quantitative {G}reen's function estimates for lattice quasi-periodic
  {S}chr\"{o}dinger operators.
\newblock {\em Sci. China Math.}, 67(5):1011--1058, 2024.

\bibitem[CW93]{CW93}
W.~Craig and C.~E. Wayne.
\newblock Newton's method and periodic solutions of nonlinear wave equations.
\newblock {\em Comm. Pure Appl. Math.}, 46(11):1409--1498, 1993.

\bibitem[Din97]{Din97}
E.~I. Dinaburg.
\newblock Some problems in the spectral theory of discrete operators with
  quasiperiodic coefficients.
\newblock {\em Uspekhi Mat. Nauk}, 52(3(315)):3--52, 1997.

\bibitem[Eli92]{Eli92}
L.~H. Eliasson.
\newblock Floquet solutions for the {$1$}-dimensional quasi-periodic
  {S}chr\"{o}dinger equation.
\newblock {\em Comm. Math. Phys.}, 146(3):447--482, 1992.

\bibitem[Eli97]{Eli97}
L.~H. Eliasson.
\newblock Discrete one-dimensional quasi-periodic {S}chr\"{o}dinger operators
  with pure point spectrum.
\newblock {\em Acta Math.}, 179(2):153--196, 1997.

\bibitem[FS83]{FS83}
J.~Fr\"{o}hlich and T.~Spencer.
\newblock Absence of diffusion in the {A}nderson tight binding model for large
  disorder or low energy.
\newblock {\em Comm. Math. Phys.}, 88(2):151--184, 1983.

\bibitem[FSW90]{FSW90}
J.~Fr\"{o}hlich, T.~Spencer, and P.~Wittwer.
\newblock Localization for a class of one-dimensional quasi-periodic
  {S}chr\"{o}dinger operators.
\newblock {\em Comm. Math. Phys.}, 132(1):5--25, 1990.

\bibitem[FV25]{FV25}
Y.~Forman and T.~VandenBoom.
\newblock Localization and {C}antor spectrum for quasiperiodic discrete
  {S}chr\"odinger operators with asymmetric, smooth, cosine-like sampling
  functions.
\newblock {\em Mem. Amer. Math. Soc.}, 312(1583):v+86, 2025.

\bibitem[GJ24]{GJ24}
L.~Ge and S.~Jitomirskaya.
\newblock {H}idden subcriticality, symplectic structure, and universality of
  sharp arithmetic spectral results for type {I} operators.
\newblock {\em arXiv:2407.08866}, 2024.

\bibitem[GS01]{GS01}
M.~Goldstein and W.~Schlag.
\newblock H\"{o}lder continuity of the integrated density of states for
  quasi-periodic {S}chr\"{o}dinger equations and averages of shifts of
  subharmonic functions.
\newblock {\em Ann. of Math. (2)}, 154(1):155--203, 2001.

\bibitem[GS08]{GS08}
M.~Goldstein and W.~Schlag.
\newblock Fine properties of the integrated density of states and a
  quantitative separation property of the {D}irichlet eigenvalues.
\newblock {\em Geom. Funct. Anal.}, 18(3):755--869, 2008.

\bibitem[GY20]{GY20}
L.~Ge and J.~You.
\newblock Arithmetic version of {A}nderson localization via reducibility.
\newblock {\em Geom. Funct. Anal.}, 30(5):1370--1401, 2020.

\bibitem[GYZ23]{GYZ23}
L.~Ge, J.~You, and Q.~Zhou.
\newblock Exponential dynamical localization: criterion and applications.
\newblock {\em Ann. Sci. \'{E}c. Norm. Sup\'{e}r. (4)}, 56(1):91--126, 2023.

\bibitem[Han19]{Han19}
R.~Han.
\newblock Shnol's theorem and the spectrum of long range operators.
\newblock {\em Proc. Amer. Math. Soc.}, 147(7):2887--2897, 2019.

\bibitem[Han24]{Han24}
R.~Han.
\newblock {S}harp localization on the first supercritical stratum for
  {L}iouville frequencies.
\newblock {\em arXiv:2405.07810}, 2024.

\bibitem[HS22]{HS22}
R.~Han and W.~Schlag.
\newblock {A}vila's acceleration via zeros of determinants, and applications to
  {S}chr\"odinger cocycles.
\newblock {\em arXiv:2212.05988}, 2022.

\bibitem[Jit94]{Jit94}
S.~Jitomirskaya.
\newblock Anderson localization for the almost {M}athieu equation: a
  nonperturbative proof.
\newblock {\em Comm. Math. Phys.}, 165(1):49--57, 1994.

\bibitem[Jit99]{Jit99}
S.~Jitomirskaya.
\newblock Metal-insulator transition for the almost {M}athieu operator.
\newblock {\em Ann. of Math. (2)}, 150(3):1159--1175, 1999.

\bibitem[JK16]{JK16}
S.~Jitomirskaya and I.~Kachkovskiy.
\newblock {$L^2$}-reducibility and localization for quasiperiodic operators.
\newblock {\em Math. Res. Lett.}, 23(2):431--444, 2016.

\bibitem[JL18]{JL18}
S.~Jitomirskaya and W.~Liu.
\newblock Universal hierarchical structure of quasiperiodic eigenfunctions.
\newblock {\em Ann. of Math. (2)}, 187(3):721--776, 2018.

\bibitem[JL24]{JL24}
S.~Jitomirskaya and W.~Liu.
\newblock Universal reflective-hierarchical structure of quasiperiodic
  eigenfunctions and sharp spectral transition in phase.
\newblock {\em J. Eur. Math. Soc. (JEMS)}, 26(8):2797--2836, 2024.

\bibitem[JLS20]{JLS20}
S.~Jitomirskaya, W.~Liu, and Y.~Shi.
\newblock Anderson localization for multi-frequency quasi-periodic operators on
  {${\Bbb Z}^D$}.
\newblock {\em Geom. Funct. Anal.}, 30(2):457--481, 2020.

\bibitem[JSY19]{JSY19}
W.~Jian, Y.~Shi, and X.~Yuan.
\newblock Anderson localization for one-frequency quasi-periodic block
  operators with long-range interactions.
\newblock {\em J. Math. Phys.}, 60(6):063504, 15, 2019.

\bibitem[Kle05]{Kle05}
S.~Klein.
\newblock Anderson localization for the discrete one-dimensional quasi-periodic
  {S}chr\"{o}dinger operator with potential defined by a {G}evrey-class
  function.
\newblock {\em J. Funct. Anal.}, 218(2):255--292, 2005.

\bibitem[Liu22]{Liu22}
W.~Liu.
\newblock Quantitative inductive estimates for {G}reen's functions of
  non-self-adjoint matrices.
\newblock {\em Anal. PDE}, 15(8):2061--2108, 2022.

\bibitem[Liu23]{Liu23}
W.~Liu.
\newblock {S}mall denominators and large numerators of quasiperiodic
  {S}chr{\"o}dinger operators.
\newblock {\em Peking Mathematical Journal}, pages 1--30, 2023.

\bibitem[Shi21]{Shi21}
Y.~Shi.
\newblock A multi-scale analysis proof of the power-law localization for random
  operators on {$\Bbb{Z}^d$}.
\newblock {\em J. Differential Equations}, 297:201--225, 2021.

\bibitem[Shi22]{Shi22}
Y.~Shi.
\newblock Spectral theory of the multi-frequency quasi-periodic operator with a
  {G}evrey type perturbation.
\newblock {\em J. Anal. Math.}, 148(1):305--338, 2022.

\bibitem[Shi23]{Shi23}
Y.~Shi.
\newblock Localization for almost-periodic operators with power-law long-range
  hopping: a {N}ash-{M}oser iteration type reducibility approach.
\newblock {\em Comm. Math. Phys.}, 402(2):1765--1806, 2023.

\bibitem[Sin87]{Sin87}
Y.~G. Sinai.
\newblock Anderson localization for one-dimensional difference
  {S}chr\"{o}dinger operator with quasiperiodic potential.
\newblock {\em J. Statist. Phys.}, 46(5-6):861--909, 1987.

\bibitem[SW86]{SW86}
B.~Simon and T.~Wolff.
\newblock Singular continuous spectrum under rank one perturbations and
  localization for random {H}amiltonians.
\newblock {\em Comm. Pure Appl. Math.}, 39(1):75--90, 1986.

\bibitem[SW23]{SW23}
Y.~Shi and L.~Wen.
\newblock Diagonalization in a quantum kicked rotor model with non-analytic
  potential.
\newblock {\em J. Differential Equations}, 355:334--368, 2023.

\end{thebibliography}

\end{document}